\documentclass[aoas,preprint]{imsart}

\usepackage{amsmath,amssymb,amsthm,bm}
\usepackage{graphicx}
\usepackage[table]{xcolor}
\usepackage{booktabs,array,tabularx,threeparttable,multirow}
\usepackage{tikz}
\usepackage{makecell}
\usetikzlibrary{arrows.meta,calc,positioning,shapes.geometric,plotmarks}
\usepackage{float}
\floatstyle{ruled}
\newfloat{algorithm}{tbp}{loa}
\floatname{algorithm}{Algorithm}
\floatstyle{plain}
\usepackage{enumitem}
\usepackage{placeins}
\usepackage{needspace}
\usepackage{microtype}
\usepackage{longtable}
\usepackage[authoryear]{natbib}
\usepackage[
  colorlinks=true,
  citecolor=blue,
  linkcolor=blue,
  urlcolor=blue
]{hyperref}

\startlocaldefs

\definecolor{rewardblue}{RGB}{0,114,178}
\definecolor{rewardorange}{RGB}{213,94,0}
\definecolor{rewardgreen}{RGB}{0,158,115}
\definecolor{rewardpurple}{RGB}{204,121,167}
\definecolor{gridgray}{RGB}{198,204,209}
\definecolor{cellgray}{RGB}{241,243,245}
\definecolor{startblue}{HTML}{2463A6}
\definecolor{destorange}{HTML}{D97706}
\definecolor{accent}{HTML}{315A78}
\definecolor{softgray}{HTML}{F4F5F6}
\definecolor{rowgray}{RGB}{230,230,230}
\definecolor{tableblue}{RGB}{0,114,178}
\definecolor{rulegray}{HTML}{7A838B}
\definecolor{positive}{HTML}{1B7F5A}

\renewcommand{\arraystretch}{1.10}
\setlist[itemize]{leftmargin=1.5em,itemsep=1pt,topsep=3pt}

\newcommand{\RT}{\operatorname{RT}}
\newcommand{\RO}{\operatorname{RO}}
\newcommand{\Fsn}{\mathcal{F}_{\mathrm{sn}}}
\newcommand{\Flinear}{\mathcal F_{\mathrm{sn}}^{\mathrm{lin}}}
\newcommand{\PP}{\mathbb P}
\newcommand{\EE}{\mathbb E}
\newcommand{\ind}{\mathbf 1}

\numberwithin{equation}{section}
\newcommand{\NLL}{\mathrm{NLL}}
\newcommand{\ARI}{\mathrm{ARI}}
\newcommand{\Occ}{\mathrm{occ}}

\newcolumntype{L}[1]{>{\raggedright\arraybackslash}p{#1}}
\newcolumntype{Y}{>{\raggedright\arraybackslash}X}
\newcolumntype{C}[1]{>{\centering\arraybackslash}p{#1}}

\theoremstyle{definition}
\newtheorem{definition}{Definition}[section]
\theoremstyle{plain}

\newtheorem{lemma}[definition]{Lemma}
\newtheorem{theorem}[definition]{Theorem}
\newtheorem{proposition}[definition]{Proposition}
\newtheorem{corollary}[definition]{Corollary}
\theoremstyle{definition}
\newtheorem{remark}[definition]{Remark}
\newtheorem{example}[definition]{Example}

\counterwithin{table}{section}
\counterwithin{figure}{section}

\endlocaldefs

\begin{document}

\begin{frontmatter}
\pdfsubject{Crossed reward-temperature mixtures for unlabeled trajectory clustering}
\title{TRAIL-IRL: Crossed Reward–Temperature Mixtures for Unlabeled Trajectory Clustering}
\runtitle{TRAIL-IRL}

\begin{aug}
\author[A]{\fnms{Yichen}~\snm{Zhu}\ead[label=e1]{yczhu25@m.fudan.edu.cn}}%
\author[A]{\fnms{Guanhua}~\snm{Fang}\ead[label=e3]{fanggh@fudan.edu.cn}}
\author[A]{\fnms{Ming}~\snm{Zheng}\ead[label=e4]{mingzheng@fudan.edu.cn}}
\address[A]{School of Management, Fudan University\printead[presep={,\ }]{e1,e3,e4}}
\end{aug}
\runauthor{}

\begin{abstract}
Different paths need not imply different preferences, trajectories
also differ in how randomly actions are chosen.
Many multi-reward inverse reinforcement learning (IRL) methods assume
common action randomness and can confound it with preference
heterogeneity.
We introduce TRAIL-IRL, which clusters unlabeled trajectories using
rewards for preferences and temperatures for action randomness.
Its crossed mixture pairs each reward with a shared set of temperature
levels, allowing action regularity to vary within preference groups.
With centered and normalized rewards in a known finite-state environment,
we establish the identifiability conditions under which the distribution of sufficiently
long trajectories determines rewards, temperatures, mixing weights,
and class counts up to relabeling.
Under the same conditions and reward constraints, common-temperature
reward mixtures with the same component count cannot reproduce a
crossed mixture with multiple temperatures.
We develop a generalized expectation--maximization algorithm and select
class counts using the Bayesian information criterion (BIC).
Simulations show improved predictive fit, reward recovery, and
trajectory classification over common-temperature mixtures with the
same component count.
For 140 bird--year trajectories from 79 white storks, BIC favors four
reward classes and three temperature levels over common-temperature
candidates.
Younger birds have higher posterior membership probabilities for
a southwestward migration class, whereas older birds have higher
probabilities for a northern retention class.
Within reward classes, higher temperatures are associated with more
variable movement directions and positions farther ahead along the
migration route, consistent with \textit{social information use}
and \textit{collective sensing}.
\end{abstract}


\end{frontmatter}

\section{Introduction}
\label{sec:introduction}

Advances in GPS tracking, biologging, and automated sensing have expanded
the availability of trajectory data in animal ecology
\citep{kays2015,nathan2022}, pedestrian mobility
\citep{brscic2013,rudenko2020}, and urban transportation
\citep{barmpounakis2020}.
A trajectory records an individual's successive locations and their
observation times \citep{zheng2015}.
These records provide information on movement routes, speeds, changes
in direction, and stopping patterns.

Movement behavior varies across individuals, and group labels are
often unavailable.
Clustering methods recover recurring patterns from the trajectories
themselves.
Geometric methods define similarity through path shape and distance
\citep{lee2007,morris2009,zheng2015}.
Regression mixtures and functional models identify groups by
modeling trajectory curves and their variation over time
\citep{gaffney1999,james2003,gaffney2007,jacques2014}.
Neural methods learn representations for trajectory clustering
and similarity comparison
\citep{yao2018,li2018trajectory,chang2023}.
These methods describe observed movement patterns, but do not
explicitly infer the behavioral objectives that may underlie them.

Inverse reinforcement learning (IRL) infers such behavioral objectives
by estimating a reward function from observed trajectories under a
decision model \citep{ng2000,abbeel2004}.
The reward function represents preferences over states and actions,
and the decision model relates action choices to immediate and
expected future rewards.
Single-reward methods, including linear programming IRL (LP-IRL)
\citep{ng2000} and maximum-entropy IRL \citep{ziebart2008},
estimate a common reward function without inferring latent
trajectory groups.
When objectives differ across individuals, reward-mixture models
jointly infer group-specific reward functions and latent group
memberships \citep{babes2011,ramponi2020,bighashdel2023}.
For example, DPM-BIRL \citep{choi2012} uses a Dirichlet process
mixture to group trajectories according to the reward functions
that explain their observed behavior.

Individuals pursuing the same objective can also differ in how they
plan and choose actions.
Existing IRL formulations model such differences through variation
in strategy preferences \citep{chen2020}, planning horizons
\citep{yao2024}, or computational budgets \citep{jacob2024}.
In this paper, we focus on \emph{action regularity}, the consistency with which
individuals favor actions that advance a given objective.
In entropy-regularized decision models, this consistency is controlled
by a \emph{temperature} parameter, which governs the concentration
of action choices under a given reward \citep{ziebart2008,geist2019}.
DPM-BIRL \citep{choi2012} uses a fixed decision-precision parameter
shared across individuals to describe this regularity.
IRLEED \citep{beliaev2025} allows decision precision to vary across
demonstrators and estimates it jointly with a shared reward and
individual reward biases, using trajectories with known demonstrator
identities.
We allow temperature to vary among individuals who share a reward
and infer action-regularity groups from unlabeled trajectories.

We propose \textbf{TRAIL-IRL} (\textbf{I}nverse \textbf{R}einforcement \textbf{L}earning for
\textbf{T}emperature--\textbf{R}eward \textbf{A}ssociation and \textbf{I}nference of \textbf{L}atent Groups)
to infer reward and action-regularity groups from unlabeled trajectories.
The crossed reward--temperature mixture, denoted by
\(\mathrm{RT}(K,H)\), combines \(K\) reward classes and \(H\)
temperature classes into \(K\times H\) latent components, with each
trajectory generated from one reward--temperature pair.
Reward parameters are shared across temperature classes, and
temperature parameters are shared across reward classes as well.
This structure allows trajectories with the same objective to differ
in action regularity, and trajectories with different objectives to
share the same level of regularity.
To characterize these two sources of variation, we develop an
algorithm that jointly estimates the reward and temperature
parameters and infers each trajectory's reward and temperature
class labels.
The number of reward classes \(K\) and the number of temperature
classes \(H\) are also selected from the observed trajectories.

However, different reward--temperature pairs can induce the same
optimal policy and hence the same trajectory distribution
\citep{ng1999,cao2021,skalse2023}.
Existing IRL theory establishes conditions for identifying a shared
reward \citep{cao2021,rolland2022}, but mixtures introduce a further
ambiguity, i.e., different sets of component distributions and mixing
weights may produce the same mixture distribution
\citep{teicher1963identifiability,mclachlan2019finite}.
To remove ambiguity in the reward--temperature representation,
we work within a centered and normalized reward class.
We then establish an identifiability theorem showing that, under
suitable conditions, the trajectory distribution determines the
numbers of reward and temperature classes, their parameters,
and the mixing weights, up to a label permutation.
Simulation studies evaluate parameter estimation, trajectory
classification, and selection of \(K\) and \(H\) in finite samples.
Table~\ref{tab:literature-comparison} compares the representative
approaches in terms of unlabeled trajectory clustering, reward
inference, and action-regularity heterogeneity.

\begin{table}[!htbp]
\centering
\begin{threeparttable}
\caption{Comparison of representative approaches to trajectory heterogeneity.
The columns indicate whether a method takes into account trajectory clustering, reward estimation, and action regularity, respectively.
}
\label{tab:literature-comparison}
\small
\setlength{\tabcolsep}{5pt}
\renewcommand{\arraystretch}{1.15}

\begin{tabularx}{\linewidth}{@{}
  >{\raggedright\arraybackslash}X
  >{\centering\arraybackslash}p{0.19\linewidth}
  >{\centering\arraybackslash}p{0.18\linewidth}
  >{\centering\arraybackslash}p{0.13\linewidth}
@{}}
\toprule
\textbf{Method and reference}
& \textbf{Trajectory clustering}
& \textbf{Reward estimation}
& \textbf{Action regularity} \\
\midrule

Trajectory clustering\newline
\citep{lee2007,james2003,yao2018}
& Yes & No & No \\
\addlinespace[4pt]

Single-reward IRL\newline
\citep{ng2000,ziebart2008}
& No & Yes & No \\
\addlinespace[4pt]

Reward-mixture IRL (DPM-BIRL)\newline
\citep{choi2012}
& Yes & Yes & No \\
\addlinespace[4pt]

IRLEED\newline
\citep{beliaev2025}
& No & Yes & Yes \\

\midrule
\textbf{TRAIL-IRL} (our work)
& \textbf{Yes} & \textbf{Yes} & \textbf{Yes} \\
\bottomrule
\end{tabularx}

\begin{tablenotes}[flushleft]
\footnotesize
\item[]
See Appendix~\ref{app:literature-review} for a detailed comparison.
\end{tablenotes}
\end{threeparttable}
\end{table}

We apply TRAIL-IRL to white-stork autumn migration to examine
variation in migratory preferences and action regularity.
The inferred reward classes differ in migration direction and
stopping tendencies.
Younger birds are more strongly associated with a southwestward
migration class, whereas older birds are more strongly associated
with a northern retention class.
This age pattern is consistent with \textit{developmental plasticity},
whereby conditions and experience during development shape migratory
strategies \citep{andrade2025}.
It is also consistent with \textit{the dominance hypothesis},
which predicts that older, competitively dominant birds can secure
local food and remain resident, whereas younger, subordinate birds
are more likely to migrate \citep{boyle2008,martins2024}.
Within reward classes, higher fitted temperatures are associated
with less direct and less directionally consistent movement.
Higher temperatures are also associated with positions farther
ahead along the migration route.
These associations are consistent with the exploration and following
behaviors described in studies of \textit{social information use}
and \textit{collective sensing}: birds ahead explore favorable
updrafts, while followers use their movements as cues
\citep{flack2018,nagy2018}.

The paper makes the following contributions.
\begin{enumerate}[
  leftmargin=1.75em,
  label=(\arabic*),
  itemsep=0.1em,
  topsep=0.15em,
  parsep=0pt
]

\item \textit{Statistical model.}
We develop a crossed reward--temperature mixture model for
clustering unlabeled trajectories by behavioral objective and
action regularity.
The model shares reward parameters across temperature classes
and temperature parameters across reward classes.

\item \textit{Identifiability theory.}
Within the centered and normalized reward class, we establish
conditions under which finite trajectory distributions identify
the numbers of reward and temperature classes, their shared
parameters, and the mixing weights, up to relabeling.
We also show that regular \(\mathrm{RT}(K,H)\) models with \(H\geq2\)
and reward-only mixtures with a common temperature and \(KH\)
components are non-nested under the conditions of
Corollary~\ref{cor:strict-non-nesting}.

\item \textit{Estimation and finite-sample evaluation.}
We develop a generalized expectation--maximization (GEM) algorithm
to estimate the reward and temperature parameters and mixing
weights, with an initialization technique.

\item \textit{Application findings.}
The white-stork analysis identifies age-related differences in
inferred migration preferences and relates fitted temperature to
movement regularity and relative position along the migration route.

\end{enumerate}

The remainder of the paper is organized as follows.
Section~\ref{sec:data} introduces MDPs, trajectory data, and the
classification problem.
Section~\ref{sec:mixture-model} develops the reward--temperature mixture
model and establishes its identification results.
Section~\ref{sec:gem} presents the estimation method.
Sections~\ref{sec:simulation} and~\ref{sec:white-stork-application}
report the simulation study and the white-stork migration application,
respectively.

\section{Preliminaries}
\label{sec:data}

\subsection{Markov decision processes}
\label{sec:mdp-preliminaries}

Everyday decisions affect both immediate benefits and future opportunities.
For example, a migrating bird must decide whether to continue flying or
stop to feed, i.e., continuing brings it closer to its destination but consumes
energy, whereas feeding delays progress while replenishing the reserves
needed for later flights.
Similarly, a taxi driver choosing a trip must consider both its fare and
where it leaves the vehicle, i.e., a higher-paying trip may be less attractive
if it ends in an area where finding the next passenger is difficult.
A Markov decision process (MDP) formalizes such sequential decisions.

A finite discounted MDP is specified by
\((\mathcal S,\mathcal A,P,f,\gamma)\)
\citep{blackwell1962discrete}.
The finite sets \(\mathcal S\) and \(\mathcal A\) are called the
\emph{state space} and \emph{action space}; they contain the possible
states and actions, respectively. The transition kernel
\(P(s'\mid s,a)\) gives the probability of reaching state \(s'\) after
taking action \(a\) at state \(s\), with
\(P(s'\mid s,a)\ge0\) and \(\sum_{s'\in\mathcal S}P(s'\mid s,a)=1\).
Writing \(S_t\) and \(A_t\) for the state and action at time \(t\),
the Markov assumption is
\(\Pr(S_{t+1}=s'\mid S_0,A_0,\ldots,S_t,A_t)
=P(s'\mid S_t,A_t)\): the current state and action determine the
distribution of the next state. The reward
\(f:\mathcal S\times\mathcal A\to\mathbb R\) assigns an immediate
value to each state--action pair, and \(\gamma\in[0,1)\) discounts
future rewards.

A stationary\footnote{Stationarity means that \(\pi(a\mid s)\) does
not depend on time: each visit to state \(s\) uses the same action
distribution.} Markov policy \(\pi(a\mid s)\) specifies the probability
of selecting action \(a\) at state \(s\). Under this policy,
\(A_t\sim\pi(\cdot\mid S_t)\) and
\(S_{t+1}\sim P(\cdot\mid S_t,A_t)\). The policy determines action
selection, and the transition kernel determines the outcome conditional
on the current state and selected action.

Under the same MDP, different criteria for policy learning can produce
different policies. The standard criterion maximizes the expected
discounted reward
\(\mathbb E_s^\pi[\sum_{t=0}^{\infty}\gamma^t f(S_t,A_t)]\), where
\(\mathbb E_s^\pi\) conditions on \(S_0=s\). This paper considers
entropy regularization, which also accounts for randomness in action
choices; the policy is specified in \eqref{eq:soft-optimal-policy}.

\begin{figure}[!htbp]
\centering
\begin{minipage}[t]{0.28\linewidth}
\vspace{0pt}\centering
\makebox[\linewidth][l]{\small (a)}\par
\includegraphics[width=\linewidth,trim=0 0 0 16bp,clip]{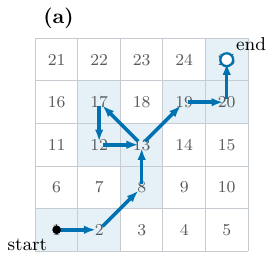}
\end{minipage}\hfill
\begin{minipage}[t]{0.25\linewidth}
\vspace{0pt}\centering
\makebox[\linewidth][l]{\small (b)}\par
\includegraphics[width=\linewidth,trim=0 0 0 17bp,clip]{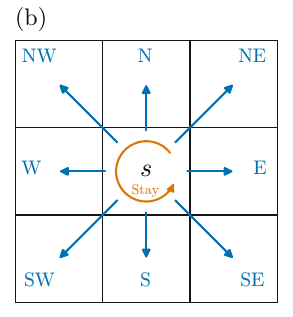}
\end{minipage}\hfill
\begin{minipage}[t]{0.43\linewidth}
\vspace{0pt}\centering
\makebox[\linewidth][l]{\small (c)}\par
\includegraphics[width=\linewidth]{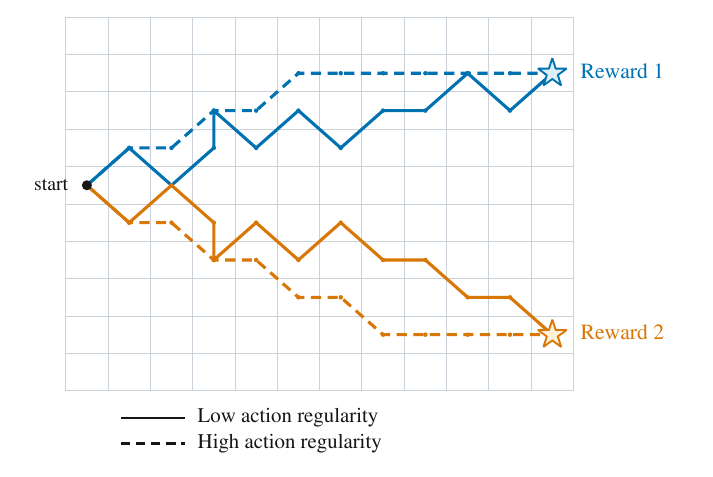}
\end{minipage}
\caption{Trajectory representations and behavioral heterogeneity.
(a) A trajectory under deterministic grid transitions. The 25 states
are numbered from left to right and from bottom to top. Arrows trace
the observed path; shaded cells are visited at least once.
(b) Eight directional actions move from the center cell to neighboring
cells. The circular arrow denotes Stay.
(c) Two reward classes, each with high and low action regularity.
Blue and orange paths represent Reward 1 and Reward 2, respectively.
Dashed lines indicate high action regularity, and solid lines indicate
low action regularity. The black dot marks the common starting point,
and the stars mark the two destinations.}
\label{fig:grid-trajectory}
\label{fig:crossed-trajectories}
\end{figure}

Figure~\ref{fig:grid-trajectory}(a,b) illustrates a deterministic grid model.
The 25 cells, numbered from 1 to 25, form the state space. The actions
are North (N), Northeast (NE), East (E), Southeast (SE), South (S),
Southwest (SW), West (W), Northwest (NW), and Stay. At an interior cell,
a directional action moves to the corresponding neighboring cell,
whereas Stay leaves the state unchanged. Action choice under \(\pi\)
can be either deterministic or stochastic.

\subsection{Finite-horizon trajectory data}
\label{sec:finite-trajectory}

A length-\(T\) trajectory is defined as
\(\tau=(s_0,a_0,\ldots,a_{T-1},s_T)\), which contains \(T\) actions
and \(T+1\) states in temporal order. For example, in
Figure~\ref{fig:grid-trajectory}(a), the trajectory is
\(\tau_{\mathrm{fig1}}=(1,\mathrm E,2,\allowbreak
\mathrm{NE},8,\allowbreak\mathrm N,13,\allowbreak
\mathrm{NW},17,\allowbreak\mathrm S,12,\allowbreak
\mathrm E,13,\allowbreak\mathrm{NE},19,\allowbreak
\mathrm E,20,\allowbreak\mathrm N,25)\),
with \(T=9\) and state 13 visited twice.

We consider independently sampled trajectory data
\(\mathcal D_n=\{\tau_i;i=1,\ldots,n\}\), where
\(\tau_i=(s_{i0},a_{i0},\ldots,s_{it},a_{it},\ldots,a_{i,T_i-1},s_{i,T_i})\)
is the \(i\)-th trajectory and has length \(T_i\). The lengths \(T_i\)
may differ.
We assume that all trajectories in \(\mathcal D_n\) share the environment
\((P,\mu_0,\gamma,\mathcal S,\mathcal A)\), where \(\mu_0\) is the
initial-state distribution. Trajectories may be generated under different
rewards. For each reward, different action regularity rules may also
induce different policies, as discussed in
Section~\ref{sec:structured-classification}. Within each trajectory,
the generating reward and action regularity rule remain fixed from
start to finish.

\subsection{Reward heterogeneity and action regularity}
\label{sec:structured-classification}

Differences among trajectories can reflect differences in their generating
policies. These policy differences may arise from different behavioral
preferences or from different levels of action regularity under the same
preferences.

Consider three migrating birds following the same migration corridor.
For illustration, suppose that at a northern location the first bird assigns
scores of \(0.8\) and \(0.2\) to moving south and staying, respectively,
whereas the second bird assigns scores of \(0.2\) and \(0.8\).
The first bird therefore prefers to continue its southward migration, while
the second prefers to remain in the north.
Their different evaluations of the same available actions can generate
different migration trajectories.
This is \emph{reward heterogeneity}: the birds differ in what they prefer.

Now suppose the first and third birds share the same reward, so that both
assign scores of \(0.8\) and \(0.2\) to moving south and staying under the
same conditions.
Their long-run behavior may nevertheless differ.
The first bird almost always chooses the higher-valued action, whereas the
third more sometimes chooses the lower-valued alternative despite sharing
the same ranking of the two actions.
Across repeated decisions, the first bird's choices are therefore more
concentrated on the actions favored by the common reward, while the third
bird's choices are more dispersed.
This is \emph{heterogeneity in action regularity}: the birds share the same
objective but differ in how consistently their behavior follows that
objective. 

The same distinction arises among three shoppers entering a mall.
The first prefers an electronics store, whereas the second prefers
a clothing store.
Their paths lead toward different destinations, illustrating
\emph{reward heterogeneity}.
The third has exactly the same preferences as the first:
both want to reach the same electronics store from the same entrance.
The first walks directly toward the store, making few detours.
The third also moves toward that store overall, but sometimes turns
into a side corridor or briefly walks away before turning back
toward it.
The two trajectories therefore share the same overall direction,
but the third contains more irregular turns and detours.
This illustrates \emph{heterogeneity in action regularity}:
the shoppers differ in how steadily they move toward the same
preferred destination.

Figure~\ref{fig:crossed-trajectories}(c) shows four schematic trajectories
combining two reward classes with two levels of action regularity.
In this schematic, the reward classes correspond to different
preferred destinations.
Within each reward class, both trajectories lead toward the same
destination, but the trajectory with lower action regularity
makes more irregular turns and detours.
This motivates the central question:

\begin{quote}
\itshape
Can we infer each trajectory's reward class and level of action
regularity from unlabeled trajectory data?
\end{quote}

\section{Reward--Temperature Mixture Model}
\label{sec:mixture-model}

\subsection{Entropy-regularized trajectory mixtures}
\label{sec:single-entropy-model}

In a known MDP, an optimal policy can be obtained from the Bellman
optimality equations, which combine immediate rewards with future
values. However, standard optimal policies select only actions with
maximal value and do not describe how agents may choose actions with
lower values. Entropy regularization accommodates such choices by
balancing expected reward against policy entropy, with a temperature
parameter controlling action regularity. This leads to the soft
Bellman formulation, which has roots in dynamic discrete-choice
models \citep{rust1987}. Maximum-entropy IRL was introduced by
\citet{ziebart2008} and extended to stochastic dynamics through
maximum causal entropy by \citet{ziebart2010}.

 To introduce maximum-entropy IRL, we use the finite MDP, reward function \(f\), and trajectory data
introduced in Sections~\ref{sec:mdp-preliminaries}--\ref{sec:finite-trajectory},
with \(|\mathcal A|\ge2\) and known \(P,\gamma,\mu_0\).
For any stationary Markov policy \(\pi\) and
\(\lambda\in\mathbb R_{>0}\), define the entropy-regularized value
starting from state \(s\) as
\[
V_{f,\lambda}^{\pi}(s)
=\EE_s^{\pi}\!\left[
\sum_{t=0}^{\infty}\gamma^t
\{f(S_t,A_t)-\lambda\log\pi(A_t\mid S_t)\}
\right].
\]
This value combines the expected discounted reward with an entropy
bonus, i.e., conditional on \(S_t\), the expectation of
\(-\log\pi(A_t\mid S_t)\) is the entropy of the action distribution
\(\pi(\cdot\mid S_t)\).\footnote{We use the convention
\(0\log0=0\) when computing entropy.}

A \emph{soft optimal policy}, denoted by \(\pi_{f,\lambda}^*\),
maximizes this value over stationary Markov policies for every
initial state \(s\); write
\(V_{f,\lambda}^*(s)=\max_\pi V_{f,\lambda}^{\pi}(s)\).
The temperature parameter \(\lambda\) controls the balance between
reward and entropy in this optimization. A larger \(\lambda\)
places greater weight on entropy, encouraging more randomized
action choices, whereas a smaller \(\lambda\) gives greater priority
to reward maximization. As \(\lambda\downarrow0\), the optimal policy
concentrates on actions that are optimal under the unregularized
reward criterion.\footnote{If the unregularized optimal action is
unique at every state, the limiting policy is deterministic.
When optimal actions are tied, randomization among them may remain.}

To obtain this policy, we first solve the soft Bellman equation
\begin{equation}
V_{f,\lambda}^*(s)
=\lambda\log\sum_{a\in\mathcal A}
\exp\!\left\{
\frac{
f(s,a)+\gamma\sum_{s'\in\mathcal S}
P(s'\mid s,a)V_{f,\lambda}^*(s')
}{\lambda}
\right\},
\label{eq:soft-bellman}
\end{equation}
which has a unique solution for the finite discounted MDP.
The optimal action value
\(Q_{f,\lambda}^*(s,a)
=f(s,a)+\gamma\sum_{s'\in\mathcal S}
P(s'\mid s,a)V_{f,\lambda}^*(s')\)
combines the immediate reward from action \(a\) with the discounted
optimal continuation value. The soft optimal policy is then
\begin{equation}
\pi_{f,\lambda}^*(a\mid s)
=\exp\!\left\{
\frac{Q_{f,\lambda}^*(s,a)-V_{f,\lambda}^*(s)}
{\lambda}
\right\}.
\label{eq:soft-optimal-policy}
\end{equation}

With \(P,\gamma,\mu_0\) fixed, each reward--temperature pair
\((f,\lambda)\) therefore determines the probability of a
length-\(T\) trajectory \(\tau\) from
Section~\ref{sec:finite-trajectory},
\begin{equation}
\PP_{f,\lambda}^{(T)}(\tau)
=\mu_0(s_0)\prod_{t=0}^{T-1}
\pi_{f,\lambda}^*(a_t\mid s_t)
P(s_{t+1}\mid s_t,a_t).
\label{eq:single-trajectory-law}
\end{equation}

The distribution \eqref{eq:single-trajectory-law} describes one
reward and one temperature.
As discussed in Section~\ref{sec:structured-classification}, both can
vary across trajectories: objectives may differ, and trajectories that
share an objective may still differ in action regularity.
We represent the two sources by a crossed mixture, in which every
reward is paired with the same set of temperatures and each trajectory
is generated by one reward--temperature pair.
For \(K,H\in\mathbb N=\{1,2,\ldots\}\), an \(\RT(K,H)\) model contains rewards \(f_1,\ldots,f_K\),
temperatures \(\lambda_1,\ldots,\lambda_H>0\), and weights
\(\omega_{kh}\ge0\) with \(\sum_{k,h}\omega_{kh}=1\).
The pair \((K,H)\) is the \emph{model order}.
For each trajectory \(\tau_i\), an unobserved latent cell \(Z_i\) is first drawn with
\(\Pr\{Z_i=(k,h)\}=\omega_{kh}\). Conditional on \(Z_i=(k,h)\),
the entire observed trajectory \(\tau_i\) is generated by the soft-optimal policy
\(\pi_{f_k,\lambda_h}^*\). For
\(\vartheta=(K,H,f_{1:K},\lambda_{1:H},\omega)\), the mixture probability of trajectory \(\tau\) is
\begin{equation}
 \PP_{\vartheta}^{(T)}(\tau)
 =\sum_{k=1}^{K}\sum_{h=1}^{H}
 \omega_{kh}\PP_{f_k,\lambda_h}^{(T)}(\tau).
 \label{eq:RT-mixture-law}
\end{equation}
The model is \emph{regular} when every \(\omega_{kh}>0\), the reward
functions are pairwise distinct, and the temperatures are pairwise distinct.
Its parameter \(\vartheta\) is then also called \emph{regular}.

\begin{remark}[Reward-only mixtures]
    The reward-only model \(\RO(K,1)\) is the special case
\(\RT(K,1)\), in which all \(K\) reward components share a common
temperature \(\lambda_1\), i.e., $H  = 1$. Component differences therefore reflect
reward heterogeneity alone.
\end{remark}

\subsection{Finite- and infinite-length identifiability}
\label{sec:mixture-identifiability-definitions}

For the fixed environment in Section~\ref{sec:single-entropy-model},
a reward class \(\mathcal F\) induces
the \emph{regular parameter family}
\[
 \mathcal R(\mathcal F)
 :=\left\{\vartheta=(K,H,f_{1:K},\lambda_{1:H},\omega):
 \begin{array}{l}
 K,H\in\mathbb N,\quad \vartheta\text{ is regular},\\
 f_k\in\mathcal F,\quad k=1,\ldots,K
 \end{array}\right\}.
\]
For fixed \(K,H\in\mathbb N\), the induced \emph{regular parameter
family of order \((K,H)\)} is \(\mathcal R_{K,H}(\mathcal F) :=\{\vartheta\in\mathcal R(\mathcal F): \vartheta\text{ has order }(K,H)\}.\)
Thus \(\mathcal R(\mathcal F)
=\bigcup_{K,H\in\mathbb N}\mathcal R_{K,H}(\mathcal F)\).

Identifiability is a central concern in mixture models: without it,
component parameters and mixing weights cannot be uniquely recovered
from the observed-data distribution, even up to label permutations
\citep{teicher1963identifiability,mclachlan2019finite}.
We therefore define finite- and infinite-length identifiability for
the RT and RO models, beginning with their label equivalence.
Two regular parameters \(\vartheta\) and \(\widetilde\vartheta\)
are \emph{label-equivalent}, written \(\vartheta\sim\widetilde\vartheta\),
if \((K,H)=(\widetilde K,\widetilde H)\) and there exist
permutations \(\sigma\) of \(\{1,\ldots,K\}\) and \(\rho\) of
\(\{1,\ldots,H\}\) such that
\begin{equation}
 \widetilde f_k=f_{\sigma(k)},\qquad
 \widetilde\lambda_h=\lambda_{\rho(h)},\qquad
 \widetilde\omega_{kh}=\omega_{\sigma(k),\rho(h)}
 \quad\text{for every }(k,h).
 \label{eq:separate-label-equivalence}
\end{equation}
\begin{definition}[Mixture identifiability]
\label{def:mixture-T-identifiability}
Let \(\mathcal R_0\subseteq\mathcal R(\mathcal F)\) be a parameter
family. It is \emph{\(T\)-identifiable} if, for every
\(\vartheta,\widetilde\vartheta\in\mathcal R_0\), \(\PP_{\vartheta}^{(T)}(\cdot) =\PP_{\widetilde\vartheta}^{(T)}(\cdot) \;\Longrightarrow\; \vartheta\sim\widetilde\vartheta.\)
It is \emph{\(\infty\)-identifiable} if, for every
\(\vartheta,\widetilde\vartheta\in\mathcal R_0\), \(\PP_{\vartheta}^{(m)}(\cdot) =\PP_{\widetilde\vartheta}^{(m)}(\cdot) \;\text{for every integer }m\ge1 \;\Longrightarrow\; \vartheta\sim\widetilde\vartheta.\)
\end{definition}

\begin{remark}[Distinguishing trajectories]
\label{rem:distinguishing-trajectories}
Since each fixed-length trajectory space is finite, \(T\)-identifiability
of \(\mathcal R_0\) is equivalent to the following statement: for every
\(\vartheta,\widetilde\vartheta\in\mathcal R_0\) with
\(\vartheta\not\sim\widetilde\vartheta\), there exists a length-\(T\)
trajectory \(\tau\) such that
\(
 \PP_{\vartheta}^{(T)}(\tau)
 \ne\PP_{\widetilde\vartheta}^{(T)}(\tau).
\)
Similarly, \(\infty\)-identifiability is equivalent to requiring that,
for every \(\vartheta,\widetilde\vartheta\in\mathcal R_0\) with
\(\vartheta\not\sim\widetilde\vartheta\), there exist an integer
\(m\ge1\) and a length-\(m\) trajectory \(\tau\) such that
\(
 \PP_{\vartheta}^{(m)}(\tau)
 \ne\PP_{\widetilde\vartheta}^{(m)}(\tau).
\)
Thus \(T\)-identifiability implies \(\infty\)-identifiability, whereas
\(\infty\)-identifiability alone does not ensure a common finite horizon
for the entire family.
\end{remark}

\subsection{Centered and normalized rewards}
\label{sec:normalized-reward-class}

Different reward functions can induce the same optimal policy.
Two transformations leave that policy unchanged.
\emph{Potential shaping} replaces \(f(s,a)\) by
\(f(s,a)+v(s)-\gamma\sum_{s'}P(s'\mid s,a)v(s')\) for an arbitrary
function \(v:\mathcal S\to\mathbb R\), and preserves the policy at a
fixed temperature \citep{ng1999,cao2021}.
\emph{Common scaling} replaces \((f,\lambda)\) by \((cf,c\lambda)\) for
some \(c>0\), and likewise preserves the policy
(Corollary~\ref{cor:reward-temperature-scaling}).
We exclude both by restricting rewards to the following class.

Choose fixed positive weights \(\eta(s)\) and \(\omega_s(a)\) satisfying
\(\sum_s\eta(s)=1\) and \(\sum_a\omega_s(a)=1\) for each state. Set
\(\nu(s,a)=\eta(s)\omega_s(a)\) and
\(\|f\|_\nu^2=\sum_{s,a}\nu(s,a)f(s,a)^2\). Define
\begin{equation}
 \Fsn=\left\{f:\mathcal S\times\mathcal A\to\mathbb R:
 \underbrace{\sum_a\omega_s(a)f(s,a)=0\ \text{for every }s}
 _{\text{statewise centering}},\quad
 \underbrace{\|f\|_\nu=1}_{\text{normalization}}\right\}.
 \label{eq:Fsn-definition}
\end{equation}
Statewise centering removes potential shaping, and the unit norm
removes common scaling.

For linear rewards, fix \(1\le p\le|\mathcal S|(|\mathcal A|-1)\)
and a feature map \(\phi:\mathcal S\times\mathcal A\to\mathbb R^p\)
such that
\begin{equation}
 \sum_a\omega_s(a)\phi(s,a)=0\quad\text{for every }s,
 \qquad
 \sum_{s,a}\nu(s,a)\phi(s,a)\phi(s,a)^{\mathsf T}=I_p.
 \label{eq:linear-feature-normalization}
\end{equation}
The second condition can be obtained by whitening any centered feature map
whose weighted Gram matrix is positive definite. Write
\begin{equation}
 f_\beta(s,a)=\phi(s,a)^{\mathsf T}\beta,
 \qquad
 \Flinear=\{f_\beta:\|\beta\|_2=1\}\subset\Fsn.
 \label{eq:linear-reward-class}
\end{equation}
Then \(\|f_\beta\|_\nu=\|\beta\|_2\) and
\(\|f_\beta-f_{\widetilde\beta}\|_\nu
=\|\beta-\widetilde\beta\|_2\). Proposition~\ref{prop:hessian-conditioning}
in Appendix~\ref{app:hessian} shows that the unit-norm restriction removes
the scaling singularity from the single-component population
likelihood Hessian at the true parameter.

\begin{remark}[Reward specification]
For clarity, the theoretical statements below use linear rewards.
Analogous identifiability results hold for nonlinear reward classes under
suitable regularity conditions. Throughout the rest of paper, we write
\(\Fsn\) for \(\Flinear\) for simplicity.
\end{remark}

\subsection{ Identifiability of $\RT(K,H)$}
\label{sec:mixture-identifiability-theorem}

We now establish conditions under which trajectory distributions
identify the rewards, temperatures, mixing weights, and model order,
up to label equivalence. We assume that every state is
reachable from every other state through transitions with positive
probability. To express this condition, define the directed support
graph \(G_P=(V,E)\) by
\begin{equation}
 V=\mathcal S,\qquad
 E=\{(s,s')\in\mathcal S^2:
 P(s'\mid s,a)>0\text{ for some }a\in\mathcal A\}.
 \label{eq:support-graph}
\end{equation}
The graph is \emph{strongly connected} if a directed path exists
from every state to every other state.
The state transition matrix induced by a stationary policy \(\pi\) is \(P_\pi(s,s') =\sum_{a\in\mathcal A}\pi(a\mid s)P(s'\mid s,a).\)
By \eqref{eq:soft-optimal-policy},
\(\pi_{f,\lambda}^*(a\mid s)>0\) for every \((s,a)\). Hence
\(P_{\pi_{f,\lambda}^*}(s,s')>0\) if and only if \((s,s')\in E\),
so strong connectivity ensures reachability under every
soft-optimal policy.

\begin{theorem}[Identification of the reward--temperature mixture]
\label{thm:mixture-identification}
Suppose that \(P,\gamma,\mu_0,\phi\) are known,
\(\gamma\in[0,1)\), \(\mu_0\) is any probability distribution on
\(\mathcal S\), \(G_P\) is strongly connected, and \(\phi\) satisfies
\eqref{eq:linear-feature-normalization}.
\begin{enumerate}[label=(\roman*),leftmargin=2em]
\item
\(\mathcal R(\Fsn)\) is \(\infty\)-identifiable.

\item
Let \(\vartheta\in\mathcal R_{K,H}(\Fsn)\) and
\(\widetilde\vartheta\in
\mathcal R_{\widetilde K,\widetilde H}(\Fsn)\). If
\begin{equation}
 T\ge(KH+\widetilde K\widetilde H)|\mathcal S|-1,
 \label{eq:cross-order-horizon}
\end{equation}
then
\[
 \PP_{\vartheta}^{(T)}(\cdot)
 =\PP_{\widetilde\vartheta}^{(T)}(\cdot)
 \quad\Longrightarrow\quad
 \vartheta\sim\widetilde\vartheta.
\]
Consequently, \(\mathcal R_{K,H}(\Fsn)\) is \(T\)-identifiable for
\(T\ge2KH|\mathcal S|-1\). For every \(R\in\mathbb N\), the family
\(\bigcup_{K,H\in\mathbb N:\,KH\le R}\mathcal R_{K,H}(\Fsn)\)
is \(T\)-identifiable for \(T\ge2R|\mathcal S|-1\), including
identification of its orders.
\end{enumerate}
\end{theorem}

The proof, including the reward--temperature equivalence classes and
normalized representatives, is given in
Appendix~\ref{app:proof-mixture-identification}.
Along an infinite trajectory, empirical state--action frequencies
determine its component policy. Normalization then identifies the
corresponding reward and temperature, and the shared reward and
temperature values determine the crossed classes.
The probabilities assigned to these component policies identify
the mixture weights.
A finite-dimensional representation of the trajectory law gives
\eqref{eq:cross-order-horizon}.
These horizon bounds are sufficient conditions. The case \(K=H=1\) includes
joint identification of a single reward and temperature.

\subsection{\texorpdfstring{\(\RT(K,H)\neq\RO(KH,1)\)}
{RT(K,H) is not RO(KH,1)}}
\label{sec:rt-ro-distinction}

For \(H\ge2\), we compare \(\RT(K,H)\) and \(\RO(KH,1)\)
models, which have the same number of latent cells.
To compare the trajectory distributions they can represent, define \(\mathfrak M_{K,H}^{(T)} :=\{\PP_\vartheta^{(T)}(\cdot): \vartheta\in\mathcal R_{K,H}(\Fsn)\}.\)

\begin{corollary}[Strong non-nesting]
\label{cor:strict-non-nesting}
Under the assumptions of Theorem~\ref{thm:mixture-identification},
suppose \(H\ge2\) and both \(\mathfrak M_{K,H}^{(T)}\) and
\(\mathfrak M_{KH,1}^{(T)}\) are nonempty.
If \(T\ge2KH|\mathcal S|-1\), then
\begin{equation*}
 \mathfrak M_{K,H}^{(T)}
 \cap\mathfrak M_{KH,1}^{(T)}
 =\varnothing.
\end{equation*}
Thus no regular \(\RO(KH,1)\) model can reproduce the trajectory
law of a regular \(\RT(K,H)\) model at this horizon.
\end{corollary}

Appendix~\ref{app:proof-strict-non-nesting} collects the proof of the corollary.
Regularity is required since the two model classes can overlap when
weights are zero or both rewards and temperatures are repeated.

The following example gives an explicit positive lower bound on
the Kullback--Leibler divergence from a regular \(\RT(2,2)\) class
to the \(\RO(4,1)\) class using trajectories of length \(T=3\).

\begin{example}[One state and three actions]
\label{ex:one-state-three-action}
Let \(\mathcal S=\{s\}\), \(\mathcal A=\{a_1,a_2,a_3\}\),
\(P(s\mid s,a)=\mu_0(s)=1\), and \(\gamma\in[0,1)\).
Take \(\eta(s)=1\), \(\omega_s(a)=1/3\), and
\[
 \phi(s,a_1)=\begin{pmatrix}\sqrt2\\0\end{pmatrix},\qquad
 \phi(s,a_2)=
 \begin{pmatrix}-1/\sqrt2\\\sqrt{3/2}\end{pmatrix},\qquad
 \phi(s,a_3)=
 \begin{pmatrix}-1/\sqrt2\\-\sqrt{3/2}\end{pmatrix}.
\]
These features satisfy \eqref{eq:linear-feature-normalization}.
Let \(P_\star^{(3)}\) be the regular \(\RT(2,2)\) law specified
in Appendix~\ref{app:example}, and define \(\tau_2=(s,a_1,s,a_2,s,a_3,s),\allowbreak\; \delta=P_\star^{(3)}(\tau_2)-\frac{36}{2197}.\)
The explicit calculations in that appendix give \(\delta>0\) and
\begin{equation}
 \inf_{Q^{(3)}\in\mathfrak M_{4,1}^{(3)}}
 D_{\mathrm{KL}}\!\left(
 P_\star^{(3)}\middle\|Q^{(3)}
 \right)
 \ge2\delta^2>0,
 \label{eq:one-state-kl-gap}
\end{equation}
where \(D_{\mathrm{KL}}\) denotes the Kullback--Leibler
divergence.\footnote{For probability mass functions \(p\) and \(q\)
on a common finite space,
\(D_{\mathrm{KL}}(p\|q)
=\sum_{x:p(x)>0}p(x)\log\{p(x)/q(x)\}\),
using natural logarithms.
The divergence is \(+\infty\) if \(p(x)>0=q(x)\) for some \(x\).}
\end{example}

\section{TRAIL-IRL Estimation}
\label{sec:gem}

For a specified order \((K,H)\), TRAIL-IRL estimates rewards,
temperatures, and mixture weights by generalized
expectation--maximization (GEM) \citep{dempster1977,wu1983}.
We then select \((K,H)\) by the Bayesian information criterion (BIC).
Appendix~\ref{app:continuous-state} extends the model to continuous
state spaces with finitely many actions and gives a GEM algorithm
based on numerical quadrature.

\subsection{Likelihood and GEM Updates}
\label{sec:gem-objective}
\label{sec:generalized-m-step}

For the trajectories \(\mathcal D_n\) in
Section~\ref{sec:finite-trajectory}, we maximize the observed-data
log likelihood over \(\vartheta\in\mathcal R_{K,H}(\Fsn)\),
using the parameter family in
Section~\ref{sec:mixture-identifiability-definitions} and the linear
rewards in \eqref{eq:linear-reward-class}.
Write \(p_{i,kh}(\vartheta)=\PP_{f_k,\lambda_h}^{(T_i)}(\tau_i)\)
for the component law in \eqref{eq:single-trajectory-law}, and let
\(z_{i,kh}=\ind\{Z_i=(k,h)\}\) and \(\mathbf z=(z_{i,kh})\)
for the latent cell \(Z_i\) in
Section~\ref{sec:single-entropy-model}.
The complete-data and observed-data log likelihoods are
\begin{align}
  \ell_n^{\mathrm c}(\vartheta;\mathbf{z})
  &=\sum_{i=1}^n\sum_{k=1}^K\sum_{h=1}^H z_{i,kh}
  \{\log\omega_{kh}+\log p_{i,kh}(\vartheta)\},
  \notag\\
  \ell_n(\vartheta)
  &=\sum_{i=1}^n\log\left\{
  \sum_{k=1}^K\sum_{h=1}^H
  \omega_{kh}p_{i,kh}(\vartheta)\right\}.
  \label{eq:observed-loglikelihood}
\end{align}
At iteration \(m\), with current estimate \(\vartheta^{(m)}\),
the E-step computes the posterior cell
probabilities, or \emph{responsibilities},
\begin{equation}
  r_{i,kh}^{(m)}
  =\Pr_{\vartheta^{(m)}}\{Z_i=(k,h)\mid\tau_i\}
  =\frac{\omega_{kh}^{(m)}p_{i,kh}(\vartheta^{(m)})}
  {\sum_{k',h'}\omega_{k'h'}^{(m)}p_{i,k'h'}(\vartheta^{(m)})}.
  \label{eq:gem-responsibility}
\end{equation}
The corresponding Q-function is the conditional expected
complete-data log likelihood:
\begin{equation}
\begin{aligned}
  Q_n(\vartheta\mid\vartheta^{(m)})
  &=\mathbb E_{\vartheta^{(m)}}
    [\ell_n^{\mathrm c}(\vartheta;\mathbf{z})\mid\mathcal D_n]\\
  &=\sum_{i,k,h}r_{i,kh}^{(m)}
  \{\log\omega_{kh}+\log p_{i,kh}(\vartheta)\}.
\end{aligned}
\label{eq:gem-Q}
\end{equation}
The M-step holds these responsibilities fixed and updates the
weights, rewards, and temperatures in turn.
For each block, we accept candidates satisfying the conditions
below and retain the current value if none is accepted.

For the weights, let \(N_{kh}^{(m)}=\sum_i r_{i,kh}^{(m)}\).
Maximization over the probability simplex gives
\begin{equation}
  \omega_{kh}^{(m+1)}=N_{kh}^{(m)}/n.
  \label{eq:mixture-weight-update}
\end{equation}
For reward \(k\), we define the posterior-weighted action counts
\begin{equation}
  c_{kh}(s,a)
  =\sum_{i=1}^n r_{i,kh}^{(m)}
    \sum_{t=0}^{T_i-1}\ind\{s_{it}=s,\ a_{it}=a\}.
  \label{eq:posterior-action-counts}
\end{equation}
With temperatures fixed at \(\lambda_{1:H}^{(m)}\), the reward
objective is
\begin{equation}
  Q_{n,k}^{\mathrm R,(m)}(\beta)
  =\sum_{h=1}^H\sum_{s,a}c_{kh}(s,a)
    \log\pi^*_{f_\beta,\lambda_h^{(m)}}(a\mid s).
  \label{eq:reward-block-objective}
\end{equation}
The gradient consists of contributions from the temperature classes,
each computed through an adjoint system
(Proposition~\ref{prop:reward-gradient},
Appendix~\ref{app:reward-gradients}).
At a trial step size \(\eta_k^{(m)}>0\), we update the coefficient as,
\begin{equation}
\begin{aligned}
 \widetilde\beta_k
 &=\beta_k^{(m)}-\eta_k^{(m)}
   \nabla_\beta[-Q_{n,k}^{\mathrm R,(m)}](\beta_k^{(m)}),\\
 \beta_k^{\mathrm{cand}}
 &=\frac{\widetilde\beta_k}{\|\widetilde\beta_k\|_2}.
\end{aligned}
\label{eq:reward-coefficient-update}
\end{equation}
The line search accepts a normalized candidate only if
\begin{equation}
 Q_{n,k}^{\mathrm R,(m)}(\beta_k^{\mathrm{cand}})
 \ge Q_{n,k}^{\mathrm R,(m)}(\beta_k^{(m)}).
 \label{eq:reward-acceptance}
\end{equation}
An accepted candidate becomes \(\beta_k^{(m+1)}\).
We then set \(f_k^{(m+1)}=f_{\beta_k^{(m+1)}}\).

With the updated rewards fixed, we update each temperature by grid
search over \(\lambda>0\) for the one-dimensional objective
\begin{equation}
 Q_{n,h}^{\lambda,(m)}(\lambda)
 =\sum_{i=1}^n\sum_{k=1}^K r_{i,kh}^{(m)}
   \log\PP_{f_k^{(m+1)},\lambda}^{(T_i)}(\tau_i).
 \label{eq:temperature-block-objective}
\end{equation}
A candidate \(\widetilde\lambda_h>0\) becomes
\(\lambda_h^{(m+1)}\) only if
\begin{equation}
 Q_{n,h}^{\lambda,(m)}(\widetilde\lambda_h)
 \ge Q_{n,h}^{\lambda,(m)}(\lambda_h^{(m)}).
 \label{eq:temperature-update}
\end{equation}
These block updates give
\begin{equation}
 Q_n(\vartheta^{(m+1)}\mid\vartheta^{(m)})
 \ge Q_n(\vartheta^{(m)}\mid\vartheta^{(m)}),
 \label{eq:gem-ascent}
\end{equation}
and hence
\(\ell_n(\vartheta^{(m+1)})\ge\ell_n(\vartheta^{(m)})\).
Appendix~\ref{app:gem-numerics} specifies the numerical acceptance
checks and parameter constraints.

The main computational cost lies in solving component policies and the
\(KH\) adjoint systems. We exploit sparse transitions in Bellman
updates and use a matrix-free Krylov method for the adjoint systems.
Appendix~\ref{app:reward-gradients} gives the implementation and
cost details.

\subsection{Initialization and Multistart Optimization}
\label{sec:epi-initialization}

GEM can reach different local solutions from different initial
values \citep{wu1983}.
We use \emph{empirical policy inversion} (EPI) to construct a
reference reward and temperature, then a
\emph{multi-start initialization scheme} (MIS) to generate
additional configurations around this reference.

EPI pools the training counts
\(N(s,a)=\sum_i\sum_{t=0}^{T_i-1}\ind\{s_{it}=s,a_{it}=a\}\),
with \(N(s)=\sum_aN(s,a)\), and estimates the empirical policy
using a smoothing constant \(\epsilon>0\),
\begin{equation}
  \widehat\pi_{\mathrm E}(a\mid s)
  =\frac{N(s,a)+\epsilon}{N(s)+\epsilon|\mathcal A|}.
  \label{eq:epi-policy}
\end{equation}
At a reference temperature \(\lambda_0>0\), we construct a centered
tabular reward whose soft-optimal policy is
\(\widehat\pi_{\mathrm E}\).
We project this reward onto the feature span and jointly rescale
its coefficient and temperature to obtain the normalized reference
\((\widehat\beta_{\mathrm E},\widehat\lambda_{\mathrm E})\), with
\(\widehat f_{\mathrm E}=f_{\widehat\beta_{\mathrm E}}\).

MIS varies reward directions on the unit sphere and temperature
spacing on the log scale. We combine the reward and temperature
configurations, assign uniform weights \(\omega_{kh}^{(0)}=1/(KH)\),
and include the EPI initialization to form the bank
\(\mathcal B_{K,H}\).
Appendix~\ref{app:epi-multistart} gives the full EPI and MIS
constructions.

Algorithm~\ref{alg:gem-overview} fits the model from these initial
values and selects the converged estimate with the largest training
log likelihood, denoted by \(\widehat\vartheta_{K,H}\).
Optional screening after short GEM runs reduces computation by
continuing only the best \(L\) eligible fits; otherwise, all
initializations are run to convergence.

\begin{algorithm}[htbp]
\caption{GEM estimation for a reward--temperature mixture}
\label{alg:gem-overview}
\small
\begin{flushleft}
\textbf{Input:} trajectories \(\mathcal D_n\), the known environment,
order \((K,H)\), the reward specification, and shortlist size \(L\).
\end{flushleft}
\begin{enumerate}[leftmargin=1.6em,itemsep=3pt,topsep=4pt]
\item Construct \(\mathcal B_{K,H}\) by EPI and MIS.
\item Run GEM from each initial value. In each cycle, compute
responsibilities and \(Q_n\) using \eqref{eq:gem-responsibility}
and \eqref{eq:gem-Q}; update the weights, rewards, and temperatures
using \eqref{eq:mixture-weight-update},
\eqref{eq:reward-coefficient-update}, and
\eqref{eq:temperature-block-objective}, subject to
\eqref{eq:reward-acceptance} and \eqref{eq:temperature-update}.
\item If screening is used, retain the \(L\) eligible fits with
the largest training log likelihoods after short GEM runs, as in
Appendix~\ref{app:gem-numerics}. Continue these fits to convergence;
otherwise, continue all fits.
\item Return the converged estimate with the largest training
log likelihood as \(\widehat\vartheta_{K,H}\).
\end{enumerate}
\begin{flushleft}
\textbf{Output:} \(\widehat\vartheta_{K,H}\).
\end{flushleft}
\end{algorithm}

\subsection{Model Selection}
\label{sec:gem-selection-classification}

Let \(\mathcal K\subset\mathbb N^2\) be a finite, nonempty set of
candidate orders with \(\mathcal R_{K,H}(\Fsn)\ne\varnothing\).
For each candidate, fit \(\widehat\vartheta_{K,H}\) and compute
the Bayesian information criterion \citep{schwarz1978}:
\begin{equation}
  \mathrm{BIC}(K,H)
  =-2\ell_n(\widehat\vartheta_{K,H})+d_{K,H}\log n,
  \qquad d_{K,H}=K(p-1)+H+KH-1.
  \label{eq:bic}
\end{equation}
Each normalized reward coefficient lies on \(\mathbb S^{p-1}\)
and contributes \(p-1\) free parameters; the temperatures contribute
\(H\), and the simplex-constrained weights contribute \(KH-1\).
The penalty uses the number \(n\) of independent trajectories.
We select
\((\widehat K,\widehat H)\in
\arg\min_{(K,H)\in\mathcal K}\mathrm{BIC}(K,H)\)
and write
\(\widehat\vartheta=\widehat\vartheta_{\widehat K,\widehat H}\).

\section{Simulation Study}
\label{sec:simulation}
\label{sec:simulation-framework}

We assess parameter recovery under correct specification, compare
\(\RT(K,H)\) with \(\RO(KH,1)\), and examine joint selection of
the reward and temperature orders.
The experiments use an \(8\times10\) grid with nine deterministic
actions (stay or move in one of eight compass directions) and
\(\gamma=0.95\).
Rewards favor moves toward specified targets and are centered over
actions and normalized to unit \(\nu\)-norm.
All trajectories within a setting have length \(T_i=T\), and models
are fitted by the procedure in Section~\ref{sec:gem}.
Appendix~\ref{app:sim-reward-spec} gives the environment and
generating parameters.

Each evaluation trajectory is assigned to the cell with the largest
fitted posterior probability in \eqref{eq:gem-responsibility}, giving
its maximum a posteriori (MAP) assignment.
The adjusted Rand index (ARI) measures agreement with the generating
cells, and \(D_f\) measures root-mean-square reward error after
optimal label matching.
Predictive fit is measured by the mixture negative log-likelihood
per action, \(\NLL_{\mathrm{eva}}\), on independent held-out
trajectories from the same generating model.
Occupancy error \(D_{\Occ}\) is the \(\ell_1\) distance between
the fitted and generating expected proportions of visits to
state--action pairs over \(T\) steps.
Higher ARI and lower \(D_f\), \(\NLL_{\mathrm{eva}}\), and
\(D_{\Occ}\) indicate better performance.

Each experiment uses 50 independent Monte Carlo replications.
Figure bands are pointwise \(95\%\) Monte Carlo confidence
intervals for mean performance.
Appendix~\ref{app:sim-metrics} gives the metric definitions and
interval calculations.

\subsection{Parameter recovery}
\label{sec:recovery}

We generate data from \(\RT(2,2)\) and fit the same order, varying
\(n\in\{50,100,200,300\}\) and
\(T\in\{10,50,100,500,1000,2000\}\).
Cell weights are uniform, and the generating temperatures are
\(\lambda_1^0=0.5\) and \(\lambda_2^0=1.5\).
Each fit is evaluated on \(N_{\mathrm{eva}}=300\) independent
held-out trajectories.

\begin{figure}[!htbp]
\centering
\includegraphics[width=\linewidth]{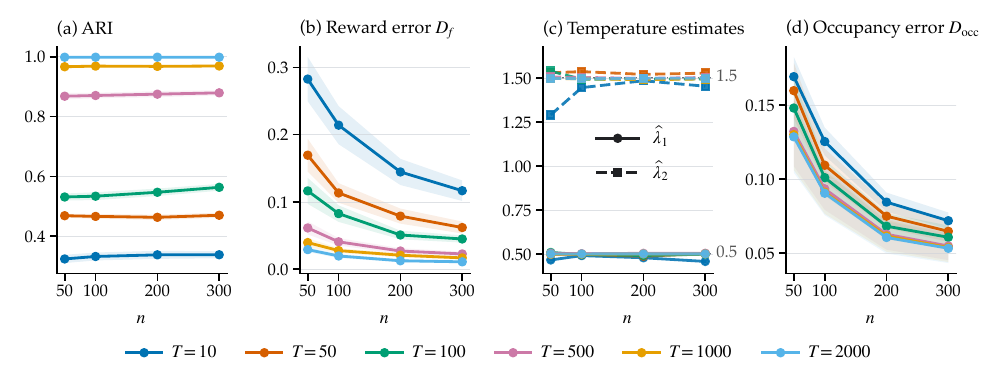}
\caption{Recovery under correctly specified \(\RT(2,2)\) across
\(n\) and \(T\): (a) cell ARI, (b) reward error \(D_f\),
(c) ordered temperature estimates, and (d) occupancy error \(D_{\Occ}\).
Colors denote \(T\). Points and shaded bands show means and their
pointwise \(95\%\) Monte Carlo confidence intervals from 50 replications.
In (c), circles with solid lines denote \(\widehat\lambda_1\),
and squares with dashed lines denote \(\widehat\lambda_2\);
dotted lines mark the generating values \(0.5\) and \(1.5\).}
\label{fig:main-exp1-recovery}
\end{figure}

At \(T=10\), larger training samples improve reward and occupancy
recovery but yield little improvement in cell classification
(Appendix Table~\ref{tab:exp1-rt2_2-complete}).
More training trajectories improve estimation of the shared
parameters, while each held-out trajectory still contains only ten
actions and can remain difficult to classify.
At fixed \(n\), longer trajectories improve both classification
and reward recovery (Figure~\ref{fig:main-exp1-recovery}).
Mean temperature estimates also approach their generating values,
and their Monte Carlo standard deviations decrease.

Appendix~\ref{app:sim-higher-results} gives complete numerical
summaries and recovery results for models with additional reward
or temperature classes.
Appendix~\ref{app:rare} examines zero and small cell weights.

\subsection{Comparison with a reward-only model}
\label{sec:separation}

To examine the consequences of Corollary~\ref{cor:strict-non-nesting},
we compare \(\RT(K,H)\) and \(\RO(KH,1)\) with the same number
of latent cells.
Data are generated from \(\RT(2,2)\), \(\RT(2,3)\),
\(\RT(2,4)\), and \(\RT(3,2)\), using the rewards,
temperatures, and uniform weights in Appendix~\ref{app:sim-reward-spec}.
For each setting, both models use the same \(n=300\) training
trajectories and \(N_{\mathrm{eva}}=300\) evaluation trajectories,
with \(T\in\{500,1000\}\).

Table~\ref{tab:rt-ro-gaps} reports paired differences
\(\Delta_{\NLL}=\NLL_{\mathrm{eva}}^{\RO}-\NLL_{\mathrm{eva}}^{\RT}\)
and \(\Delta_{\Occ}=D_{\Occ}^{\RO}-D_{\Occ}^{\RT}\);
positive values favor RT.
We report medians and interquartile intervals to reduce the
influence of atypical fits that may arise from suboptimal local optima.

\begin{table}[!htbp]
\centering
\begin{threeparttable}
\caption{Paired comparison of \(\RT(K,H)\) and \(\RO(KH,1)\):
median \([Q_{25},Q_{75}]\) of RO-minus-RT differences.}
\label{tab:rt-ro-gaps}
\small
\setlength{\tabcolsep}{6pt}
\renewcommand{\arraystretch}{1.12}
\begin{tabular}{@{}lccc@{}}
\toprule
& & \multicolumn{2}{c}{RO minus RT} \\
\cmidrule(lr){3-4}
True model & \(T\)
& \(\Delta_{\NLL}\;(\times 10^{-3})\)
& \(\Delta_{\Occ}\;(\times 10^{-2})\) \\
\midrule
\multirow{2}{*}{\(\RT(2,2)\)}
& 500  & \(1.827\;[1.635,\,1.935]\) & \(1.62\;[0.66,\,2.65]\) \\
& 1000 & \(2.163\;[2.061,\,2.254]\) & \(1.66\;[0.57,\,2.49]\) \\
\addlinespace[3pt]
\multirow{2}{*}{\(\RT(2,3)\)}
& 500  & \(0.915\;[0.838,\,1.012]\) & \(0.35\;[0.13,\,1.04]\) \\
& 1000 & \(1.153\;[1.091,\,1.264]\) & \(0.68\;[0.28,\,1.45]\) \\
\addlinespace[3pt]
\multirow{2}{*}{\(\RT(2,4)\)}
& 500  & \(0.638\;[0.573,\,0.716]\) & \(0.32\;[0.13,\,0.75]\) \\
& 1000 & \(0.869\;[0.769,\,0.942]\) & \(0.35\;[0.10,\,0.69]\) \\
\addlinespace[3pt]
\multirow{2}{*}{\(\RT(3,2)\)}
& 500  & \(1.763\;[1.675,\,1.912]\) & \(0.83\;[0.33,\,1.79]\) \\
& 1000 & \(2.141\;[2.040,\,2.258]\) & \(0.83\;[0.12,\,1.42]\) \\
\bottomrule
\end{tabular}
\end{threeparttable}
\end{table}

All reported medians and interquartile intervals favor RT for
predictive fit and occupancy recovery.
Appendix~\ref{app:rtro} also shows better cell classification
and reward recovery for RT.
These comparisons suggest that additional reward classes do not
fully compensate for omitting temperature heterogeneity in these settings.

\subsection{Selection of \texorpdfstring{\(K\) and \(H\)}{K and H}}
\label{sec:selection}

We evaluate the BIC rule in \eqref{eq:bic} under five generating
models: \(\RT(1,1)\), \(\RO(3,1)\), \(\RT(2,2)\),
\(\RT(2,3)\), and \(\RT(3,2)\).
Each sample contains \(n=300\) trajectories of length \(T=1000\),
and candidate orders satisfy \(K,H\in\{1,2,3,4\}\).
Rewards follow Appendix~\ref{app:sim-reward-spec}; positive cell
weights and temperatures are specified in
Appendix~\ref{app:sim-order-design}.

\begin{table}[!htbp]
\centering
\caption{Order-selection counts (percentages) under standard BIC.}
\label{tab:bic-selection}
\small
\setlength{\tabcolsep}{9pt}
\renewcommand{\arraystretch}{1.10}
\begin{tabular}{@{}lccc@{}}
\toprule
True model
& Exact order
& Correct \(K\)
& Correct \(H\) \\
\midrule
\(\RT(1,1)\)
& \(50/50\ (100\%)\)
& \(50/50\ (100\%)\)
& \(50/50\ (100\%)\) \\
\(\RO(3,1)\)
& \(49/50\ (98\%)\)
& \(50/50\ (100\%)\)
& \(49/50\ (98\%)\) \\
\(\RT(2,2)\)
& \(50/50\ (100\%)\)
& \(50/50\ (100\%)\)
& \(50/50\ (100\%)\) \\
\(\RT(2,3)\)
& \(49/50\ (98\%)\)
& \(49/50\ (98\%)\)
& \(50/50\ (100\%)\) \\
\(\RT(3,2)\)
& \(45/50\ (90\%)\)
& \(48/50\ (96\%)\)
& \(47/50\ (94\%)\) \\
\bottomrule
\end{tabular}
\end{table}

The frequency of selecting both generating orders ranges from
\(90\%\) to \(100\%\) across these settings
(Table~\ref{tab:bic-selection}).
Appendix~\ref{app:bic-details} reports BIC differences across
candidate orders.

\FloatBarrier

\section{White-Stork Migration Application}
\label{sec:white-stork-application}

\subsection{Data and representation of migration decisions}
\label{sec:stork-data}

White-stork autumn trajectories differ in migration direction,
daily displacement, and time spent in the same area.
We use TRAIL-IRL to interpret different migratory preferences from
variation in action regularity under shared preferences.
We also examine associations of reward-class membership with age
and of inferred temperature with relative position along the
migration route.

\begin{figure}[!htbp]
\centering
\includegraphics[width=0.75\linewidth]{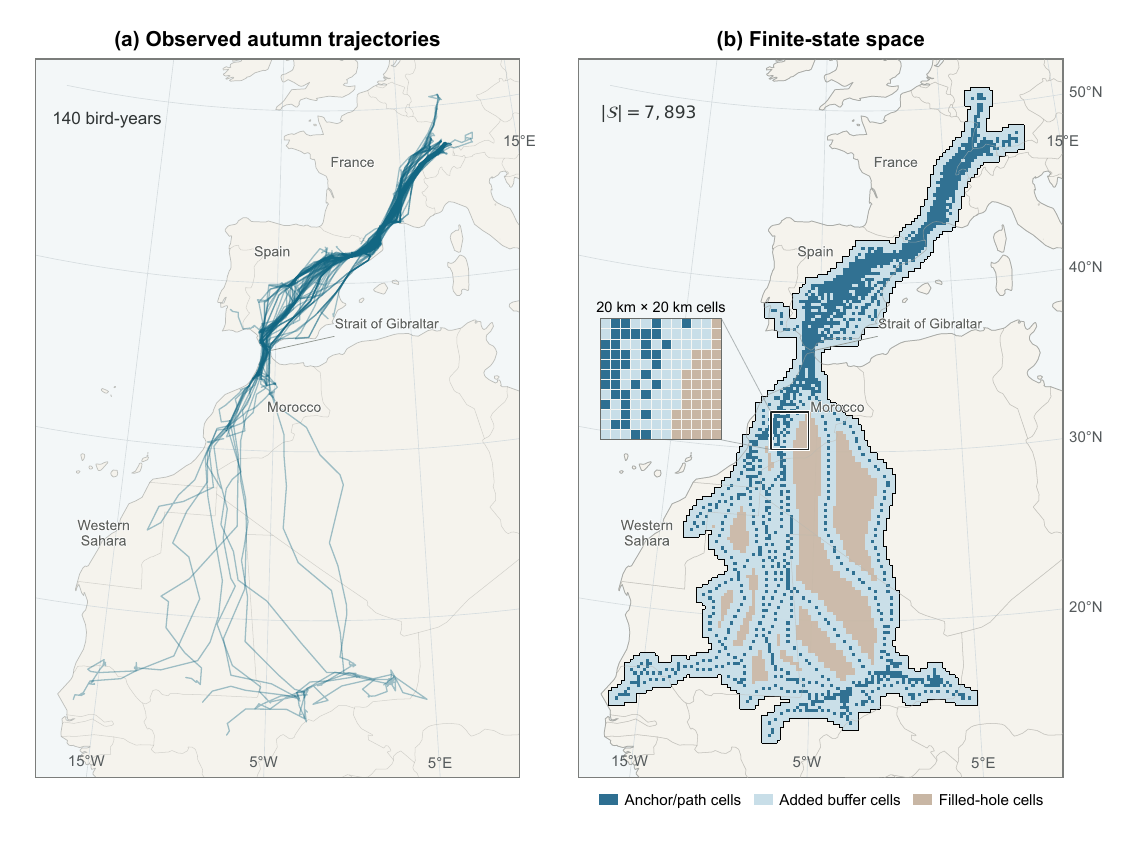}
\caption{(a) The 140 bird--year
trajectories used in the analysis. (b) The $20$ km grid covers the observed corridor, a
$40$ km outer buffer, and filled interior regions.}
\label{fig:stork-state-space}
\end{figure}

The Movebank LifeTrack White Stork SW Germany study records GPS
locations of individually tagged white storks (\textit{Ciconia ciconia}),
together with timestamps and bird characteristics
\citep{movebank_white_stork}.
We analyze observations from August~1 to October~31 each year,
taking each bird's trajectory during this period as a bird--year.
Following the daily sampling and filtering procedures in
Appendix~\ref{app:stork-anchors}, the sample comprises
140 bird--years from 79 birds, with 10,343 observed transitions
between daily locations recorded one day apart.
Figure~\ref{fig:stork-state-space}(a) shows the observed migration
corridor through western Europe and Africa.

We construct a grid map by extending the observed migration corridor and
dividing the resulting region into \(20\,\mathrm{km}\times20\,\mathrm{km}\)
cells.
Each cell represents one state, giving \(|\mathcal S|=7{,}893\).
Figure~\ref{fig:stork-state-space}(b) shows the complete state space,
comprising anchor/path cells, added buffer cells, and filled-hole cells.
Construction details are given in Appendix~\ref{app:stork-anchors}.

Each observed daily displacement is encoded as one of eight
compass directions or Stay, with Stay indicating that the two
daily locations lie in the same grid cell.
The shared stochastic kernel \(\widehat P\) gives the distribution
of the next-day cell conditional on the current cell and action.
We estimate this kernel from the pooled transitions as described
in Appendix~\ref{app:stork-kernel} and hold it fixed during
mixture fitting and model selection.
The discount factor is set to \(\gamma=0.95\).

We use normalized linear rewards with five features,
the expectations of the two next-day coordinates and of their
squares under \(\widehat P\), together with a Stay indicator,
constructed as detailed in Appendix~\ref{app:stork-features}.
These features capture spatial preferences and an additional
preference for remaining in the current cell.
We assume that the reward and temperature remain fixed throughout
each bird--year trajectory.

\subsection{Fitted reward--temperature structure}
\label{sec:stork-model-selection}

We fit candidate models with
\((K,H)\in\{1,\ldots,5\}\times\{1,\ldots,4\}\) by GEM.
Table~\ref{tab:stork-bic} reports BIC differences relative to the
minimum, attained by \(\RT(4,3)\).
The selected model contains four reward classes and three shared
temperature levels, with
\((\widehat\lambda_1,\widehat\lambda_2,\widehat\lambda_3)
=(0.544,0.801,1.048)\).
We refer to these as the low, medium, and high temperature levels,
respectively.

Moreover, at every reward-class count considered,
Table~\ref{tab:stork-bic} favors multiple temperature levels over
a common temperature.
This consistent pattern indicates that a finer reward classification
alone does not fully capture the heterogeneity in the migration
trajectories.
The fitted \(\RT\) model identifies an additional dimension of
this heterogeneity, i.e., shared levels of action regularity that extend
across reward classes.

\begin{table}[!htbp]
\centering
\caption{BIC differences across reward and temperature orders.
Each entry is the candidate BIC minus the minimum over the displayed
orders; smaller values are preferred. The $H=1$ column contains the
reward-only candidates.}
\label{tab:stork-bic}
\small
\setlength{\tabcolsep}{11pt}
\renewcommand{\arraystretch}{1.10}
\begin{tabular}{@{}lrrrr@{}}
\toprule
& $H=1$ & $H=2$ & $H=3$ & $H=4$ \\
\midrule
$K=1$ & 710.68 & 281.33 & 220.21 & 178.21 \\
$K=2$ & 290.81 & 82.42  & 43.69  & 42.78  \\
$K=3$ & 179.08 & 32.33  & 1.75   & 21.89  \\
$K=4$ & 198.16 & 24.68  & \textbf{0.00} & 13.78 \\
$K=5$ & 203.96 & 47.40  & 24.82  & 5.73 \\
\bottomrule
\end{tabular}
\end{table}

\subsection{Migratory preferences and their association with age}
\label{sec:stork-interpretation}

\begin{figure*}[!htbp]
\centering
\includegraphics[width=\linewidth]{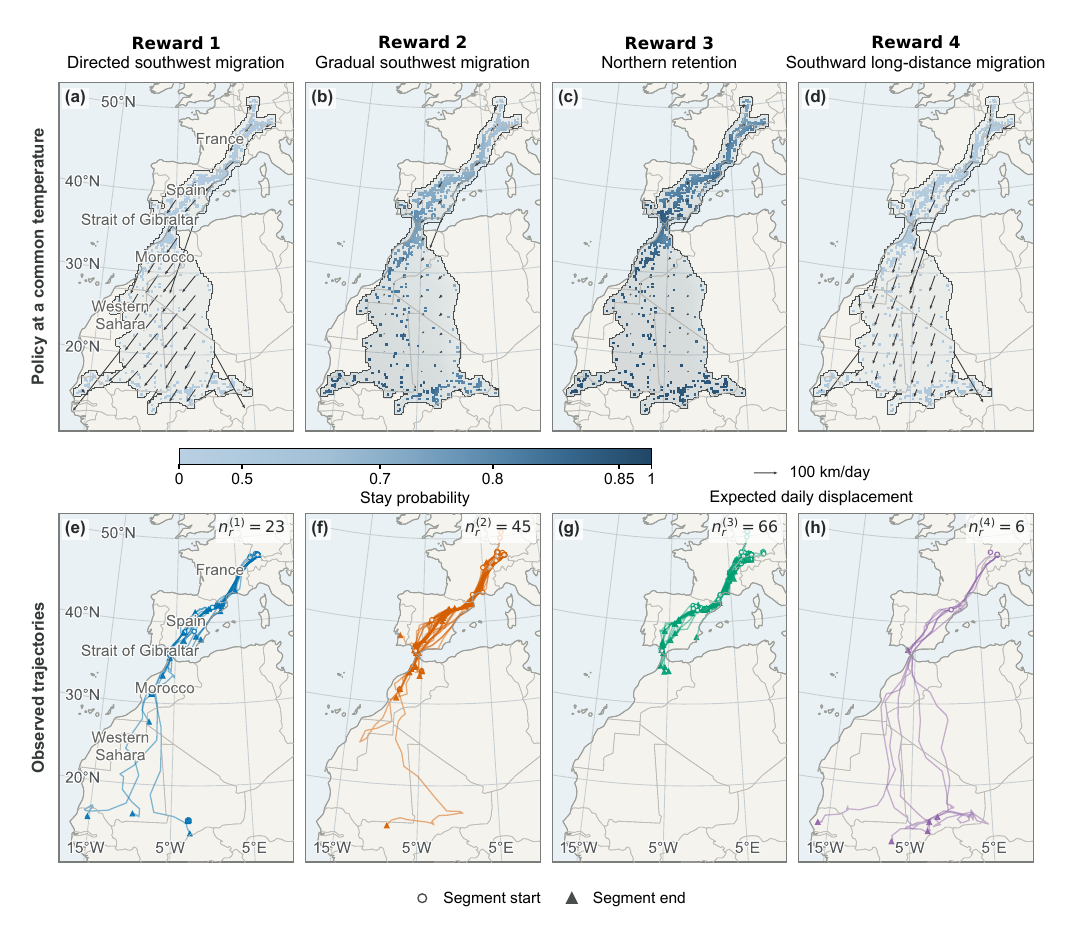}
\caption{Fitted policies and observed trajectories for Rewards~1--4.
In the top row, cell colors show stay probabilities and arrows
indicate expected daily displacement vectors (km/day).
These quantities are defined in \eqref{eq:stork-reference-policy-maps}
and evaluated at the common reference temperature
\(\lambda_\star=\widehat{\lambda}_2=0.801\).
The four policy maps share a nonlinear color scale.
Cells with no observed daily locations are faded.
The bottom row shows trajectories grouped by the marginal MAP
reward assignments defined in Appendix~\ref{app:stork-segments}.
Here, \(n_r^{(k)}\) denotes the number of bird--years assigned
to reward class \(k\), for \(k=1,\ldots,4\).
Lines connect locations within observed segments; open circles
and triangles mark segment starts and ends, respectively.}
\label{fig:stork-reward-maps}
\end{figure*}

We interpret the four fitted rewards by comparing their induced
policies at the common temperature
\(\lambda_\star=\widehat\lambda_2=0.801\) in
Figure~\ref{fig:stork-reward-maps}.
With temperature and the transition kernel held fixed, differences
between these policies reflect the fitted rewards.
Background colors show Stay probabilities, and arrows show expected
daily displacement vectors.\footnote{Each arrow averages the vector
from the center of the current grid cell to the center of the cell
occupied one day later over possible actions and destinations,
using the displayed policy and the shared transition kernel.
Arrows are expressed in km/day.}

The lower panels show observed trajectories grouped by their most
probable reward class.
Numerical summaries of these trajectories use \(\rho_{ik}\), the
posterior probability that bird--year \(i\) belongs to reward class
\(k\), as weights, following Appendix~\ref{app:stork-segments}.
In these summaries, daily displacement is the distance between
locations recorded one day apart.

Rewards~1 and~2 both favor southwestward movement.
At the common temperature, Reward~2 assigns higher Stay probabilities
along much of the northern corridor than Reward~1.
The observed Reward~1 group follows more direct paths and has a
greater mean daily displacement, approximately \(46\) km compared
with \(29\) km for Reward~2. This contrast in stopping and route directness is consistent with
evidence that juvenile white storks with longer stopovers make greater
route detours \citep{marcelino2023}.

Rewards~3 and~4 distinguish northern retention from migration to
more southerly locations.
The observed Reward~3 group combines frequent Stay actions with
limited daily displacement; the posterior-weighted proportion
of bird--years ending north of \(35^\circ\mathrm N\) is \(93.5\%\).
The Reward~4 group extends farther south, with a mean minimum
observed latitude near \(18^\circ\mathrm N\).
Comparable variation, from residency to short- and long-distance
migration, has been documented in white storks
\citep{flack2016,gilbert2016}.
At the common temperature, the displacement vectors for Reward~4
through North Africa and the Sahara point closer to due south than
those for Reward~1.

\paragraph{Association with age}
Southwestward migration under Reward~1 is associated with younger
birds, whereas northern retention under Reward~3 is associated with
older birds.
We define migration age as migration year minus birth year, with
age zero denoting the first autumn migration.
The posterior-weighted mean ages are \(0.17\) years for Reward~1
and \(2.40\) years for Reward~3.
We fitted linear regressions of posterior reward membership
probabilities on age, adjusting for calendar year and sex.
Each additional year of age was associated with a decrease of
\(5.5\) percentage points in the posterior probability of membership
in Reward~1
(\(95\%\) confidence interval for the decrease: \([1.8,9.2]\)
percentage points; FDR-adjusted \(p=0.0067\)).
For Reward~3, each additional year of age was associated with an
increase of \(10.8\) percentage points
(\(95\%\) confidence interval for the increase: \([5.3,16.3]\)
percentage points; FDR-adjusted \(p=0.0007\)).
Appendix~\ref{app:stork-biology} provides summaries for all four
reward classes and specifies the regression and inference procedures.

These age associations are consistent with the \textit{developmental
plasticity} explanation proposed by \citet{andrade2025} for
Portuguese white storks.
They argue that conditions and experience during development
shape adult migratory strategies.
The proportion of birds migrating to Africa declined from
\(98\%\) in the first year of life to \(67\%\) in the second
and \(33\%\) in the third.
Adult strategies were comparatively stable, with changes in only
\(3\) of \(113\) consecutive-year transitions.
Our age pattern is also consistent with \textit{the dominance
hypothesis}, outlined by \citet{boyle2008}, which predicts that
competitively dominant individuals, typically older birds, can
secure the local food needed to remain resident.
\citet{martins2024} provide evidence for this competitive advantage
in white storks: adults outcompeted juveniles at landfills and
consumed nearly four times as many food units in the core
feeding areas.
These findings suggest that age-related differences in access
to food may help explain the greater northern retention
of older birds.

\subsection{Action regularity and relative migration position}
\label{sec:stork-action-regularity}

Within each reward class, we examine how observed movement differs
among the fitted temperature levels.
We compare path directness, directional concentration, and Stay
proportion.
Path directness measures net displacement relative to distance
traveled, whereas directional concentration measures the alignment
of non-Stay movement directions.

Table~\ref{tab:stork-temperature} shows that path directness and
directional concentration decrease with temperature among the
reported groups within Rewards~1--3.
Directional concentration also separates the temperature levels
across reward classes: every reported low-temperature cell has a
higher mean than every medium-temperature cell, and every
medium-temperature cell has a higher mean than every
high-temperature cell.
Stay proportions also decline with temperature.
Thus, more frequent movement between grid cells accompanies
less direct and less directionally consistent paths.

\paragraph{Relative migration position}
We examine whether temperature is associated with relative migration
position.
We introduce a \textit{front score}, which measures how far ahead a bird
typically lies relative to others during migration.
Specifically, we rank the positions of birds in the same MAP reward
class on the same calendar date.
The ranking follows the class's migration direction, estimated by
averaging net displacements across all bird--years using their posterior
class probabilities as weights.
For bird--year \(i\), let \(R_{it}\) be its rank from back to front
among the \(n_{it}\) birds on its \(t\)-th comparison date, and let
\(m_i\) be the number of such dates.
We define \(F_i = \frac{1}{m_i} \sum_{t=1}^{m_i} \frac{R_{it}-1}{n_{it}-1}.\)
\footnote{Birds with the same projected position along the class's
migration direction receive the average of the ranks they would
otherwise occupy. For example, two birds tied for ranks \(3\) and \(4\)
each receive rank \(3.5\).}
The score ranges within $[0,1]$, where higher values indicate more advanced positions.
Appendix~\ref{app:stork-position} gives the further construction details.

We fitted a linear regression of front score on posterior mean
temperature, adjusting for reward class, calendar year, sex,
migration age, and first-migration status.
Each one-unit increase in posterior mean temperature was associated
with an increase of \(0.492\) in front score
(\(95\%\) confidence interval \([0.133,0.850]\);
FDR-adjusted \(p=0.021\)).
Together, the regression and movement summaries link higher
inferred temperature to lower directional concentration and
more forward migration positions.
Appendix~\ref{app:stork-position} describes the regression
specification and inference procedures.

These patterns are consistent with \textit{social information use}
and \textit{collective sensing}, as described in white storks by
\citet{flack2018,nagy2018}.
Social information use allows birds to locate favourable updrafts
by responding to the movements of flockmates.
\citet{flack2018} found that forward position in a flock of \(27\)
juvenile storks was positively correlated with circling
irregularity within thermals
(Pearson's \(r=0.570\), \(p=0.002\)).
They interpreted the irregular paths of birds ahead as a search
for favourable lift, while followers used these movements as
cues to locate thermal centres and circled more regularly.
\citet{nagy2018} examined \textit{collective sensing}, whereby individuals
sample different parts of a thermal and respond to one another's
movements, helping the group locate and exploit rising air.
They observed coordinated changes in circling direction that
helped birds remain within thermals.
These findings suggest that active search for favourable flight
conditions can accompany less regular movement in birds
occupying forward positions.

\begin{table*}[!htbp]
\centering
\caption{Observed trajectory summaries by reward class and temperature level.
Except for MAP counts, entries are means of bird--year summaries
weighted by posterior membership probabilities for the corresponding
reward--temperature group.
Reward labels match those in Figure~\ref{fig:stork-reward-maps}.}
\label{tab:stork-temperature}
\small
\setlength{\tabcolsep}{7pt}
\renewcommand{\arraystretch}{1.10}
\begin{tabular}{@{}lrrrrrr@{}}
\toprule
& \multicolumn{3}{c}{Reward 1}
& \multicolumn{3}{c}{Reward 2} \\
\cmidrule(lr){2-4}\cmidrule(lr){5-7}
Quantity & Low & Medium & High & Low & Medium & High \\
\midrule
MAP bird--years
& 17 & 4 & 2 & \(0^{\dagger}\) & 5 & 40 \\
Stay proportion
& 0.612 & 0.474 & 0.390 & --- & 0.744 & 0.605 \\
Mean daily displacement (km)
& 43.1 & 57.6 & 46.0 & --- & 23.2 & 29.7 \\
Path directness
& 0.919 & 0.845 & 0.744 & --- & 0.821 & 0.779 \\
Directional concentration
& 0.799 & 0.619 & 0.485 & --- & 0.636 & 0.480 \\
\addlinespace[5pt]
\midrule
& \multicolumn{3}{c}{Reward 3}
& \multicolumn{3}{c}{Reward 4} \\
\cmidrule(lr){2-4}\cmidrule(lr){5-7}
Quantity & Low & Medium & High & Low & Medium & High \\
\midrule
MAP bird--years
& 7 & 44 & 15 & \(0^{\dagger}\) & \(0^{\dagger}\) & 6 \\
Stay proportion
& 0.928 & 0.809 & 0.660 & --- & --- & 0.394 \\
Mean daily displacement (km)
& 8.4 & 15.1 & 18.1 & --- & --- & 59.6 \\
Path directness
& 0.751 & 0.743 & 0.645 & --- & --- & 0.678 \\
Directional concentration
& 0.857 & 0.524 & 0.321 & --- & --- & 0.465 \\
\bottomrule
\end{tabular}

\begin{minipage}{0.98\textwidth}
\footnotesize
\emph{Notes:} Low, medium, and high refer to the fitted temperatures
\(\widehat\lambda_1=0.544\), \(\widehat\lambda_2=0.801\), and
\(\widehat\lambda_3=1.048\), respectively.
MAP counts are based on joint reward--temperature assignments.
The symbol \(\dagger\) marks groups with no joint MAP assignments
and total posterior mass below \(0.01\) bird--years.
Movement summaries for these groups are omitted.
Definitions of the trajectory summaries are given in
Appendix~\ref{app:stork-metrics}.
\end{minipage}
\end{table*}

\section{Conclusion}
\label{sec:conclusion}
This paper studies how to cluster unlabeled trajectories while distinguishing heterogeneity in behavioral objectives from heterogeneity in action regularity. This distinction connects trajectory groups to the objectives individuals pursue and the consistency with which their actions follow those objectives. Joint recovery is challenging because group memberships are unknown, reward representations are nonunique, and reward scale is confounded with temperature. We propose TRAIL-IRL, a crossed reward–temperature mixture that jointly infers behavioral preferences, action regularity, and latent memberships. Statewise centering and reward normalization resolve shaping and scale ambiguities, and our theory establishes conditions under which finite trajectories identify the mixture structure. Estimation combines a generalized EM algorithm with empirical policy inversion for initialization and BIC for selecting the numbers of reward and temperature classes.

Simulation studies show that longer trajectories improve parameter recovery and classification, and that TRAIL-IRL outperforms reward-only mixtures with the same number of components in prediction and classification. The white-stork analysis identifies four migratory preference classes and three levels of action regularity. Younger birds are more strongly associated with southwestward migration, whereas older birds are more strongly associated with northern retention. Within reward classes, higher temperatures are associated with less directionally consistent movement and positions farther ahead along the migration route. These findings reveal distinct dimensions of migration behavior through their associations with age and relative migration position.

The current model assumes that rewards and temperatures remain fixed within each trajectory and that individuals share a common transition mechanism. Future work could incorporate weather, habitat, and interactions among individuals, while allowing preferences and action regularity to change over time. Such extensions would enable the study of how migratory behavior adapts to environmental conditions and social information.

\FloatBarrier

\bibliographystyle{spr-ims-nameyear}
\bibliography{ref_aoas}

\clearpage

\begin{center}
{\Large\bfseries
Supplementary material for\\[3pt]
TRAIL-IRL: Crossed Reward--Temperature Mixtures
for Unlabeled Trajectory Clustering\par}
\end{center}

Appendix~\ref{app:literature-review} compares TRAIL-IRL with related
trajectory clustering and inverse reinforcement learning methods.
Appendix~\ref{app:proof-mixture-identification} proves the mixture
identifiability theorem and the RT--RO non-nesting result.
Appendix~\ref{app:hessian} examines the local curvature of the
single-component population negative log likelihood under reward
normalization.
Appendix~\ref{app:example} develops the one-state, three-action
example and proves its Kullback--Leibler lower bound.
Appendix~\ref{app:gem-computation} gives the reward-gradient
derivations, initialization procedures, and numerical details for GEM.
Appendix~\ref{app:simulation-details} provides the simulation
specifications and additional results on recovery, zero and rare
mixture cells, RT--RO comparisons, and model selection.
Appendix~\ref{app:stork-details} describes the white-stork data
processing, model fitting, trajectory summaries, and regression
analyses of migration age and relative migration position.
Appendix~\ref{app:continuous-state} extends the model to continuous
states with finite actions, gives a single-component identification
result, and develops GEM updates based on numerical quadrature.

\begin{center}
\small
\textbf{Main notation}
\par\smallskip

\renewcommand{\arraystretch}{1.15}
\setlength{\tabcolsep}{8pt}

\begin{tabularx}{\linewidth}{@{}l
  >{\raggedright\arraybackslash}X@{}}
\toprule
Symbol & Meaning \\
\midrule

\(\mathcal S,\mathcal A\)
& State and action spaces \\

\(P(s'\mid s,a)\)
& Probability of reaching \(s'\) after taking action \(a\)
in state \(s\) \\

\(\mu_0,\gamma\)
& Initial-state distribution and discount factor \\

\(n\)
& Number of observed trajectories \\

\(\tau_i,T_i\)
& Observed trajectory \(i\) and the number of actions it contains \\

\addlinespace[5pt]

\(K,H\)
& Numbers of reward classes and temperature classes \\

\(f_k\)
& Reward function for class \(k\) \\

\(\lambda_h\)
& Temperature for class \(h\) \\

\(\omega_{kh}\)
& Probability that a trajectory belongs to cell \((k,h)\) \\

\(\pi^*_{f_k,\lambda_h}\)
& Soft-optimal policy under reward \(f_k\)
and temperature \(\lambda_h\) \\

\(\Fsn\)
& Class of centered and normalized rewards \\

\(\vartheta\)
& Model order, rewards, temperatures, and mixing weights \\

\addlinespace[5pt]

\(\ell_n\)
& Observed-data log likelihood \\

\(r_{i,kh}^{(m)}\)
& Posterior probability that trajectory \(i\) belongs to
cell \((k,h)\), evaluated at GEM iteration \(m\) \\

\(Q_n\)
& Conditional expected complete-data log likelihood \\

\bottomrule
\end{tabularx}
\end{center}

\appendix
\counterwithin{algorithm}{section}

\section{Comparison with related methods}
\label{app:literature-review}

In Table~\ref{tab:literature-comparison-full},
\emph{Unlabeled clustering} indicates whether a method infers
trajectory groups without supplied group labels.
\emph{Objective inference} indicates whether rewards or goals
are inferred from observed behavior.
\emph{Heterogeneity modeled} lists the sources of variation
represented in each formulation.
\emph{Identifiability} summarizes uniqueness guarantees for
the inferred objectives and behavioral parameters under the
assumptions of each cited work.
``Not established'' means that the cited work provides no such
guarantee; ``N/A'' applies to methods without objective inference.

\begin{table}[!htbp]
\centering
\caption{Detailed comparison of representative trajectory clustering
and inverse reinforcement learning methods.}
\label{tab:literature-comparison-full}
\small
\setlength{\tabcolsep}{4pt}
\renewcommand{\arraystretch}{1.15}

\begin{tabularx}{\linewidth}{@{}
  >{\hsize=1.30\hsize\linewidth=\hsize\raggedright\arraybackslash}X
  >{\hsize=.70\hsize\linewidth=\hsize\centering\arraybackslash}X
  >{\hsize=.70\hsize\linewidth=\hsize\centering\arraybackslash}X
  >{\hsize=1.10\hsize\linewidth=\hsize\raggedright\arraybackslash}X
  >{\hsize=1.20\hsize\linewidth=\hsize\raggedright\arraybackslash}X
@{}}
\toprule
\textbf{Method and reference}
& \textbf{Unlabeled clustering}
& \textbf{Objective inference}
& \textbf{Heterogeneity modeled}
& \textbf{Identifiability} \\
\midrule

\multicolumn{5}{@{}l}{\textit{Trajectory clustering}} \\
\addlinespace[2pt]

TRACLUS\newline
\citet{lee2007}
& Yes
& No
& Segment geometry
& N/A \\
\addlinespace[3pt]

Regression mixtures\newline
\citet{gaffney1999}
& Yes
& No
& Curve shapes
& N/A \\
\addlinespace[3pt]

FCM\newline
\citet{james2003}
& Yes
& No
& Curve shapes
& N/A \\
\addlinespace[3pt]

Neural clustering\newline
\citet{yao2018}
& Yes
& No
& Trajectory patterns in learned representations
& N/A \\

\midrule
\multicolumn{5}{@{}l}{\textit{Single-reward IRL}} \\
\addlinespace[2pt]

LP-IRL\newline
\citet{ng2000}
& No
& Yes
& Not modeled
& Not established \\
\addlinespace[3pt]

MaxEnt IRL\newline
\citet{ziebart2008}
& No
& Yes
& Not modeled
& Not established \\

\midrule
\multicolumn{5}{@{}l}{\textit{Heterogeneous IRL}} \\
\addlinespace[2pt]

DPM-BIRL\newline
\citet{choi2012}
& Yes
& Yes
& Rewards
& Not established \\
\addlinespace[3pt]

MSRD\newline
\citet{chen2020}
& No
& Yes
& Strategy preferences
& Not established \\
\addlinespace[3pt]

MPLP-IRL /\newline
MPMCE-IRL\newline
\citet{yao2024}
& No
& Yes
& Planning horizons
& Feasible sets; conditional reward uniqueness \\
\addlinespace[3pt]

L-IBM\newline
\citet{jacob2024}
& No
& Yes
& Computational budgets
& Not established \\
\addlinespace[3pt]

IRLEED\newline
\citet{beliaev2025}
& No
& Yes
& Reward biases and decision precision
& Not established \\

\midrule
\multicolumn{5}{@{}l}{\textit{IRL identifiability}} \\
\addlinespace[2pt]

Reward identification\newline
\citet{cao2021}
& No
& Yes
& Known dynamics or discount factors
& Shared reward up to an additive constant \\
\addlinespace[3pt]

Multi-expert identification\newline
\citet{rolland2022}
& No
& Yes
& Known dynamics or discount factors
& Shared reward or linear reward parameters \\

\midrule
\textbf{TRAIL-IRL}\newline
This work
& Yes
& Yes
& Rewards and action regularity
& Mixture orders, shared parameters, and weights \\

\bottomrule
\end{tabularx}
\end{table}

\section{Proof of Theorem~\ref{thm:mixture-identification} and Corollary~\ref{cor:strict-non-nesting}}
\label{app:proof-mixture-identification}

We first show that a policy uniquely determines its reward and temperature
under the centering and normalization constraints. We then identify the
mixture components and their weights from trajectory distributions at all
finite horizons. Finally, a finite-dimensional representation of these
distributions gives a sufficient trajectory length for identification.

\subsection{Soft Bellman equation and reward equivalence}
\label{app:bellman-equivalence}

For a function $v:\mathcal S\to\mathbb R$, write
$P_av(s)=\sum_{s'\in\mathcal S}P(s'\mid s,a)v(s')$.
The soft Bellman operator is
\[
 (\mathcal B_{f,\lambda}v)(s)
 =\lambda\log\sum_{a\in\mathcal A}
 \exp\!\left\{\frac{f(s,a)+\gamma P_av(s)}{\lambda}\right\}.
\]

\begin{lemma}[Unique solution of the soft Bellman equation]
\label{lem:soft-bellman}
For every reward $f:\mathcal S\times\mathcal A\to\mathbb R$
and $\lambda>0$, the operator $\mathcal B_{f,\lambda}$ has a
unique fixed point $V^*_{f,\lambda}$.
\end{lemma}

\begin{proof}
Because $P(\cdot\mid s,a)$ is a probability distribution,
\[
 P_av(s)\le P_au(s)+\|v-u\|_\infty
 \qquad\text{for every }(s,a).
\]
Substituting this bound into the Bellman operator gives
\[
 (\mathcal B_{f,\lambda}v)(s)
 \le(\mathcal B_{f,\lambda}u)(s)
    +\gamma\|v-u\|_\infty.
\]
Interchanging $u$ and $v$ therefore yields
\[
 \|\mathcal B_{f,\lambda}v-\mathcal B_{f,\lambda}u\|_\infty
 \le\gamma\|v-u\|_\infty.
\]
Since $\gamma<1$, the contraction mapping theorem gives a unique
fixed point in $\mathbb R^{\mathcal S}$.
\end{proof}

The fixed point determines the policy in
\eqref{eq:soft-optimal-policy}. The next proposition verifies that
this policy attains the optimal value at every state.

\begin{proposition}[Unique entropy-regularized optimal policy]
\label{prop:soft-policy-optimality}
For every reward $f$ and $\lambda>0$, the policy
$\pi^*_{f,\lambda}$ in \eqref{eq:soft-optimal-policy} is strictly
positive and satisfies
\[
 V_{f,\lambda}^{\pi^*_{f,\lambda}}(s)
 =V^*_{f,\lambda}(s)
 =\max_{\pi}V_{f,\lambda}^{\pi}(s)
 \qquad\text{for every }s\in\mathcal S,
\]
where the maximum is over stationary policies.
It is the unique stationary policy attaining these maxima
at all states.
\end{proposition}

\begin{proof}
Write $V^*=V^*_{f,\lambda}$ and $\pi^*=\pi^*_{f,\lambda}$.
The Bellman equation gives
$\sum_a\pi^*(a\mid s)=1$, and the exponential formula makes
every action probability positive.

To compare $\pi^*$ with an arbitrary stationary policy $\pi$,
substitute
$\lambda\log\pi^*(a\mid s)
=f(s,a)+\gamma P_aV^*(s)-V^*(s)$ to obtain
\[
 V^*(s)
 -\sum_a\pi(a\mid s)
 \{f(s,a)+\gamma P_aV^*(s)-\lambda\log\pi(a\mid s)\}
 =\lambda D_{\mathrm{KL}}\!\left(
   \pi(\cdot\mid s)\middle\|\pi^*(\cdot\mid s)\right).
\]
Here $D_{\mathrm{KL}}(r\|q)
=\sum_{a:r(a)>0}r(a)\log\{r(a)/q(a)\}$, with $0\log0=0$.
This divergence is nonnegative and vanishes exactly when $r=q$.

Apply the identity along a trajectory generated by $\pi$,
take expectations, and sum with discount weights $\gamma^t$.
The continuation values telescope. Since $V^*$ is bounded,
the terminal term $\gamma^m\EE_s^\pi V^*(S_m)$ vanishes as
$m\to\infty$, leaving
\[
 V^*(s)-V_{f,\lambda}^{\pi}(s)
 =\lambda\sum_{t=0}^{\infty}\gamma^t
 \EE_s^\pi D_{\mathrm{KL}}\!\left(
   \pi(\cdot\mid S_t)\middle\|\pi^*(\cdot\mid S_t)\right)
 \ge0.
\]
All terms are finite because the state and action spaces are finite
and $\pi^*$ is strictly positive.

For $\pi=\pi^*$, the right-hand side is zero, so $\pi^*$ attains
$V^*$ at every state. If another stationary policy attains the
same value at every state, the $t=0$ term must vanish for each
initial state $s$. Hence
$\pi(\cdot\mid s)=\pi^*(\cdot\mid s)$ for every $s$.
\end{proof}

\begin{corollary}[Common scaling of reward and temperature]
\label{cor:reward-temperature-scaling}
For every reward $f$, temperature $\lambda>0$, and constant $c>0$,
\begin{equation}
 V^*_{cf,c\lambda}=cV^*_{f,\lambda},
 \qquad
 \pi^*_{cf,c\lambda}=\pi^*_{f,\lambda}.
 \label{eq:appendix-scale-equivalence}
\end{equation}
\end{corollary}

\begin{proof}
For every stationary policy $\pi$, the value definition gives
$V_{cf,c\lambda}^{\pi}=cV_{f,\lambda}^{\pi}$.
Multiplication by $c>0$ scales the optimal value and preserves
the unique maximizing policy.
\end{proof}

We next characterize the rewards that induce a given policy.
The following identity is Theorem~1 of \citet{cao2021},
stated in our notation.

\begin{lemma}[Rewards inducing a given policy]
\label{lem:reward-equivalence}
Let $\pi$ be a strictly positive stationary policy and
let $\lambda>0$. A reward $f$ satisfies
$\pi^*_{f,\lambda}=\pi$ if and only if there exists a function
$v:\mathcal S\to\mathbb R$ such that
\begin{equation}
 f(s,a)
 =\lambda\log\pi(a\mid s)+v(s)-\gamma P_av(s)
 \qquad\text{for every }(s,a).
 \label{eq:appendix-policy-inversion}
\end{equation}
In this representation, $V^*_{f,\lambda}=v$.
\end{lemma}

\begin{proof}
Suppose first that $\pi^*_{f,\lambda}=\pi$.
Taking logarithms in \eqref{eq:soft-optimal-policy} gives
\[
 f(s,a)
 =\lambda\log\pi(a\mid s)
  +V^*_{f,\lambda}(s)-\gamma P_aV^*_{f,\lambda}(s),
\]
which is \eqref{eq:appendix-policy-inversion} with
$v=V^*_{f,\lambda}$.

Conversely, suppose $f$ has the form
\eqref{eq:appendix-policy-inversion}. Substituting it into
the Bellman operator gives
\[
 (\mathcal B_{f,\lambda}v)(s)
 =\lambda\log\sum_a
   \exp\!\left\{\log\pi(a\mid s)+\frac{v(s)}{\lambda}\right\}
 =v(s),
\]
because $\sum_a\pi(a\mid s)=1$.
Thus $v$ is a fixed point, and Lemma~\ref{lem:soft-bellman}
implies $v=V^*_{f,\lambda}$.
The policy formula then yields
\[
 \pi^*_{f,\lambda}(a\mid s)
 =\exp\!\left\{
   \frac{f(s,a)+\gamma P_av(s)-v(s)}{\lambda}\right\}
 =\pi(a\mid s).
\]
\end{proof}

Allowing the temperature to vary, the reward--temperature pairs
that induce $\pi$ form the equivalence class
\begin{equation}
 \mathcal E(\pi)
 =\left\{(f,\lambda):
 \begin{array}{l}
 \lambda>0,\quad
 f(s,a)=\lambda\log\pi(a\mid s)+v(s)-\gamma P_av(s)\\
 \text{for some }v:\mathcal S\to\mathbb R
 \text{ and every }(s,a)\in\mathcal S\times\mathcal A
 \end{array}\right\}.
 \label{eq:appendix-equivalence-class}
\end{equation}
For fixed $\pi$ and $\lambda$, each choice of $v$ determines
a reward with optimal value $v$ and optimal policy $\pi$.
The policy alone therefore does not determine the reward.

\begin{remark}[Degrees of freedom]
Given a strictly positive stationary policy $\pi$, we must specify
$|\mathcal S|+1$ additional independent scalars to determine
$(f,\lambda)$ uniquely. These can be taken as the temperature
$\lambda$ and the $|\mathcal S|$ values $v(s)$ in
\eqref{eq:appendix-equivalence-class}.
This motivates the constraints in \eqref{eq:Fsn-definition}:
statewise centering removes the $|\mathcal S|$ degrees of freedom
associated with the potential function, and the unit-norm
constraint fixes the common scale of reward and temperature.
\end{remark}

\subsection{Centered representatives and policy identification}
\label{app:normalized-identification}

The reward class $\Fsn$ used in
Section~\ref{sec:normalized-reward-class} imposes statewise
centering and unit $\nu$-norm on the linear rewards $f_\beta$.
We show that every pair $(f_\beta,\lambda)$ with $\beta\ne0$
and $\lambda>0$ has a unique equivalent representative in
$\Fsn\times(0,\infty)$.
Existence ensures that imposing these constraints preserves the
policy generated by the original pair. Uniqueness ensures that
the policy identifies the reward and temperature once these
constraints hold.

To prove existence, we first construct a statewise centered
reward that induces the same policy at the original temperature. To satisfy the centering condition in
\eqref{eq:Fsn-definition}, we seek a potential adjustment
$f(s,a)-h(s)+\gamma P_ah(s)$ whose weighted mean over actions
is zero at every state.
Using the weights $\omega_s(a)$ in
\eqref{eq:Fsn-definition}, define the action-averaged
transition matrix and reward by
\[
 P_\omega(s,s')
 =\sum_a\omega_s(a)P(s'\mid s,a),
 \qquad
 \bar f(s)=\sum_a\omega_s(a)f(s,a).
\]
The centering condition then becomes
$(I-\gamma P_\omega)h=\bar f$.
If this equation admits a solution $h$, then the adjusted reward
$f(s,a)-h(s)+\gamma P_ah(s)$ is statewise centered and induces
the same optimal policy as $f$ at the original temperature.
Since $P_\omega$ is stochastic and $\gamma<1$, the inverse
$(I-\gamma P_\omega)^{-1}
=\sum_{j=0}^{\infty}\gamma^jP_\omega^j$ exists.
Thus the required potential is uniquely determined, and we define
\begin{equation}
 h_f=(I-\gamma P_\omega)^{-1}\bar f,
 \qquad
 (Cf)(s,a)=f(s,a)-h_f(s)+\gamma P_ah_f(s).
 \label{eq:appendix-canonicalization}
\end{equation}
The next proposition verifies that this construction satisfies
the centering condition while remaining in the same policy
equivalence class.

\begin{proposition}[Centering without changing the policy]
\label{prop:policy-preserving-centering}
For every reward $f$ and temperature $\lambda>0$, the reward
$Cf$ is statewise centered and satisfies
$\pi^*_{Cf,\lambda}=\pi^*_{f,\lambda}$.
\end{proposition}

\begin{proof}
The defining equation for $h_f$ gives
\[
 \sum_a\omega_s(a)(Cf)(s,a)
 =\bar f(s)-h_f(s)+\gamma(P_\omega h_f)(s)=0.
\]
To verify that the policy is unchanged, write
$\pi=\pi^*_{f,\lambda}$ and $V=V^*_{f,\lambda}$.
By Lemma~\ref{lem:reward-equivalence},
$f(s,a)=\lambda\log\pi(a\mid s)+V(s)-\gamma P_aV(s)$.
Substituting this expression into the definition of $Cf$ gives
\[
 (Cf)(s,a)
 =\lambda\log\pi(a\mid s)
  +(V-h_f)(s)-\gamma P_a(V-h_f)(s).
\]
Lemma~\ref{lem:reward-equivalence} therefore implies that
$Cf$ induces the policy $\pi$ at temperature $\lambda$.
\end{proof}

The first constraint in \eqref{eq:Fsn-definition} is now
satisfied. To impose unit norm, we can divide both $Cf$ and
$\lambda$ by $\|Cf\|_\nu$, provided this norm is positive.
We first determine when the centered reward is zero.

\begin{lemma}[Zero centered reward and the uniform policy]
\label{lem:uniform-policy-characterization}
For every reward $f$ and temperature $\lambda>0$,
$Cf=0$ if and only if
$\pi^*_{f,\lambda}(a\mid s)=1/|\mathcal A|$
for every $(s,a)$.
\end{lemma}

\begin{proof}
The zero reward has constant optimal value
$\lambda\log|\mathcal A|/(1-\gamma)$ and the uniform policy.
Thus $Cf=0$ implies that $\pi^*_{Cf,\lambda}$ is uniform.
Proposition~\ref{prop:policy-preserving-centering} gives
the same conclusion for $\pi^*_{f,\lambda}$.

Conversely, suppose $\pi^*_{f,\lambda}$ is uniform.
The centered reward $Cf$ induces the same policy.
Applying Lemma~\ref{lem:reward-equivalence} with
$v=V^*_{Cf,\lambda}$ gives
\[
 (Cf)(s,a)
 =-\lambda\log|\mathcal A|+v(s)-\gamma P_av(s).
\]
Averaging over actions and using the centering of $Cf$ yields
\[
 (I-\gamma P_\omega)v
 =\lambda\log|\mathcal A|\mathbf1.
\]
Since $P_\omega\mathbf1=\mathbf1$, the unique solution is
$v=\lambda\log|\mathcal A|\mathbf1/(1-\gamma)$.
Substitution into the expression for $Cf$ gives $Cf=0$.
\end{proof}

Thus every policy that is nonuniform at some state admits
a nonzero centered reward. For such a policy, the common
scaling property in \eqref{eq:appendix-scale-equivalence}
allows us to impose the second constraint in
\eqref{eq:Fsn-definition}.

\begin{corollary}[Existence of a normalized representative]
\label{lem:canonical-representative}
Let $f$ be a reward and $\lambda>0$.
If $\pi^*_{f,\lambda}$ is nonuniform at some state, then
\begin{equation}
 \left(
 \frac{Cf}{\|Cf\|_\nu},
 \frac{\lambda}{\|Cf\|_\nu}
 \right)
 \label{eq:appendix-normalized-representative}
\end{equation}
has a statewise centered reward of unit $\nu$-norm
and induces the policy $\pi^*_{f,\lambda}$.
\end{corollary}

\begin{proof}
Lemma~\ref{lem:uniform-policy-characterization} gives
$Cf\ne0$, so the pair is well defined.
Proposition~\ref{prop:policy-preserving-centering} shows that
$Cf$ is centered and preserves the policy.
Dividing the reward and temperature by $\|Cf\|_\nu$
gives unit reward norm and preserves the policy by
\eqref{eq:appendix-scale-equivalence}.
\end{proof}

This construction establishes existence under the two
normalization constraints. To prove uniqueness, we compare two pairs in
$\Fsn\times(0,\infty)$ that induce the same policy.
We first show that centering alone identifies the reward
divided by its temperature.

\begin{proposition}[Identification of the reward-to-temperature ratio]
\label{prop:centered-ratio-identification}
Let $f,g$ be statewise centered rewards and let
$\lambda,\kappa>0$. If
$\pi^*_{f,\lambda}=\pi^*_{g,\kappa}$, then
$f/\lambda=g/\kappa$.
\end{proposition}

\begin{proof}
Apply Lemma~\ref{lem:reward-equivalence} to the common policy.
Dividing the two reward representations by their respective
temperatures cancels the log-policy terms.
With
$\Delta V=V^*_{f,\lambda}/\lambda-V^*_{g,\kappa}/\kappa$,
we obtain
\[
 \frac{f(s,a)}{\lambda}-\frac{g(s,a)}{\kappa}
 =\Delta V(s)-\gamma P_a\Delta V(s).
\]
Both rewards are centered, so averaging over actions gives
$(I-\gamma P_\omega)\Delta V=0$.
The invertibility of $I-\gamma P_\omega$ implies
$\Delta V=0$, and hence $f/\lambda=g/\kappa$.
\end{proof}

The common ratio leaves only the reward--temperature scale
undetermined. Under the unit-norm constraint,
$\|f/\lambda\|_\nu=1/\lambda$, so this ratio also determines
the temperature and then the reward.

\begin{theorem}[Uniqueness under reward normalization]
\label{lem:policy-identification}
Let $f,g$ be statewise centered rewards with unit
$\nu$-norm, as specified in \eqref{eq:Fsn-definition},
and let $\lambda,\kappa>0$. Then
\[
 \pi^*_{f,\lambda}=\pi^*_{g,\kappa}
 \quad\Longrightarrow\quad
 f=g,\qquad \lambda=\kappa.
\]
\end{theorem}

\begin{proof}
Proposition~\ref{prop:centered-ratio-identification} gives
$f/\lambda=g/\kappa$.
Taking $\nu$-norms and using $\|f\|_\nu=\|g\|_\nu=1$ yields
$\lambda^{-1}=\kappa^{-1}$.
Thus $\lambda=\kappa$, and the equality of the scaled rewards
gives $f=g$.
\end{proof}

These results establish existence and uniqueness of a centered,
unit-norm representative for every reward--temperature pair
whose policy is nonuniform at some state.
We now apply them to the linear class $\Fsn$ used in the
mixture model. The feature conditions in
\eqref{eq:linear-feature-normalization} make every $f_\beta$
centered and give $\|f_\beta\|_\nu=\|\beta\|_2$.
Consequently, the representative can be obtained by normalizing
the coefficient vector and adjusting the temperature by the
same factor.

\begin{corollary}[Normalization of a linear reward]
\label{cor:linear-normalized-representative}
Assume \eqref{eq:linear-feature-normalization}.
For every $\beta\in\mathbb R^p\setminus\{0\}$ and $\lambda>0$,
the pair
\[
 \left(
 f_{\beta/\|\beta\|_2},
 \frac{\lambda}{\|\beta\|_2}
 \right)
 \in\Fsn\times(0,\infty)
\]
induces the same policy as $(f_\beta,\lambda)$.
\end{corollary}

\begin{proof}
Feature centering gives $\bar f_\beta=0$, so
$Cf_\beta=f_\beta$.
Whitening gives $\|Cf_\beta\|_\nu=\|\beta\|_2>0$.
The representative in
\eqref{eq:appendix-normalized-representative} is therefore
the displayed pair, whose reward belongs to $\Fsn$.
\end{proof}

The representative thus remains in the specified linear
feature space. To identify its coefficients from the policy,
we also need equality of reward functions to imply equality
of coefficients. The whitening condition supplies this
property.

\begin{corollary}[Identification of linear coefficients and temperature]
\label{cor:linear-policy-identification}
Under \eqref{eq:linear-feature-normalization}, the map
\[
 (\beta,\lambda)\longmapsto\pi^*_{f_\beta,\lambda}
\]
is injective on
$\{\beta\in\mathbb R^p:\|\beta\|_2=1\}\times(0,\infty)$.
\end{corollary}

\begin{proof}
Suppose two parameter pairs in this domain induce the same
policy. Their rewards are centered and have unit $\nu$-norm.
Theorem~\ref{lem:policy-identification} therefore identifies
the temperatures and gives $f_\beta=f_{\widetilde\beta}$.
By \eqref{eq:linear-feature-normalization},
\[
 \|\beta-\widetilde\beta\|_2
 =\|f_\beta-f_{\widetilde\beta}\|_\nu=0.
\]
Hence the coefficients also agree.
\end{proof}

Each component policy therefore uniquely determines its
normalized reward coefficients and temperature.
We next establish conditions under which trajectory
distributions identify these component policies.

\subsection{Finite-horizon identification for \texorpdfstring{$\RT(1,1)$}{RT(1,1)}}
\label{app:single-horizon}

Corollary~\ref{cor:linear-policy-identification} shows that
the policy of an $\RT(1,1)$ model uniquely determines its
normalized reward coefficients and temperature.
We now establish when the length-$T$ trajectory distribution
determines this policy.
The results below allow any probability distribution $\mu_0$
on $\mathcal S$; the required horizon depends on reachability
from its support.

To recover the policy $\pi(a\mid s)$ from the length-$T$ trajectory distribution,
it suffices that every state has positive probability of being
visited at some time before $T$.

We express this requirement using the support graph $G_P$
defined in \eqref{eq:support-graph}.
For $u,v\in\mathcal S$, let
$\operatorname{dist}_{G_P}(u,v)$ denote the length of a shortest
directed path from $u$ to $v$:
\[
 \operatorname{dist}_{G_P}(u,v)
 :=\min\left\{
 m\in\mathbb Z_{\ge0}:
 \begin{array}{l}
 \exists\,s_0,\ldots,s_m\in\mathcal S
 \text{ with }s_0=u,\ s_m=v,\\
 (s_{j-1},s_j)\in E
 \text{ for every }j=1,\ldots,m
 \end{array}
 \right\}.
\]
We use the convention $\min\varnothing=\infty$ when no path
exists. Allowing $m=0$ gives
$\operatorname{dist}_{G_P}(u,u)=0$.
Writing $\operatorname{supp}(\mu_0)=\{s\in\mathcal S:\mu_0(s)>0\}$,
define the largest shortest-path distance from this support by
\begin{equation}
 d_0:=\max_{v\in\mathcal S}
       \min_{u\in\operatorname{supp}(\mu_0)}
       \operatorname{dist}_{G_P}(u,v).
 \label{eq:appendix-reachability-depth}
\end{equation}

\begin{lemma}[Reachability before time $T$]
\label{lem:single-horizon}
Let $T\ge1$ be an integer and let $\pi$ be a stationary policy
satisfying $\pi(a\mid s)>0$ for every
$(s,a)\in\mathcal S\times\mathcal A$.
Under $\pi$ with initial distribution $\mu_0$, the condition
\[
 \forall v\in\mathcal S,\qquad
 \exists t\in\{0,\ldots,T-1\}:\quad
 \Pr(S_t=v)>0
\]
holds if and only if $T\ge d_0+1$.
\end{lemma}

\begin{proof}
If $\Pr(S_t=v)>0$, then some path of $t$ transitions from
a state in $\operatorname{supp}(\mu_0)$ to $v$ has positive
probability.
Every edge of that path belongs to $G_P$.

Conversely, take a directed path in $G_P$ from a state in
$\operatorname{supp}(\mu_0)$ to $v$.
Each edge can be realized by an action with positive transition
probability.
Since $\pi$ assigns positive probability to every action,
the path has positive probability under $\pi$.
Consequently,
\[
 \Pr(S_t=v)>0\text{ for some }t<T
 \quad\Longleftrightarrow\quad
 \min_{u\in\operatorname{supp}(\mu_0)}
 \operatorname{dist}_{G_P}(u,v)\le T-1.
\]
Requiring this inequality for every $v\in\mathcal S$ gives
$d_0\le T-1$.
\end{proof}

When $T\ge d_0+1$, each state has a positive-probability
conditioning event $\{S_t=s\}$ for some $t<T$.
The trajectory distribution then determines the policy
at every state through conditional action probabilities.

\begin{proposition}[Policy identification from a trajectory distribution]
\label{prop:single-component-policy-identification}
Fix $P$ and $\mu_0$.
Let $f,g:\mathcal S\times\mathcal A\to\mathbb R$ and
$\lambda,\kappa>0$.
If $d_0<\infty$ and $T\ge d_0+1$, then
\[
 \PP_{f,\lambda}^{(T)}(\cdot)
 =\PP_{g,\kappa}^{(T)}(\cdot)
 \quad\Longrightarrow\quad
 \pi^*_{f,\lambda}=\pi^*_{g,\kappa}.
\]
\end{proposition}

\begin{proof}
Suppose $\PP_{f,\lambda}^{(T)}=\PP_{g,\kappa}^{(T)}$.
Marginalizing over all coordinates except $(S_t,A_t)$ gives
\[
 \PP_{f,\lambda}^{(T)}(S_t=s,A_t=a)
 =\PP_{g,\kappa}^{(T)}(S_t=s,A_t=a)
\]
for every $t<T$ and $(s,a)\in\mathcal S\times\mathcal A$.
Summing over $a$ also gives
\[
 \PP_{f,\lambda}^{(T)}(S_t=s)
 =\PP_{g,\kappa}^{(T)}(S_t=s).
\]

By \eqref{eq:soft-optimal-policy},
$\pi^*_{f,\lambda}(a\mid s)>0$ for every $(s,a)$.
Since $T\ge d_0+1$, Lemma~\ref{lem:single-horizon}
ensures that, for each $s\in\mathcal S$, there exists
$t\in\{0,\ldots,T-1\}$ with
$\PP_{f,\lambda}^{(T)}(S_t=s)>0$.
Thus both denominators below are strictly positive.
For this $t$ and every $a\in\mathcal A$, stationarity gives
\begin{align*}
 \pi^*_{f,\lambda}(a\mid s)
 &=\frac{\PP_{f,\lambda}^{(T)}(S_t=s,A_t=a)}
         {\PP_{f,\lambda}^{(T)}(S_t=s)}\\
 &=\frac{\PP_{g,\kappa}^{(T)}(S_t=s,A_t=a)}
         {\PP_{g,\kappa}^{(T)}(S_t=s)}
 =\pi^*_{g,\kappa}(a\mid s).
\end{align*}
Hence $\pi^*_{f,\lambda}=\pi^*_{g,\kappa}$.
\end{proof}

For rewards in $\Fsn$, equality of policies also identifies
the normalized coefficients and temperatures by
Corollary~\ref{cor:linear-policy-identification}.
We thus obtain finite-horizon identification for $\RT(1,1)$.

\begin{corollary}[Parameter identification for $\RT(1,1)$]
\label{cor:single-component-identification}
Assume \eqref{eq:linear-feature-normalization},
$d_0<\infty$, and $T\ge d_0+1$.
For
$(f_\beta,\lambda),(f_{\widetilde\beta},\kappa)
\in\Fsn\times(0,\infty)$,
\[
 \PP_{f_\beta,\lambda}^{(T)}(\cdot)
 =\PP_{f_{\widetilde\beta},\kappa}^{(T)}(\cdot)
 \quad\Longrightarrow\quad
 \beta=\widetilde\beta,\qquad \lambda=\kappa.
\]
\end{corollary}

\begin{proof}
Proposition~\ref{prop:single-component-policy-identification}
gives $\pi^*_{f_\beta,\lambda}
=\pi^*_{f_{\widetilde\beta},\kappa}$.
Corollary~\ref{cor:linear-policy-identification} then yields
$\beta=\widetilde\beta$ and $\lambda=\kappa$.
\end{proof}

\begin{remark}[Horizon bounds]
Under the strong connectivity assumption of
Theorem~\ref{thm:mixture-identification},
a shortest directed path from any supported initial state
to any other state has at most $|\mathcal S|-1$ edges.
Thus $d_0\le|\mathcal S|-1$, and $T\ge|\mathcal S|$
suffices for identification in $\RT(1,1)$ for any initial
distribution $\mu_0$.
In the special case $\mu_0(s)>0$ for every $s\in\mathcal S$,
we have $d_0=0$, so $T\ge1$ suffices.
\end{remark}

Lemma~\ref{lem:single-horizon} also implies
$\sum_{t=0}^{T-1}\Pr(S_t=s)>0$ for every state when
$T\ge d_0+1$.
Appendix~\ref{app:hessian} uses these positive expected
visit counts to establish positive definiteness of the
$\RT(1,1)$ likelihood Hessian under normalization.

For a mixture, the conditional distribution of $A_t$ given
$S_t=s$ generally averages several component policies.
Identifying those policies therefore requires separating
the components.
The next subsection establishes this separation through
empirical state--action frequencies along long trajectories.

\subsection{Identification of mixture components}
\label{app:persistent-policies}

We now establish how trajectory distributions at all finite
horizons identify the component reward--temperature pairs and
their mixing weights in Theorem~\ref{thm:mixture-identification}.
In the model \eqref{eq:RT-mixture-law}, the cell $(k,h)$ is
selected once and remains fixed throughout the trajectory.
This persistence allows us to distinguish component policies
through their limiting empirical state--action frequencies.

We first establish the finite-state Markov chain results needed
to describe these limits.
The results hold for any fixed initial distribution $\mu_0$
on $\mathcal S$.

\begin{definition}[Irreducibility]
\label{def:irreducible-transition-matrix}
A transition matrix $Q$ on a finite nonempty set $\mathcal X$
is called irreducible if every state can be reached from
every other state with positive probability; that is,
\[
 \forall x,y\in\mathcal X,\qquad
 \exists n\in\mathbb{Z}_{>0}
 \quad\text{such that}\quad
 (Q^n)_{xy}>0.
\]
\end{definition}

\begin{lemma}[Stationary distribution of a finite irreducible chain]
\label{lem:finite-chain-stationary}
Let $Q$ be an irreducible transition matrix on a finite nonempty
set $\mathcal X$. There is a unique probability vector $d$ such
that
\[
 d^{\mathsf T}Q=d^{\mathsf T}.
\]
Moreover, $d(x)>0$ for every $x\in\mathcal X$.
The vector $d$ is called the stationary distribution of $Q$.
\end{lemma}

\begin{proof}
For any probability vector $v$, define
\[
 d_n^{\mathsf T}
 :=\frac1n\sum_{t=0}^{n-1}v^{\mathsf T}Q^t.
\]
Compactness of the probability simplex gives a convergent
subsequence with limit $d$. Since
\[
 d_n^{\mathsf T}Q-d_n^{\mathsf T}
 =\frac{v^{\mathsf T}Q^n-v^{\mathsf T}}{n}
 \longrightarrow0,
\]
the limit satisfies $d^{\mathsf T}Q=d^{\mathsf T}$.

Choose $x$ with $d(x)>0$. For every $y\in\mathcal X$,
irreducibility gives an integer $m\ge0$ such that
$(Q^m)_{xy}>0$. Hence
\[
 d(y)
 =(d^{\mathsf T}Q^m)_y
 \ge d(x)(Q^m)_{xy}>0.
\]

To prove uniqueness, suppose $Qh=h$ and choose a state where
$h$ attains its maximum. The identity
\[
 h(x)=\sum_y Q_{xy}h(y)
\]
forces every $y$ with $Q_{xy}>0$ to attain the same maximum.
Applying this argument along directed paths shows that $h$
is constant. Therefore
\[
 \ker(I-Q)=\operatorname{span}\{\mathbf1\},
 \qquad
 \operatorname{rank}(I-Q)=|\mathcal X|-1.
\]
The left nullspace of $I-Q$ is thus one-dimensional.
The normalization $d^{\mathsf T}\mathbf1=1$ makes the stationary
probability vector unique.
\end{proof}

The stationary distribution describes the empirical frequencies
even when the chain starts outside stationarity. The next lemma
establishes the convergence needed to separate the mixture
components.

\begin{lemma}[Convergence of empirical frequencies]
\label{lem:finite-chain-occupancy}
Let $Q$ be an irreducible transition matrix on a finite nonempty
set $\mathcal X$, and let $d$ be its stationary distribution.
Let $\rho$ be any probability distribution on $\mathcal X$.
Consider a Markov chain $(X_t)_{t\ge0}$ with initial distribution
$\rho$ and transition matrix $Q$.
Write $\EE_{\rho,Q}$ for expectation under the law of this chain.
Then
\[
 \lim_{n\to\infty}
 \EE_{\rho,Q}\!\left[
   \left(
     \frac1n\sum_{t=0}^{n-1}\mathbf1\{X_t=x\}-d(x)
   \right)^2
 \right]
 =0
 \qquad\text{for every }x\in\mathcal X.
\]
\end{lemma}

\begin{proof}
The proof of Lemma~\ref{lem:finite-chain-stationary} gives
$\operatorname{rank}(I-Q)=|\mathcal X|-1$.
Together with $d^{\mathsf T}(I-Q)=0$, this implies
\begin{equation}
 \operatorname{range}(I-Q)
 =\{g\in\mathbb R^{\mathcal X}:d^{\mathsf T}g=0\}.
 \label{eq:poisson-range}
\end{equation}
Indeed, the left-hand side is contained in the right-hand side,
and both spaces have dimension $|\mathcal X|-1$.

Fix $x\in\mathcal X$ and define
\[
 g(y):=\mathbf1\{y=x\}-d(x).
\]
Since $d^{\mathsf T}g=0$, equation~\eqref{eq:poisson-range}
provides a vector $h$ satisfying
\[
 (I-Q)h=g.
\]
Define
\[
 \mathcal H_t:=\sigma(X_0,\ldots,X_t),
 \qquad
 D_{t+1}:=h(X_{t+1})-(Qh)(X_t).
\]
The Markov property gives
\[
 \EE[D_{t+1}\mid\mathcal H_t]=0,
 \qquad
 |D_{t+1}|\le2\|h\|_\infty.
\]
For $t<u$, the variable $D_{t+1}$ is
$\mathcal H_u$-measurable, so
\[
 \EE[D_{t+1}D_{u+1}]
 =\EE\!\left[
   D_{t+1}\EE[D_{u+1}\mid\mathcal H_u]
 \right]
 =0.
\]
Consequently,
\[
 \EE\!\left[
   \left(\sum_{t=0}^{n-1}D_{t+1}\right)^2
 \right]
 =\sum_{t=0}^{n-1}\EE[D_{t+1}^2]
 \le4n\|h\|_\infty^2.
\]

The equation $(I-Q)h=g$ also gives the telescoping identity
\[
 \sum_{t=0}^{n-1}g(X_t)
 =h(X_0)-h(X_n)+\sum_{t=0}^{n-1}D_{t+1}.
\]
Using $(a+b)^2\le2a^2+2b^2$, we obtain
\[
 \EE\!\left[
   \left(\frac1n\sum_{t=0}^{n-1}g(X_t)\right)^2
 \right]
 \le
 \frac{8\|h\|_\infty^2}{n^2}
 +\frac{8\|h\|_\infty^2}{n}
 \longrightarrow0.
\]
Substituting the definition of $g$ proves the claim.
\end{proof}

We apply these results to the state--action process within each
mixture component. For
$\vartheta\in\mathcal R_{K,H}(\Fsn)$, write
\[
 \pi_{kh}:=\pi^*_{f_k,\lambda_h}.
\]
Conditional on cell $(k,h)$, the process
$X_t:=(S_t,A_t)$ is a Markov chain with transition matrix
\begin{equation}
 Q_{kh}\bigl((s,a),(s',a')\bigr)
 :=P(s'\mid s,a)\pi_{kh}(a'\mid s').
 \label{eq:cell-pair-kernel}
\end{equation}
Its initial distribution assigns probability
$\mu_0(s)\pi_{kh}(a\mid s)$ to $(s,a)$.
To apply Lemma~\ref{lem:finite-chain-occupancy}, it remains to
verify that $Q_{kh}$ is irreducible.

\begin{lemma}[Irreducibility of the component state--action chain]
\label{lem:cell-chain-irreducibility}
If the support graph $G_P$ in \eqref{eq:support-graph} is
strongly connected, then $Q_{kh}$ is irreducible for every
cell $(k,h)$.
\end{lemma}

\begin{proof}
Equation~\eqref{eq:soft-optimal-policy} gives
\[
 \pi_{kh}(a\mid s)>0
 \qquad\text{for every }(s,a)\in\mathcal S\times\mathcal A.
\]
Fix $(s,a),(u,b)\in\mathcal S\times\mathcal A$.
Choose $v_0$ such that $P(v_0\mid s,a)>0$.
Strong connectivity provides a directed path
$v_0,\ldots,v_\ell=u$ in $G_P$, where $\ell$ may be zero.
For each $j<\ell$, choose an action $c_j$ satisfying
\[
 P(v_{j+1}\mid v_j,c_j)>0.
\]
The definition of $Q_{kh}$ then gives
\begin{align*}
 &(Q_{kh}^{\ell+1})\bigl((s,a),(u,b)\bigr)\\
 &\quad\ge
 P(v_0\mid s,a)
 \left[
   \prod_{j=0}^{\ell-1}
   \pi_{kh}(c_j\mid v_j)P(v_{j+1}\mid v_j,c_j)
 \right]
 \pi_{kh}(b\mid u)
 >0.
\end{align*}
Thus every state--action pair is reachable from every other
pair under $Q_{kh}$.
\end{proof}

Assume henceforth that $G_P$ is strongly connected.
By Lemmas~\ref{lem:finite-chain-stationary}
and~\ref{lem:cell-chain-irreducibility}, each $Q_{kh}$ has a
unique stationary probability vector $d_{kh}$ satisfying
\[
 d_{kh}(s,a)>0
 \qquad\text{for every }(s,a)\in\mathcal S\times\mathcal A.
\]
For a length-$n$ trajectory, define its empirical
state--action frequencies by
\[
 \widehat d_n(s,a)
 :=\frac1n\sum_{t=0}^{n-1}
   \mathbf1\{(S_t,A_t)=(s,a)\}.
\]
Lemma~\ref{lem:finite-chain-occupancy}, applied to each coordinate,
gives mean-square convergence of $\widehat d_n$ to $d_{kh}$
under cell $(k,h)$. In particular,
\begin{equation}
 \lim_{n\to\infty}
 \PP_{f_k,\lambda_h}^{(n)}
 \left\{\|\widehat d_n-d_{kh}\|_2\ge\varepsilon\right\}
 =0
 \qquad\text{for every }\varepsilon>0.
 \label{eq:empirical-occupancy-limit}
\end{equation}
These conclusions hold for every initial distribution $\mu_0$;
neither full support nor aperiodicity is required.

To use the limiting frequencies for identification, we next
show that $d_{kh}$ determines the component policy.

\begin{proposition}[Policy recovery from stationary frequencies]
\label{prop:policy-from-stationary-occupancy}
For every cell $(k,h)$,
\begin{equation}
 \pi_{kh}(a\mid s)
 =\frac{d_{kh}(s,a)}
        {\sum_{b\in\mathcal A}d_{kh}(s,b)}
 \qquad
 \text{for every }(s,a)\in\mathcal S\times\mathcal A.
 \label{eq:policy-from-stationary-occupancy}
\end{equation}
\end{proposition}

\begin{proof}
Stationarity of $d_{kh}$ and
\eqref{eq:cell-pair-kernel} imply
\[
 d_{kh}(s,a)
 =\pi_{kh}(a\mid s)
   \sum_{r\in\mathcal S}\sum_{b\in\mathcal A}
   d_{kh}(r,b)P(s\mid r,b).
\]
Summing over $a$ gives
\[
 \sum_{a\in\mathcal A}d_{kh}(s,a)
 =\sum_{r\in\mathcal S}\sum_{b\in\mathcal A}
   d_{kh}(r,b)P(s\mid r,b).
\]
The left-hand side is strictly positive because $d_{kh}$ is
strictly positive. Dividing the first identity by the second
proves \eqref{eq:policy-from-stationary-occupancy}.
\end{proof}

It remains to identify the vectors $d_{kh}$ and their mixing
weights from the trajectory distributions. The statistic
$\widehat d_n$ is observable, and
\eqref{eq:empirical-occupancy-limit} describes its behavior
under each component. We can therefore compare the probabilities
that $\widehat d_n$ lies near each possible limit.

\begin{lemma}[Identification of persistent mixtures]
\label{lem:persistent-cell-identification}
Under the assumptions of
Theorem~\ref{thm:mixture-identification}, let
$\vartheta\in\mathcal R_{K,H}(\Fsn)$ and
$\widetilde\vartheta
\in\mathcal R_{\widetilde K,\widetilde H}(\Fsn)$.
If
\[
 \PP_\vartheta^{(m)}(\cdot)
 =\PP_{\widetilde\vartheta}^{(m)}(\cdot)
 \qquad\text{for every integer }m\ge1,
\]
then $\vartheta\sim\widetilde\vartheta$, where label equivalence
is defined in \eqref{eq:separate-label-equivalence}.
\end{lemma}

\begin{proof}
Since $\widehat d_n$ is determined by a length-$n$ trajectory,
the assumed equality of trajectory distributions gives
\[
 \PP_\vartheta^{(n)}\{\widehat d_n\in B\}
 =
 \PP_{\widetilde\vartheta}^{(n)}\{\widehat d_n\in B\}
\]
for every $n\ge1$ and every
$B\subseteq\mathbb R^{\mathcal S\times\mathcal A}$.
We use these equalities to identify the component stationary
distributions and their mixing weights.

Let $d_{kh}$ and
$\widetilde d_{\widetilde k\widetilde h}$ denote the stationary
distributions of the state--action chains for the components
of $\vartheta$ and $\widetilde\vartheta$, respectively.
By \eqref{eq:empirical-occupancy-limit}, conditional on each
component, $\widehat d_n$ converges in probability to that
component's stationary distribution.
Equation~\eqref{eq:policy-from-stationary-occupancy} and
Theorem~\ref{lem:policy-identification} show that each stationary
distribution determines its reward--temperature pair.
The pairwise distinct rewards and temperatures required by
regularity in
Section~\ref{sec:mixture-identifiability-definitions}
therefore imply that the component stationary distributions
are pairwise distinct within each model.

We recover each component weight from the probability that
$\widehat d_n$ lies near its stationary distribution.
To compare these probabilities across the two models, define
\begin{align*}
 \mathcal D
 &:=\{d_{kh}:1\le k\le K,\ 1\le h\le H\}\\
 &\quad{}\cup
 \{\widetilde d_{\widetilde k\widetilde h}:
   1\le\widetilde k\le\widetilde K,\
   1\le\widetilde h\le\widetilde H\},
\end{align*}
and enumerate its distinct elements as $d_1,\ldots,d_J$.
If $J>1$, choose a radius $\varepsilon>0$ such that
\[
 \varepsilon
 <\frac12\min_{j\ne\ell}\|d_j-d_\ell\|_2.
\]
If $J=1$, choose any $\varepsilon>0$.
For each $j$, define
\[
 B_j:=
 \left\{
   x\in\mathbb R^{\mathcal S\times\mathcal A}:
   \|x-d_j\|_2<\varepsilon
 \right\}.
\]
Each $B_j$ contains exactly one element of $\mathcal D$.

Under a component with $d_{kh}=d_j$,
equation~\eqref{eq:empirical-occupancy-limit} gives
$\PP_{f_k,\lambda_h}^{(n)}
 \{\widehat d_n\in B_j\}\to1$.
Under a component with $d_{kh}\ne d_j$, the triangle inequality
gives
\[
 \widehat d_n\in B_j
 \quad\Longrightarrow\quad
 \|\widehat d_n-d_{kh}\|_2
 \ge
 \|d_j-d_{kh}\|_2-\|\widehat d_n-d_j\|_2
 >\varepsilon.
\]
The same convergence result then gives
$\PP_{f_k,\lambda_h}^{(n)}
 \{\widehat d_n\in B_j\}\to0$.
Consequently,
\[
 \lim_{n\to\infty}
 \PP_{f_k,\lambda_h}^{(n)}
 \{\widehat d_n\in B_j\}
 =\mathbf1\{d_{kh}=d_j\}.
\]
The corresponding limit holds for every component of
$\widetilde\vartheta$.

Applying the mixture formula \eqref{eq:RT-mixture-law}
and the equality of the distributions of $\widehat d_n$,
we obtain
\begin{equation}
 \begin{aligned}
 \sum_{k,h:\,d_{kh}=d_j}\omega_{kh}
 &=
 \lim_{n\to\infty}
 \PP_\vartheta^{(n)}\{\widehat d_n\in B_j\}\\
 &=
 \lim_{n\to\infty}
 \PP_{\widetilde\vartheta}^{(n)}
 \{\widehat d_n\in B_j\}\\
 &=
 \sum_{\widetilde k,\widetilde h:
       \,\widetilde d_{\widetilde k\widetilde h}=d_j}
 \widetilde\omega_{\widetilde k\widetilde h}.
 \end{aligned}
 \label{eq:matched-occupancy-masses}
\end{equation}
Each $d_j$ occurs in at least one model.
Since regularity requires all mixing weights to be strictly
positive, \eqref{eq:matched-occupancy-masses} implies that
$d_j$ occurs in both models.
Distinctness of the stationary distributions within each
model then identifies exactly one component in each model
with stationary distribution $d_j$, and their weights agree.

For each such pair of components,
equation~\eqref{eq:policy-from-stationary-occupancy}
identifies their policies, and
Theorem~\ref{lem:policy-identification} identifies their
rewards and temperatures.
Thus the two models have the same reward--temperature pairs
and the same weight for each pair.

It remains to express these matches using the separate
reward and temperature permutations in
\eqref{eq:separate-label-equivalence}.
Taking the two coordinates of the matched pairs gives
\[
 \{f_1,\ldots,f_K\}
 =
 \{\widetilde f_1,\ldots,\widetilde f_{\widetilde K}\},
 \qquad
 \{\lambda_1,\ldots,\lambda_H\}
 =
 \{\widetilde\lambda_1,\ldots,
   \widetilde\lambda_{\widetilde H}\}.
\]
Pairwise distinctness implies $K=\widetilde K$ and
$H=\widetilde H$.
Hence there are permutations $\varpi$ of $\{1,\ldots,K\}$
and $\varkappa$ of $\{1,\ldots,H\}$ such that
\[
 \widetilde f_k=f_{\varpi(k)},
 \qquad
 \widetilde\lambda_h=\lambda_{\varkappa(h)}.
\]
The unique component of $\vartheta$ with pair
$(\widetilde f_k,\widetilde\lambda_h)$ is
$(\varpi(k),\varkappa(h))$.
Equality of the weights attached to this pair gives
\[
 \widetilde\omega_{kh}
 =\omega_{\varpi(k),\varkappa(h)}
 \qquad\text{for every }(k,h).
\]
Therefore $\vartheta\sim\widetilde\vartheta$ by
\eqref{eq:separate-label-equivalence}.
\end{proof}

Lemma~\ref{lem:persistent-cell-identification} establishes
label equivalence when the trajectory distributions agree
at every finite horizon.
We next show that agreement at one sufficiently long horizon
implies agreement at all finite horizons.

\subsection{Reduction to a finite horizon}
\label{app:finite-horizon-reduction}

Lemma~\ref{lem:persistent-cell-identification} identifies the
mixture parameters, up to label equivalence, from the trajectory
distributions at all finite horizons.
We now derive a finite horizon that suffices for identification.
The bound follows from finite-dimensional linear representations
of trajectory probabilities.
For two such representations, equality on sequences up to a
length determined by their dimensions implies equality on all
finite sequences, allowing us to apply
Lemma~\ref{lem:persistent-cell-identification}.

\begin{definition}[Finite-dimensional linear representation]
\label{def:linear-representation}
Let $\mathcal Y$ be a finite set. For each integer $m\ge1$,
define the set of length-$m$ sequences by
\[
 \mathcal Y^m
 :=\{(y_1,\ldots,y_m):y_i\in\mathcal Y,\ i=1,\ldots,m\}.
\]
Denote the sequence of length zero by $()$, and define
\[
 \mathcal Y^0:=\{()\},
 \qquad
 \mathcal Y^*:=\bigcup_{m=0}^{\infty}\mathcal Y^m.
\]
For $w\in\mathcal Y^m$, write $|w|:=m$.

Let $F:\mathcal Y^*\to\mathbb R$ assign a real value $F(w)$
to each finite sequence $w$.
For a positive integer $D$, consider vectors
$a,c\in\mathbb R^D$ and a matrix
$B_y\in\mathbb R^{D\times D}$ for each $y\in\mathcal Y$.
Collect these objects in the tuple
\[
 \mathsf L:=\bigl(a,(B_y)_{y\in\mathcal Y},c\bigr).
\]
For each sequence $w\in\mathcal Y^*$, define
\[
 B_w:=
 \begin{cases}
  I_D, & w=(),\\
  B_{y_1}\cdots B_{y_m},
    & w=(y_1,\ldots,y_m),\quad m\ge1,
 \end{cases}
\]
where $I_D$ is the $D\times D$ identity matrix.

We call $\mathsf L$ a $D$-dimensional linear representation
of $F$ if
\[
 F(w)=a^{\mathsf T}B_wc
 \qquad\text{for every }w\in\mathcal Y^*.
\]
\end{definition}

To establish equality of two functions with finite-dimensional
linear representations, it suffices to compare their values
on all sequences up to a finite length
\citep[Section~3]{kiefer2013}.
The following lemma gives a sufficient length bound in terms
of the dimensions of their representations.

\begin{lemma}[Finite equivalence of linear representations]
\label{lem:finite-linear-equivalence}
Let $F_1,F_2:\mathcal Y^*\to\mathbb R$ have linear
representations $\mathsf L_1,\mathsf L_2$ of dimensions
$D_1,D_2$, respectively.
If
\[
 F_1(w)=F_2(w)
 \qquad\text{for every }w\in\mathcal Y^*
 \text{ with }|w|\le D_1+D_2-1,
\]
then $F_1=F_2$ on $\mathcal Y^*$.
\end{lemma}

\begin{proof}
We first construct a linear representation of $F_1-F_2$.
For $j=1,2$, write
$\mathsf L_j=(a_j,(B_{j,y})_{y\in\mathcal Y},c_j)$.
Set $D:=D_1+D_2$ and define
\[
 a^{\mathsf T}:=(a_1^{\mathsf T},-a_2^{\mathsf T}),
 \qquad
 B_y:=
 \begin{pmatrix}
  B_{1,y}&0\\
  0&B_{2,y}
 \end{pmatrix},
 \qquad
 c:=
 \begin{pmatrix}
  c_1\\
  c_2
 \end{pmatrix}.
\]
Then
\[
 F_1(w)-F_2(w)=a^{\mathsf T}B_wc
 \qquad\text{for every }w\in\mathcal Y^*.
\]
If $c=0$, this difference is zero for every $w$.
Assume henceforth that $c\ne0$.

By hypothesis, $a^{\mathsf T}B_wc=0$ whenever $|w|\le D-1$.
To extend this equality to all sequences, it suffices to show
that the vectors $B_wc$ with $|w|\le D-1$ span those obtained
from sequences of arbitrary length.
For each integer $n\ge0$, define
\[
 \mathcal U_n
 :=\operatorname{span}
   \{B_wc:w\in\mathcal Y^*,\ |w|\le n\}.
\]
The empty sequence gives
$\mathcal U_0=\operatorname{span}\{c\}$, so
$\dim\mathcal U_0=1$.
Adding one entry at the beginning of a sequence gives
\[
 \mathcal U_{n+1}
 =\mathcal U_n+\sum_{y\in\mathcal Y}B_y\mathcal U_n.
\]
In particular,
$\mathcal U_n\subseteq\mathcal U_{n+1}\subseteq\mathbb R^D$.

If $\mathcal U_{n+1}=\mathcal U_n$, the displayed identity
implies $B_y\mathcal U_n\subseteq\mathcal U_n$ for every
$y\in\mathcal Y$.
It follows that
\[
 \mathcal U_{n+r}=\mathcal U_n
 \qquad\text{for every integer }r\ge0.
\]
Thus, once two consecutive subspaces are equal, no later
sequence produces a vector outside that subspace.
Each strict inclusion increases the dimension by at least one.
Since the dimensions start at $1$ and cannot exceed $D$,
the subspaces must stabilize by $n=D-1$.
Hence
\[
 \mathcal U_{D-1}
 =\operatorname{span}\{B_wc:w\in\mathcal Y^*\}.
\]

The hypothesis gives $a^{\mathsf T}x=0$ for every generator
$x=B_wc$ of $\mathcal U_{D-1}$.
By linearity, this equality holds throughout $\mathcal U_{D-1}$
and therefore for every vector $B_wc$.
Consequently,
\[
 F_1(w)-F_2(w)=a^{\mathsf T}B_wc=0
 \qquad\text{for every }w\in\mathcal Y^*.
\]
\end{proof}

To apply Lemma~\ref{lem:finite-linear-equivalence}, we encode
trajectories as finite sequences of transition triples.
Let $\mathcal Y:=\mathcal S\times\mathcal A\times\mathcal S$.
Consider a nonempty sequence $w=(y_0,\ldots,y_{m-1})$,
where $y_t=(u_t,a_t,v_t)$.
We call $w$ consistent if the destination state of each
transition equals the starting state of the next:
\[
 v_t=u_{t+1}
 \qquad\text{for every }0\le t<m-1.
\]
For a consistent sequence, the states $s_0,\ldots,s_m$
are uniquely determined by $y_t=(s_t,a_t,s_{t+1})$.
Define the corresponding trajectory by
\[
 \tau(w):=(s_0,a_0,\ldots,s_{m-1},a_{m-1},s_m).
\]

For $\vartheta\in\mathcal R_{K,H}(\Fsn)$, extend its trajectory
probabilities to a function
$F_\vartheta:\mathcal Y^*\to\mathbb R$ by defining
\[
 F_\vartheta(w):=
 \begin{cases}
  1,
    & w=(),\\
  \PP_\vartheta^{(|w|)}(\tau(w)),
    & |w|\ge1\text{ and }w\text{ is consistent},\\
  0,
    & \text{otherwise}.
 \end{cases}
\]
The function $F_\vartheta$ is therefore defined on all finite
sequences, as required by
Lemma~\ref{lem:finite-linear-equivalence}.

We next show that the mixture trajectory law in
\eqref{eq:RT-mixture-law} admits a linear representation
of dimension $KH|\mathcal S|$.

\begin{proposition}[Linear representation of trajectory distributions]
\label{lem:trajectory-linear-representation}
Let $\vartheta\in\mathcal R_{K,H}(\Fsn)$ and set
$D_\vartheta:=KH|\mathcal S|$.
The function $F_\vartheta$ has a $D_\vartheta$-dimensional
linear representation of the form
\[
 \mathsf L_\vartheta
 :=\bigl(\alpha_\vartheta,
         (B_{\vartheta,y})_{y\in\mathcal Y},
         \mathbf1\bigr),
\]
where $\mathbf1\in\mathbb R^{D_\vartheta}$ is the vector
of ones. This representation satisfies
\begin{equation}
 F_\vartheta(w)
 =\alpha_\vartheta^{\mathsf T}B_{\vartheta,w}\mathbf1
 \qquad\text{for every }w\in\mathcal Y^*.
 \label{eq:trajectory-linear-representation}
\end{equation}
\end{proposition}

\begin{proof}
Write $\pi_{kh}:=\pi^*_{f_k,\lambda_h}$.
Index the coordinates of $\mathbb R^{D_\vartheta}$ by
$(k,h,s)$ and define the initial vector by
\[
 (\alpha_\vartheta)_{(k,h,s)}
 :=\omega_{kh}\mu_0(s).
\]
For each transition triple $y=(u,a,v)\in\mathcal Y$, define
the matrix $B_{\vartheta,y}$ by
\begin{equation}
 (B_{\vartheta,y})_{(k,h,s),(k',h',s')}
 :=
 \mathbf1\{k=k',\ h=h',\ s=u,\ s'=v\}
 \pi_{kh}(a\mid u)P(v\mid u,a).
 \label{eq:trajectory-letter-matrix}
\end{equation}
Figure~\ref{fig:trajectory-letter-matrix} shows its block
diagonal structure, with one block for each component $(k,h)$.
\begin{figure}[htbp]
\centering
\begin{tikzpicture}[font=\small,>=Latex]
  \draw[fill=black!4,draw=black!35]
    (0,2.9) rectangle (1.1,4);
  \draw[fill=blue!10,draw=blue!65!black,thick]
    (1.45,1.45) rectangle (2.55,2.55);
  \draw[fill=black!4,draw=black!35]
    (2.9,0) rectangle (4,1.1);

  \node at (0.55,3.45) {$(1,1)$};
  \node at (2,2) {$(k,h)$};
  \node at (3.45,0.55) {$(K,H)$};
  \node at (1.275,2.725) {$\ddots$};
  \node at (2.725,1.275) {$\ddots$};

  \node at (2.8,3.3) {$\mathbf0$};
  \node at (1.2,0.7) {$\mathbf0$};

  \draw[thick]
    (-0.12,0)--(-0.3,0)--(-0.3,4)--(-0.12,4);
  \draw[thick]
    (4.12,0)--(4.3,0)--(4.3,4)--(4.12,4);
  \node at (2,4.45) {$B_{\vartheta,y}$};
  \node at (2,-0.4) {$D_\vartheta\times D_\vartheta$};

  \node (detail) at (9,2) {
    \setlength{\arraycolsep}{3pt}
    \renewcommand{\arraystretch}{1.25}
    $\begin{pmatrix}
       0 & \cdots & 0 & \cdots & 0\\
       \vdots & & \vdots & & \vdots\\
       0 & \cdots &
       \textcolor{blue!65!black}{
         \pi_{kh}(a\mid u)P(v\mid u,a)
       }
       & \cdots & 0\\
       \vdots & & \vdots & & \vdots\\
       0 & \cdots & 0 & \cdots & 0
     \end{pmatrix}$
  };

  \node[above=2mm of detail.north] {$v$};
  \node[left=1mm of detail.west] {$u$};
  \node[above=10mm of detail.north] {Component $(k,h)$};
  \node[below=3mm of detail.south]
    {$|\mathcal S|\times|\mathcal S|$};

  \draw[->,thick,blue!65!black]
    (4.55,2)--($(detail.west)+(-0.55,0)$);
\end{tikzpicture}
\caption{Block structure of $B_{\vartheta,y}$ for $y=(u,a,v)$.
Each diagonal block is labeled by its component $(k,h)$;
all off-diagonal blocks are zero.
Within each diagonal block, only the entry in row $u$ and
column $v$ can be nonzero.}
\label{fig:trajectory-letter-matrix}
\end{figure}
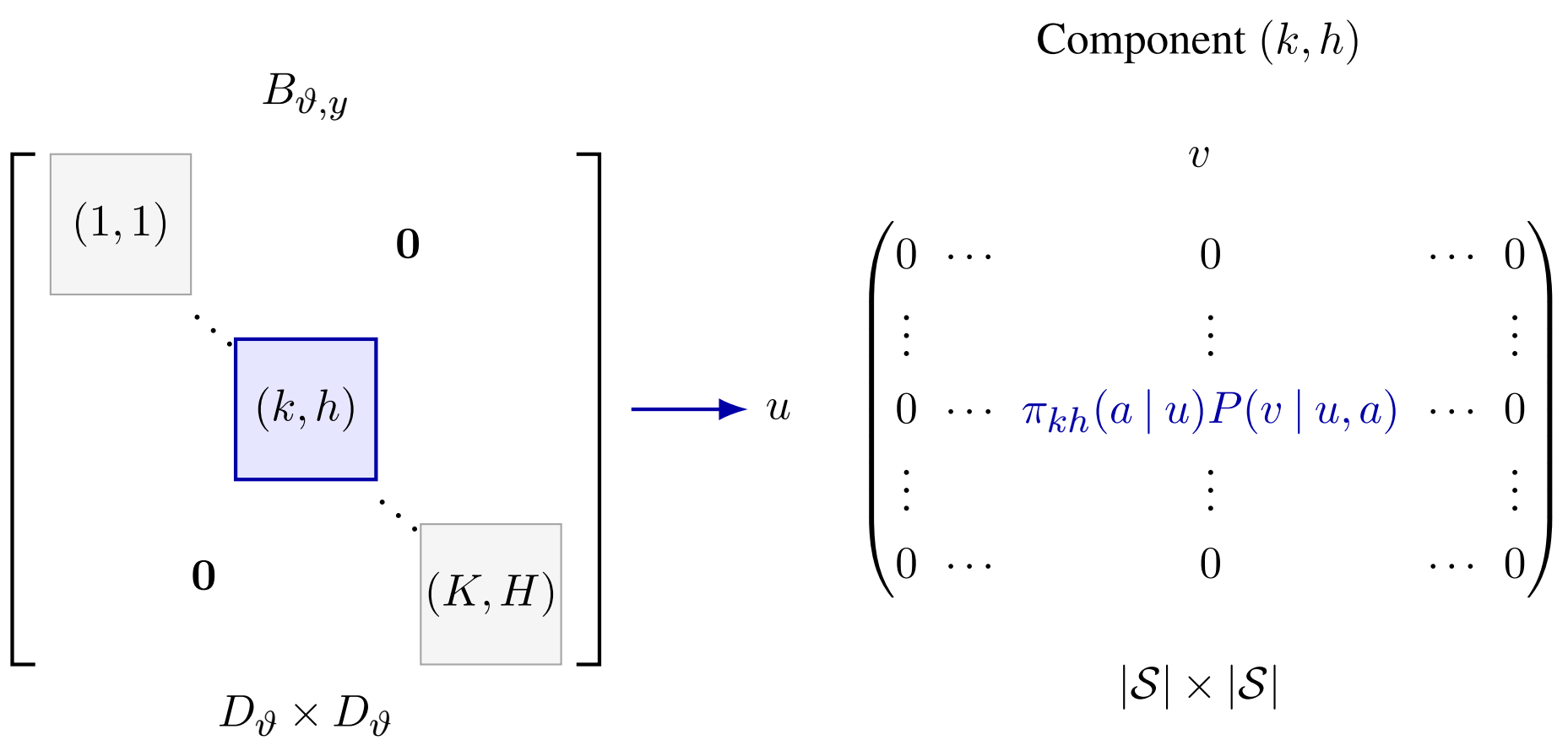
For $w=(y_0,\ldots,y_{m-1})$, use the product convention
\[
 B_{\vartheta,w}
 :=B_{\vartheta,y_0}\cdots B_{\vartheta,y_{m-1}},
 \qquad
 B_{\vartheta,()}:=I_{D_\vartheta}.
\]
We verify \eqref{eq:trajectory-linear-representation}
for consistent sequences, inconsistent sequences, and
the empty sequence.

Suppose $w=(y_0,\ldots,y_{m-1})$ is nonempty and consistent,
with $y_t=(s_t,a_t,s_{t+1})$.
The indicator constraints in
\eqref{eq:trajectory-letter-matrix} keep the component
indices $(k,h)$ fixed and determine the intermediate
state indices. Therefore,
\begin{align*}
 \alpha_\vartheta^{\mathsf T}B_{\vartheta,w}\mathbf1
 &=
 \sum_{k=1}^K\sum_{h=1}^H
 \omega_{kh}\mu_0(s_0)
 \prod_{t=0}^{m-1}
 \pi_{kh}(a_t\mid s_t)P(s_{t+1}\mid s_t,a_t)\\
 &=\PP_\vartheta^{(m)}(\tau(w))
 =F_\vartheta(w),
\end{align*}
where the second equality follows from
\eqref{eq:RT-mixture-law}.

Now suppose $w$ is inconsistent.
Writing $y_t=(u_t,a_t,v_t)$, there is an index $t$ such that
$v_t\ne u_{t+1}$.
A nonzero entry in
$B_{\vartheta,y_t}B_{\vartheta,y_{t+1}}$
would require the shared state index to equal both $v_t$
and $u_{t+1}$.
Hence
\[
 B_{\vartheta,y_t}B_{\vartheta,y_{t+1}}=0,
 \qquad
 B_{\vartheta,w}=0,
\]
and consequently
\[
 \alpha_\vartheta^{\mathsf T}B_{\vartheta,w}\mathbf1
 =0=F_\vartheta(w).
\]

For the empty sequence, $B_{\vartheta,()}=I_{D_\vartheta}$.
Since the mixing weights and the initial probabilities
each sum to one,
\[
 \alpha_\vartheta^{\mathsf T}B_{\vartheta,()}\mathbf1
 =
 \sum_{k=1}^K\sum_{h=1}^H\sum_{s\in\mathcal S}
 \omega_{kh}\mu_0(s)
 =1
 =F_\vartheta(()).
\]
\end{proof}

We now apply Lemma~\ref{lem:finite-linear-equivalence} to
the trajectory representations.
The following corollary gives a horizon at which equality
of the trajectory distributions implies equality at every
finite horizon.

\begin{corollary}[Equality at every finite horizon]
\label{cor:finite-horizon-equivalence}
Let $\vartheta\in\mathcal R_{K,H}(\Fsn)$ and
$\widetilde\vartheta\in
\mathcal R_{\widetilde K,\widetilde H}(\Fsn)$.
For any integer
\[
 T\ge(KH+\widetilde K\widetilde H)|\mathcal S|-1,
\]
we have
\[
 \PP_\vartheta^{(T)}(\cdot)
 =\PP_{\widetilde\vartheta}^{(T)}(\cdot)
 \quad\Longrightarrow\quad
 \PP_\vartheta^{(m)}(\cdot)
 =\PP_{\widetilde\vartheta}^{(m)}(\cdot)
 \qquad\text{for every integer }m\ge1.
\]
\end{corollary}

\begin{proof}
Suppose
$\PP_\vartheta^{(T)}=\PP_{\widetilde\vartheta}^{(T)}$.
For $1\le m\le T$, denote the length-$m$ prefix of
$\tau_T=(s_0,a_0,\ldots,s_{T-1},a_{T-1},s_T)$ by
\[
 \tau_T|_m:=(s_0,a_0,\ldots,s_{m-1},a_{m-1},s_m).
\]
Marginalizing the common length-$T$ distribution gives,
for every length-$m$ trajectory $\tau_m$,
\begin{equation}
 \begin{aligned}
 \PP_\vartheta^{(m)}(\tau_m)
 &=
 \sum_{\tau_T:\,\tau_T|_m=\tau_m}
 \PP_\vartheta^{(T)}(\tau_T)\\
 &=
 \sum_{\tau_T:\,\tau_T|_m=\tau_m}
 \PP_{\widetilde\vartheta}^{(T)}(\tau_T)
 =
 \PP_{\widetilde\vartheta}^{(m)}(\tau_m).
 \end{aligned}
 \label{eq:trajectory-marginalization}
\end{equation}
Thus $F_\vartheta$ and $F_{\widetilde\vartheta}$ agree on
every nonempty consistent sequence of length at most $T$.
Their values also agree on inconsistent sequences and
the empty sequence, where both functions equal zero and
one, respectively.
Hence
\[
 F_\vartheta(w)=F_{\widetilde\vartheta}(w)
 \qquad\text{for every }w\in\mathcal Y^*
 \text{ with }|w|\le T.
\]

By Proposition~\ref{lem:trajectory-linear-representation},
these functions have linear representations of dimensions
$KH|\mathcal S|$ and
$\widetilde K\widetilde H|\mathcal S|$, respectively.
The bound on $T$ and
Lemma~\ref{lem:finite-linear-equivalence} therefore give
\[
 F_\vartheta(w)=F_{\widetilde\vartheta}(w)
 \qquad\text{for every }w\in\mathcal Y^*.
\]
For any integer $m\ge1$ and any length-$m$ trajectory
$\tau_m$, let $w\in\mathcal Y^m$ be its consistent sequence
of transition triples. Then
\[
 \PP_\vartheta^{(m)}(\tau_m)
 =F_\vartheta(w)
 =F_{\widetilde\vartheta}(w)
 =\PP_{\widetilde\vartheta}^{(m)}(\tau_m),
\]
which proves the claim.
\end{proof}

Under the assumptions of
Theorem~\ref{thm:mixture-identification},
Corollary~\ref{cor:finite-horizon-equivalence} provides
the equality at all finite horizons required by
Lemma~\ref{lem:persistent-cell-identification}.
That lemma then gives
$\vartheta\sim\widetilde\vartheta$, establishing
finite-horizon identifiability.

\subsection{Completion of the theorem}
\label{app:mixture-theorem-completion}

\begin{proof}[Proof of Theorem~\ref{thm:mixture-identification}]
For part~(i), let
$\vartheta,\widetilde\vartheta\in\mathcal R(\Fsn)$ and suppose
\[
 \PP_\vartheta^{(m)}(\cdot)
 =\PP_{\widetilde\vartheta}^{(m)}(\cdot)
 \qquad\text{for every integer }m\ge1.
\]
Lemma~\ref{lem:persistent-cell-identification} gives
$\vartheta\sim\widetilde\vartheta$ in the sense of
\eqref{eq:separate-label-equivalence}.
Thus $\mathcal R(\Fsn)$ is $\infty$-identifiable.

For part~(ii), let
$\vartheta\in\mathcal R_{K,H}(\Fsn)$ and
$\widetilde\vartheta\in
\mathcal R_{\widetilde K,\widetilde H}(\Fsn)$ satisfy
\(
 \PP_\vartheta^{(T)}(\cdot)
 =\PP_{\widetilde\vartheta}^{(T)}(\cdot).
\)
If
\(
 T\ge(KH+\widetilde K\widetilde H)|\mathcal S|-1,
\)
Corollary~\ref{cor:finite-horizon-equivalence} gives
\(
 \PP_\vartheta^{(m)}(\cdot)
 =\PP_{\widetilde\vartheta}^{(m)}(\cdot)
 \,\,\text{for every integer }m\ge1.
\)
Part~(i) then yields $\vartheta\sim\widetilde\vartheta$.

When both parameters belong to $\mathcal R_{K,H}(\Fsn)$,
the bound reduces to $T\ge2KH|\mathcal S|-1$.
Hence $\mathcal R_{K,H}(\Fsn)$ is $T$-identifiable for
every such horizon.

Finally, any two parameters in
\[
 \bigcup_{\substack{K,H\in\mathbb N\\KH\le R}}
 \mathcal R_{K,H}(\Fsn)
\]
satisfy $KH+\widetilde K\widetilde H\le2R$.
The horizon $T\ge2R|\mathcal S|-1$ therefore suffices
for every pair in this family, proving its
$T$-identifiability.
\end{proof}

\subsection{Proof of Corollary~\ref{cor:strict-non-nesting}}
\label{app:proof-strict-non-nesting}

\begin{proof}
Suppose that
$\mathfrak M_{K,H}^{(T)}
\cap\mathfrak M_{KH,1}^{(T)}\ne\varnothing$.
Then there exist
$\vartheta\in\mathcal R_{K,H}(\Fsn)$ and
$\widetilde\vartheta\in\mathcal R_{KH,1}(\Fsn)$
such that
$\PP_\vartheta^{(T)}(\cdot)
=\PP_{\widetilde\vartheta}^{(T)}(\cdot)$.

Both models have $KH$ components, so the horizon condition
in \eqref{eq:cross-order-horizon} reduces to
$T\ge2KH|\mathcal S|-1$, which holds by assumption.
Theorem~\ref{thm:mixture-identification}(ii) therefore gives
$\vartheta\sim\widetilde\vartheta$.
By the definition of label equivalence in
\eqref{eq:separate-label-equivalence}, the two models must
have the same number of temperature classes.
This requires $H=1$, contradicting $H\ge2$.

The two distribution classes are therefore disjoint.
Since both are nonempty by assumption, neither is
contained in the other.
\end{proof}

\section{Local curvature under reward normalization}
\label{app:hessian}

This section examines how the constraint $\|\beta\|_2=1$
affects the local curvature of the population negative
log likelihood for a single reward--temperature pair.
Recall that the linear reward is
$f_\beta(s,a)=\beta^{\mathsf T}\phi(s,a)$.
By \eqref{eq:appendix-scale-equivalence},
$\PP_{f_\beta,\lambda}^{(T)}
=\PP_{f_\beta/\lambda,1}^{(T)}$.
Since
$f_\beta(s,a)/\lambda
=(\beta/\lambda)^{\mathsf T}\phi(s,a)$,
the trajectory distribution depends on $(\beta,\lambda)$
only through the ratio $\theta:=\beta/\lambda$.

Assume \eqref{eq:linear-feature-normalization} holds.
Fix $P$ and an initial distribution $\mu_0$ for which
the reachability depth $d_0$ in
\eqref{eq:appendix-reachability-depth} is finite.
Let $T\ge d_0+1$ be an fixed integer.

Let $(\beta_0,\lambda_0)$ be the true parameter, with
$\|\beta_0\|_2=1$ and $\lambda_0>0$, and set
$\theta_0:=\beta_0/\lambda_0$.
Write $\EE_{\theta_0}$ and $\Pr_{\theta_0}$ for expectation
and probability under
$\tau\sim\PP_{f_{\beta_0},\lambda_0}^{(T)}$.
For $\theta=\beta/\lambda$, define the population negative
log likelihood by
$L_T(\theta):=-\EE_{\theta_0}
[\log\PP_{f_\beta,\lambda}^{(T)}(\tau)]$.
Scaling equivalence ensures that this definition depends
only on $\theta$.

We first examine the curvature of $L_T$ at the true
parameter $\theta_0$.

\begin{lemma}[Positive definiteness of the likelihood Hessian]
\label{lem:effective-parameter-hessian}
The function $L_T$ is twice continuously differentiable
with respect to $\theta$.
At $\theta_0$, its gradient satisfies
$\nabla_\theta L_T(\theta_0)=0$, and its Hessian
$I_0:=\nabla_\theta^2L_T(\theta_0)$ is positive definite.
\end{lemma}

\begin{proof}
Write $\pi_\theta:=\pi^*_{\theta^{\mathsf T}\phi,1}$ and
$V_\theta:=V^*_{\theta^{\mathsf T}\phi,1}$, where
$\theta^{\mathsf T}\phi$ denotes the scaled reward
$f_\beta/\lambda$ for $\theta=\beta/\lambda$.
Define
$G(\theta,V):=V-\mathcal B_{\theta^{\mathsf T}\phi,1}V$.
The map $G$ is smooth and satisfies $G(\theta,V_\theta)=0$.
Its derivative with respect to $V$ is
$D_VG(\theta,V_\theta)=I-\gamma P_{\pi_\theta}$, where
$P_{\pi_\theta}(s,s')
:=\sum_a\pi_\theta(a\mid s)P(s'\mid s,a)$.
Since $P_{\pi_\theta}$ is stochastic and $\gamma<1$,
this derivative is invertible.
The implicit function theorem therefore implies that
$V_\theta$ is smooth in $\theta$.
Equation~\eqref{eq:soft-optimal-policy} gives the same
regularity for $\pi_\theta$.

Every $\pi_\theta$ is strictly positive, so the set of
trajectories with positive probability depends only on
$P$ and $\mu_0$.
On this finite set, define
$\ell_\theta(\tau)
:=\log\PP_{\theta^{\mathsf T}\phi,1}^{(T)}(\tau)$.
The function $L_T$ is therefore smooth.
Differentiating
$\sum_\tau\exp\{\ell_\theta(\tau)\}=1$
once and twice with respect to $\theta$ gives
$\nabla_\theta L_T(\theta_0)=0$ and
\begin{equation}
 I_0
 =
 \left.
 \EE_{\theta_0}\!\left[
   \bigl(\nabla_\theta\ell_\theta(\tau)\bigr)
   \bigl(\nabla_\theta\ell_\theta(\tau)\bigr)^{\mathsf T}
 \right]
 \right|_{\theta=\theta_0}.
 \label{eq:appendix-fisher-identity}
\end{equation}

To prove positive definiteness, fix $u\in\mathbb R^p$.
Define the directional derivatives
$g_u(s,a):=
\left.u^{\mathsf T}\nabla_\theta
\log\pi_\theta(a\mid s)\right|_{\theta=\theta_0}$
and
$\dot V_u(s):=
\left.u^{\mathsf T}\nabla_\theta
V_\theta(s)\right|_{\theta=\theta_0}$.
Differentiating \eqref{eq:soft-optimal-policy} yields
\begin{equation}
 g_u(s,a)
 =
 \phi(s,a)^{\mathsf T}u
 +\gamma P_a\dot V_u(s)-\dot V_u(s).
 \label{eq:appendix-policy-score}
\end{equation}
The trajectory factorization gives
$\left.u^{\mathsf T}\nabla_\theta\ell_\theta(\tau)
\right|_{\theta=\theta_0}
=\sum_{t=0}^{T-1}g_u(S_t,A_t)$.

Let
$\mathcal H_t:=\sigma(S_0,A_0,\ldots,A_{t-1},S_t)$
be the history before action $A_t$.
Differentiating $\sum_a\pi_\theta(a\mid s)=1$
in direction $u$ at $\theta_0$ gives
$\sum_a\pi_{\theta_0}(a\mid s)g_u(s,a)=0$.
Hence
$\EE_{\theta_0}[g_u(S_t,A_t)\mid\mathcal H_t]=0$.
For $t<v$, the variable $g_u(S_t,A_t)$ is
$\mathcal H_v$-measurable, so conditioning on
$\mathcal H_v$ gives
$\EE_{\theta_0}[g_u(S_t,A_t)g_u(S_v,A_v)]=0$.

Define the expected number of visits to state $s$
before time $T$ by
$d_T(s):=\sum_{t=0}^{T-1}\Pr_{\theta_0}(S_t=s)$.
The vanishing cross terms and
\eqref{eq:appendix-fisher-identity} imply
\begin{equation}
 \begin{aligned}
 u^{\mathsf T}I_0u
 &=
 \sum_{t=0}^{T-1}
 \EE_{\theta_0}[g_u(S_t,A_t)^2]\\
 &=
 \sum_s d_T(s)\sum_a
 \pi_{\theta_0}(a\mid s)g_u(s,a)^2.
 \end{aligned}
 \label{eq:appendix-fisher-visits}
\end{equation}
Since $\pi_{\theta_0}$ is strictly positive,
Lemma~\ref{lem:single-horizon} and $T\ge d_0+1$
give $d_T(s)>0$ for every state.
Thus, if $u^{\mathsf T}I_0u=0$, every term
$g_u(s,a)$ must vanish.

Under this equality, averaging
\eqref{eq:appendix-policy-score} with the weights
$\omega_s(a)$ and using statewise feature centering
gives $(I-\gamma P_\omega)\dot V_u=0$.
Invertibility of $I-\gamma P_\omega$ implies
$\dot V_u=0$, and
\eqref{eq:appendix-policy-score} then gives
$\phi(s,a)^{\mathsf T}u=0$ for every $(s,a)$.
The whitening condition in
\eqref{eq:linear-feature-normalization} yields
$\|u\|_2^2
=\sum_{s,a}\nu(s,a)\{\phi(s,a)^{\mathsf T}u\}^2=0$.
Therefore $u=0$, proving $I_0\succ0$.
\end{proof}

We now express the same objective in the coordinates
$(\beta,\rho)$, where $\rho:=\log\lambda$, and allow
$\beta$ to vary without the unit-norm constraint.
Define
$\widetilde L_T(\beta,\rho):=L_T(e^{-\rho}\beta)$
for $(\beta,\rho)\in\mathbb R^p\times\mathbb R$.
Let $\rho_0:=\log\lambda_0$ and
$H_0:=\nabla_{(\beta,\rho)}^2
\widetilde L_T(\beta_0,\rho_0)$.
For every $b\in\mathbb R$, the rescaling
$(\beta,\rho)\mapsto(e^b\beta,\rho+b)$ preserves
$e^{-\rho}\beta$ and hence leaves $\widetilde L_T$ unchanged.
The following lemma identifies the resulting Hessian
null space.

\begin{lemma}[Hessian null space before normalization]
\label{lem:scale-hessian-nullspace}
The Hessian $H_0$ satisfies
$\ker H_0=\operatorname{span}\{(\beta_0,1)\}$.
\end{lemma}

\begin{proof}
The Jacobian of $(\beta,\rho)\mapsto e^{-\rho}\beta$
at $(\beta_0,\rho_0)$ is
$J_0:=\lambda_0^{-1}[I_p,-\beta_0]$.
By Lemma~\ref{lem:effective-parameter-hessian},
$\nabla_\theta L_T(\theta_0)=0$.
The Hessian chain rule therefore gives
$H_0=J_0^{\mathsf T}I_0J_0$.
For any $(u,r)\in\mathbb R^p\times\mathbb R$,
\begin{equation}
 (u,r)^{\mathsf T}H_0(u,r)
 =
 \lambda_0^{-2}
 (u-r\beta_0)^{\mathsf T}I_0(u-r\beta_0).
 \label{eq:appendix-hessian-quadratic}
\end{equation}
Since $I_0\succ0$, we have
$\ker H_0=\ker J_0
=\{(r\beta_0,r):r\in\mathbb R\}$,
which proves the claim.
\end{proof}

The constraint $\|\beta\|_2=1$ restricts first-order
perturbations at $(\beta_0,\rho_0)$ to the tangent space
$\mathcal T_0:=
\{(u,r)\in\mathbb R^p\times\mathbb R:
\beta_0^{\mathsf T}u=0\}$.
We equip $\mathcal T_0$ with the Euclidean metric inherited
from $\mathbb R^{p+1}$.
Since
$\nabla_{(\beta,\rho)}
\widetilde L_T(\beta_0,\rho_0)=0$,
the quadratic form of $H_0$ on $\mathcal T_0$
determines the second derivative of the objective along
smooth feasible curves through $(\beta_0,\rho_0)$.

\begin{proposition}[Hessian conditioning under normalization]
\label{prop:hessian-conditioning}
The restricted Hessian $H_0|_{\mathcal T_0}$ is positive
definite, and
\begin{equation}
 \kappa(H_0|_{\mathcal T_0})=\kappa(I_0)<\infty.
 \label{eq:hessian-conditioning}
\end{equation}
Here $H_0|_{\mathcal T_0}$ is the matrix of the restricted
quadratic form in any orthonormal basis of $\mathcal T_0$,
and $\kappa(M):=\lambda_{\max}(M)/\lambda_{\min}(M)$
for a positive definite matrix $M$.
\end{proposition}

\begin{proof}
Define the linear map $U_0:\mathcal T_0\to\mathbb R^p$
by $U_0(u,r):=u-r\beta_0$.
For every $(u,r)\in\mathcal T_0$, the identities
$\beta_0^{\mathsf T}u=0$ and $\|\beta_0\|_2=1$ give
$\|U_0(u,r)\|_2^2=\|u\|_2^2+r^2$.
Thus $U_0$ is an isometry.
Both spaces have dimension $p$, so $U_0$ is bijective.

Equation~\eqref{eq:appendix-hessian-quadratic} therefore
shows that $H_0|_{\mathcal T_0}$ is orthogonally equivalent
to $\lambda_0^{-2}I_0$.
Its eigenvalues are $\lambda_0^{-2}$ times those of $I_0$.
Since $I_0\succ0$ and the common factor cancels in their
ratio, the restricted Hessian is positive definite and
satisfies \eqref{eq:hessian-conditioning}.
\end{proof}

The unit-norm constraint removes the Hessian's zero direction
associated with jointly scaling the reward and temperature.
After normalization, the condition number is $\kappa(I_0)$,
which may still be large.
This equality applies to the coordinates
$(\beta,\log\lambda)$ with the Euclidean metric specified above.

The horizon condition ensures that each state can be visited
with positive probability at some time before $T$.
If $G_P$ is strongly connected, $T\ge|\mathcal S|$ suffices
for any initial distribution $\mu_0$.
If $\mu_0$ has full support, every state already has positive
probability at time zero, so any $T\ge1$ suffices.

\section{Details for Example~\ref{ex:one-state-three-action}}
\label{app:example}

We construct the $\RT(2,2)$ law in
Example~\ref{ex:one-state-three-action} and prove the
KL lower bound in \eqref{eq:one-state-kl-gap}.
The argument compares two trajectory probabilities.
For every competing $\RO(4,1)$ law, at least one of these
probabilities differs from its RT value by a fixed positive
amount.

In the one-state environment, the continuation value is
the same for every action and cancels from
\eqref{eq:soft-optimal-policy}.
Thus
$\pi^*_{f,\lambda}(a_i\mid s)
=\exp\{f(s,a_i)/\lambda\}/
\sum_{j=1}^3\exp\{f(s,a_j)/\lambda\}$
for every $\gamma\in[0,1)$.
Conditional on a component, successive actions are independent
with this common distribution.

Consider the trajectories
$\tau_1:=(s,a_1,s,a_1,s,a_1,s)$ and
$\tau_2:=(s,a_1,s,a_2,s,a_3,s)$,
and define $A:=\{\tau_1\}$ and $B:=\{\tau_2\}$.
The first repeats $a_1$ three times; the second selects
each action once in the specified order.
Under a component policy $\pi$, their probabilities are
$\pi(a_1\mid s)^3$ and
$\prod_{i=1}^3\pi(a_i\mid s)$, respectively.
We will construct an RT law that assigns sufficiently large
probabilities to both events, and show that a common
temperature prevents an RO model from matching them both.

Choose $\beta_1:=(1,0)^{\mathsf T}$ and
$\beta_2:=(-1/2,\sqrt3/2)^{\mathsf T}$.
For the features in
Example~\ref{ex:one-state-three-action}, these rewards
favor $a_1$ and $a_2$, respectively.
In each case, the preferred action has reward $\sqrt2$
and the other two actions have reward $-1/\sqrt2$.
Choose
$\lambda_1:=3/(\sqrt2\log98)$ and
$\lambda_2:=3/\{\sqrt2\log(4/3)\}$.
These temperatures make the odds of the preferred action
relative to either alternative equal to $98$ and $4/3$,
respectively.
Writing $\pi_{kh}:=\pi^*_{f_{\beta_k},\lambda_h}$,
the four component policies are
\[
\begin{array}{c|ccc}
 (k,h)
 & \pi_{kh}(a_1\mid s)
 & \pi_{kh}(a_2\mid s)
 & \pi_{kh}(a_3\mid s)\\
 \hline
 (1,1) & 49/50 & 1/100 & 1/100\\
 (1,2) & 2/5   & 3/10  & 3/10\\
 (2,1) & 1/100 & 49/50 & 1/100\\
 (2,2) & 3/10  & 2/5   & 3/10
\end{array}
\]

Set $\omega_{1h}:=9/20$ and $\omega_{2h}:=1/20$
for $h=1,2$, and define
$P_\star^{(3)}
:=\sum_{k=1}^2\sum_{h=1}^2
\omega_{kh}\PP_{f_{\beta_k},\lambda_h}^{(3)}$.
Both coefficient vectors have unit norm, so the feature
normalization gives $f_{\beta_1},f_{\beta_2}\in\Fsn$.
The rewards and temperatures are pairwise distinct,
and the weights are positive and sum to one.
Hence this is a regular $\RT(2,2)$ law.
Averaging the component probabilities gives
$p_1:=P_\star^{(3)}(A)$
and
$p_2:=P_\star^{(3)}(B)$.

We next bound these event probabilities under a single
normalized reward at a fixed temperature.
The bounds depend only on the temperature, so they also
apply after averaging over any RO mixture with that
common temperature.

\begin{lemma}[Probability bounds at a fixed temperature]
\label{lem:example-fixed-temperature-bounds}
Let $g\in\Fsn$ and $\widetilde\lambda>0$.
Define $q_i:=\pi^*_{g,\widetilde\lambda}(a_i\mid s)$
and $r:=\exp\{3/(\sqrt2\widetilde\lambda)\}>1$.
Then
\begin{align}
 q_1
 &\le \frac{r}{r+2},
 \label{eq:example-first-bound}\\
 q_1q_2q_3
 &\le \frac{r^2}{(1+2r)^3}.
 \label{eq:example-product-bound}
\end{align}
\end{lemma}

\begin{proof}
The features satisfy
$\frac13\sum_i\phi(s,a_i)=0$ and
$\frac13\sum_i\phi(s,a_i)\phi(s,a_i)^{\mathsf T}=I_2$.
The feature matrix with rows $\phi(s,a_i)^{\mathsf T}$
therefore has rank two, and its columns span the plane
$\{u\in\mathbb R^3:\sum_i u_i=0\}$.
Consequently, the possible reward vectors in $\Fsn$
are exactly
\begin{equation}
 \left\{
 u\in\mathbb R^3:
 \sum_{i=1}^3u_i=0,\quad
 \sum_{i=1}^3u_i^2=3
 \right\}.
 \label{eq:example-reward-circle}
\end{equation}

Set $x_i:=g(s,a_i)/\widetilde\lambda$ and
$c:=1/(\sqrt2\widetilde\lambda)$, so $r=e^{3c}$.
Equation~\eqref{eq:example-reward-circle} gives
$\sum_i x_i=0$ and $\sum_i x_i^2=6c^2$.
Since
$x_2^2+x_3^2\ge(x_2+x_3)^2/2=x_1^2/2$,
we have $x_1\le2c$.
Convexity of the exponential gives
$e^{x_2}+e^{x_3}\ge2e^{-x_1/2}$.
Therefore
$q_1\le e^{3x_1/2}/(e^{3x_1/2}+2)\le r/(r+2)$,
proving \eqref{eq:example-first-bound}.

For the product bound, define $Z(x):=\sum_i e^{x_i}$.
Because $\sum_i x_i=0$, we have
$q_1q_2q_3=Z(x)^{-3}$.
It therefore suffices to minimize $Z$ subject to
$\sum_i x_i=0$ and $\sum_i x_i^2=6c^2$.

The constraint set is compact.
Its constraint gradients are linearly independent,
because $c>0$ prevents the coordinates from all being equal.
At a minimizer, the Lagrange multiplier equations give
$e^{x_i}=\xi+\zeta x_i$ for some real scalars $\xi,\zeta$.
Strict convexity of $t\mapsto e^t-\xi-\zeta t$
implies that this equation has at most two distinct solutions.
Thus the minimizing vector has two distinct coordinate values.
The constraints then imply that, up to permutation, it is
either $(2c,-c,-c)$ or $(-2c,c,c)$.
The corresponding values of $Z$ satisfy
\[
 (e^{2c}+2e^{-c})-(e^{-2c}+2e^c)
 =
 (e^c-e^{-c})(e^c+e^{-c}-2)>0.
\]
Hence the minimum is $e^{-2c}+2e^c$, and
$q_1q_2q_3\le(e^{-2c}+2e^c)^{-3}
=r^2/(1+2r)^3$.
This proves \eqref{eq:example-product-bound}.
\end{proof}

The parameter $r$ increases as the common temperature
decreases.
The upper bound for $A$ increases with $r$, whereas
the upper bound for $B$ decreases.
We use this relation to show that every common temperature
leaves a positive discrepancy for at least one event.

\begin{proof}[Proof of \eqref{eq:one-state-kl-gap}]
Fix any $Q^{(3)}\in\mathfrak M_{4,1}^{(3)}$.
Let $\widetilde\lambda>0$ be its common temperature and
set $r:=\exp\{3/(\sqrt2\widetilde\lambda)\}$.
Every component reward belongs to $\Fsn$.
Applying Lemma~\ref{lem:example-fixed-temperature-bounds}
to each component and averaging over the mixture weights
gives
\[
 Q^{(3)}(A)\le\left(\frac{r}{r+2}\right)^3,
 \qquad
 Q^{(3)}(B)\le\frac{r^2}{(1+2r)^3}.
\]

We split the argument into the cases $1<r\le6$ and $r>6$.
At $r=6$, both upper bounds lie below the corresponding
RT probabilities: $27/64<p_1$ and $36/2197<p_2$.
As in Example~\ref{ex:one-state-three-action}, define
$\delta:=p_2-36/2197>0$.

If $1<r\le6$, monotonicity of $r/(r+2)$ gives
$P_\star^{(3)}(A)-Q^{(3)}(A)
\ge p_1-27/64>\delta$.
If $r>6$, the identity
$\frac{d}{dr}\{r^2/(1+2r)^3\}
=2r(1-r)/(1+2r)^4<0$
gives
$P_\star^{(3)}(B)-Q^{(3)}(B)
>p_2-36/2197=\delta$.
Thus there is an event $E\in\{A,B\}$ such that
$P_\star^{(3)}(E)-Q^{(3)}(E)\ge\delta$.

It remains to convert this event probability gap into
a KL lower bound.
Set $p:=P_\star^{(3)}(E)$ and $q:=Q^{(3)}(E)$.
Strict positivity of the component policies gives
$p,q\in(0,1)$.
For $x\in(0,1)$, define
$b_q(x):=x\log(x/q)+(1-x)\log\{(1-x)/(1-q)\}$.
Decomposing the KL divergence over $E$ and $E^c$
and dropping the nonnegative conditional divergences gives
$D_{\mathrm{KL}}(P_\star^{(3)}\Vert Q^{(3)})\ge b_q(p)$.
Since $b_q(q)=b_q'(q)=0$ and
$b_q''(x)=1/\{x(1-x)\}\ge4$,
Taylor's theorem yields $b_q(p)\ge2(p-q)^2$.
Consequently,
\[
 D_{\mathrm{KL}}(P_\star^{(3)}\Vert Q^{(3)})
 \ge2(p-q)^2
 \ge2\delta^2.
\]
The bound holds for every
$Q^{(3)}\in\mathfrak M_{4,1}^{(3)}$.
Taking the infimum proves \eqref{eq:one-state-kl-gap}.
\end{proof}

\begin{remark}
The probability bounds require only normalized rewards,
a common positive temperature, and nonnegative mixture
weights summing to one.
The same KL lower bound therefore holds for four-component
RO mixtures with zero weights or repeated rewards.
\end{remark}

\section{Computational Details for GEM}
\label{app:gem-computation}

This appendix provides the derivations and numerical details for
the GEM procedure in Section~\ref{sec:gem}.
Section~\ref{app:reward-gradients} derives the reward gradient
and describes the sparse structure of the adjoint systems.
Section~\ref{app:epi-multistart} explains how EPI and MIS construct
the initialization bank.
Section~\ref{app:gem-numerics} specifies the numerical update rules,
stopping criteria, and procedures for screening initial values
and selecting the final estimate.

\subsection{Reward Gradients and Sparse Computation}
\label{app:reward-gradients}

The gradient of a shared reward combines contributions from the
temperature classes. For each class, an adjoint system accounts
for the dependence of the soft-optimal policy on the reward.

\begin{proposition}[Reward gradient]
\label{prop:reward-gradient}
Fix reward class \(k\), the responsibilities \(r_{i,kh}^{(m)}\),
and temperatures \(\lambda_h=\lambda_h^{(m)}>0\).
Suppose \(0\le\gamma<1\).
For each \(h\), define
\(c_{kh}^{S}(s)=\sum_a c_{kh}(s,a)\),
\(c_{kh}^{P}(s')=\sum_{s,a}c_{kh}(s,a)P(s'\mid s,a)\), and
\(P_{kh}^{\pi}(s,s')
=\sum_a\pi^*_{f_k,\lambda_h}(a\mid s)P(s'\mid s,a)\).
Let \(z_{kh}\) be the unique solution of the adjoint system
\begin{equation}
 (I-\gamma P_{kh}^{\pi})^{\mathsf T}z_{kh}
 =\gamma c_{kh}^{P}-c_{kh}^{S}.
 \label{eq:reward-adjoint-system}
\end{equation}
The unconstrained reward gradient sums the contributions from
the \(H\) temperature classes,
\begin{equation}
\begin{aligned}
 G_{kh}(s,a)
 &=\lambda_h^{-1}
 \{c_{kh}(s,a)+\pi^*_{f_k,\lambda_h}(a\mid s)z_{kh}(s)\},\\
 \frac{\partial Q_n(\vartheta\mid\vartheta^{(m)})}
      {\partial f_k(s,a)}
 &=G_k(s,a)=\sum_{h=1}^H G_{kh}(s,a).
\end{aligned}
\label{eq:tabular-reward-score}
\end{equation}
For \(f_\beta=\phi^{\mathsf T}\beta\), the chain rule gives
\begin{equation}
 \nabla_\beta Q_{n,k}^{\mathrm R,(m)}(\beta)
 =\sum_{s,a}\phi(s,a)G_k(s,a),
 \label{eq:linear-reward-score}
\end{equation}
with \(G_k\) evaluated at \(f_k=f_\beta\).
\end{proposition}

\begin{proof}[Proof of Proposition~\ref{prop:reward-gradient}]
With the responsibilities and temperatures fixed, the counts
\(c_{kh}\) in \eqref{eq:posterior-action-counts} remain constant.
For each temperature class \(h\), define
\[
Q_{n,kh}(f)
:=\sum_{s,a}c_{kh}(s,a)
  \log\pi^*_{f,\lambda_h}(a\mid s).
\]
The reward objective in \eqref{eq:reward-block-objective} is
\(\sum_{h=1}^H Q_{n,kh}(f_\beta)\), so it suffices to
differentiate a single term and sum over \(h\).

Fix \(h\), and write
\(\pi_{kh}=\pi^*_{f_k,\lambda_h}\) and
\(V_{kh}=V^*_{f_k,\lambda_h}\).
As in the proof of Lemma~\ref{lem:effective-parameter-hessian},
the residual \(V-\mathcal B_{f,\lambda_h}V\) is smooth, and its
Jacobian with respect to \(V\) at \((f_k,V_{kh})\) is
\(I-\gamma P_{kh}^{\pi}\).
This matrix is invertible because \(P_{kh}^{\pi}\) is stochastic
and \(\gamma<1\).
The implicit function theorem therefore gives differentiability
of \(V^*_{f,\lambda_h}\) with respect to \(f\).

For an arbitrary direction
\(d:\mathcal S\times\mathcal A\to\mathbb R\), define
\[
\dot V_{kh}(s)
:=
\left.
\frac{d}{d\varepsilon}
V^*_{f_k+\varepsilon d,\lambda_h}(s)
\right|_{\varepsilon=0}.
\]
Differentiating \eqref{eq:soft-bellman} gives
\[
\bigl[(I-\gamma P_{kh}^{\pi})\dot V_{kh}\bigr](s)
=
\sum_a\pi_{kh}(a\mid s)d(s,a).
\]
The derivative of the policy formula
\eqref{eq:soft-optimal-policy} is therefore
\[
\left.
\frac{d}{d\varepsilon}
\log\pi^*_{f_k+\varepsilon d,\lambda_h}(a\mid s)
\right|_{\varepsilon=0}
=
\frac{1}{\lambda_h}
\left\{
d(s,a)
+\gamma\sum_{s'}P(s'\mid s,a)\dot V_{kh}(s')
-\dot V_{kh}(s)
\right\}.
\]
Multiplying by \(c_{kh}(s,a)\) and summing, with
\(c_{kh}^{S}\) and \(c_{kh}^{P}\) defined in the proposition, yields
\[
\left.
\frac{d}{d\varepsilon}Q_{n,kh}(f_k+\varepsilon d)
\right|_{\varepsilon=0}
=
\frac{1}{\lambda_h}
\left\{
\sum_{s,a}c_{kh}(s,a)d(s,a)
+
(\gamma c_{kh}^{P}-c_{kh}^{S})^{\mathsf T}\dot V_{kh}
\right\}.
\]

The adjoint solution \(z_{kh}\) in
\eqref{eq:reward-adjoint-system} eliminates \(\dot V_{kh}\):
the two linear systems imply
\[
\begin{aligned}
(\gamma c_{kh}^{P}-c_{kh}^{S})^{\mathsf T}\dot V_{kh}
&=
z_{kh}^{\mathsf T}(I-\gamma P_{kh}^{\pi})\dot V_{kh}\\
&=
\sum_{s,a}z_{kh}(s)\pi_{kh}(a\mid s)d(s,a).
\end{aligned}
\]
Substituting this identity into the objective derivative gives
\[
\left.
\frac{d}{d\varepsilon}Q_{n,kh}(f_k+\varepsilon d)
\right|_{\varepsilon=0}
=
\sum_{s,a}
\frac{
c_{kh}(s,a)+\pi_{kh}(a\mid s)z_{kh}(s)
}{\lambda_h}\,d(s,a).
\]
Since \(d\) is arbitrary, the coefficient of \(d(s,a)\)
is the unconstrained gradient \(G_{kh}(s,a)\) in
\eqref{eq:tabular-reward-score}.

Summing over \(h\) gives the gradient \(G_k\) for the shared
reward \(f_k\).
For the linear reward in \eqref{eq:linear-reward-class},
\(\nabla_\beta f_\beta(s,a)=\phi(s,a)\).
The chain rule then gives \eqref{eq:linear-reward-score}.
\end{proof}

\paragraph{Sparse computation}
\label{rem:adjoint-sparsity}
The matrix \(P_{kh}^{\pi}\) assigns positive probability only to
transitions allowed by \(P\). When each state has few possible
destinations, both \(P_{kh}^{\pi}\) and the adjoint coefficient
matrix are sparse. On a grid restricted to the current cell and
its eight neighbors, for example, each row of \(P_{kh}^{\pi}\)
has at most nine nonzero entries.

We solve \eqref{eq:reward-adjoint-system} for each reward--temperature
pair with a matrix-free Krylov method. For a vector \(v\) indexed
by \(\mathcal S\), the required matrix--vector product is
\[
\bigl[(I-\gamma P_{kh}^{\pi})^{\mathsf T}v\bigr](s')
=
v(s')
-\gamma\sum_{s,a}
\pi^*_{f_k,\lambda_h}(a\mid s)
P(s'\mid s,a)v(s).
\]
We evaluate the sum over the nonzero transitions of \(P\), without
assembling the coefficient matrix.

\begin{remark}[Computational cost]
\label{rem:gem-complexity}
Let \(S=|\mathcal S|\), \(A=|\mathcal A|\), and
\(N=\sum_{i=1}^n T_i\).
Given the component policies, computing the responsibilities and
accumulating the quantities in \eqref{eq:gem-Q} cost \(O(KHN)\),
and the weight update costs \(O(nKH)\).
Reward gradients require \(KH\) adjoint systems of dimension \(S\).
Candidate reward and temperature updates also require recomputing
the corresponding policies.
For dense transitions, direct factorization can require \(O(S^3)\)
per system, and a policy with \(J\) Bellman iterations costs
\(O(JAS^2)\).

For sparse transitions, let
\[
d_{\mathrm{sa}}
=
\max_{s,a}
\left|\operatorname{supp}P(\cdot\mid s,a)\right|.
\]
Our simulation grid has \(d_{\mathrm{sa}}=1\), and the fitted
white-stork transition kernel has \(d_{\mathrm{sa}}\leq273\),
both much smaller than \(S\).
Each Bellman iteration and each matrix--vector product above
cost \(O(ASd_{\mathrm{sa}})\).
Computing a policy therefore costs \(O(JASd_{\mathrm{sa}})\),
and each adjoint solve costs
\(O(M_{\mathrm{Krylov}}ASd_{\mathrm{sa}})\), where
\(M_{\mathrm{Krylov}}\) is the number of Krylov iterations.
\end{remark}

\subsection{Construction of the Initialization Bank}
\label{app:epi-multistart}

We construct the initialization bank in
Section~\ref{sec:epi-initialization} from the pooled empirical
policy. EPI converts this policy into a reference reward and
temperature; MIS then varies reward directions and temperature
spacing around that reference.

\paragraph{Constructing a centered reward}
At the reference temperature \(\lambda_0\), we seek a centered
reward that induces the empirical policy
\(\widehat\pi_{\mathrm E}\) in \eqref{eq:epi-policy}.
The centering constraint in the reward representation
\eqref{eq:appendix-policy-inversion} determines the value function.
Define
\(c_{\mathrm E}(s):=\sum_a\omega_s(a)
\log\widehat\pi_{\mathrm E}(a\mid s)\).
Using \(P_\omega\) from
Appendix~\ref{app:normalized-identification}, we solve
\begin{equation}
  (I-\gamma P_\omega)\widehat v_{\mathrm E}
  =-\lambda_0c_{\mathrm E}.
  \label{eq:epi-value-inversion}
\end{equation}
The solution is unique because \(P_\omega\) is stochastic
and \(\gamma<1\).
The corresponding reward is
\begin{equation}
  \widetilde f_{\mathrm E}(s,a)
  :=\lambda_0\log\widehat\pi_{\mathrm E}(a\mid s)
   +\widehat v_{\mathrm E}(s)
   -\gamma\sum_{s'}P(s'\mid s,a)\widehat v_{\mathrm E}(s').
  \label{eq:epi-reward-inversion}
\end{equation}
This reward satisfies the centering constraint, and
Lemma~\ref{lem:reward-equivalence} gives
\(\pi^*_{\widetilde f_{\mathrm E},\lambda_0}
=\widehat\pi_{\mathrm E}\).

\paragraph{Projection and normalization}
We project \(\widetilde f_{\mathrm E}\) onto the prescribed
linear feature space, since the centered tabular reward need not
belong to it. Under \eqref{eq:linear-feature-normalization},
the weighted least-squares coefficient is
\(b_{\mathrm E}:=\sum_{s,a}\nu(s,a)\phi(s,a)
\widetilde f_{\mathrm E}(s,a)\).
We normalize this coefficient and rescale the temperature by
the same factor.
For a fixed tolerance \(\delta_{\mathrm E}>0\), if
\(\|b_{\mathrm E}\|_2>\delta_{\mathrm E}\), set
\begin{equation}
  \widehat\beta_{\mathrm E}
  :=\frac{b_{\mathrm E}}{\|b_{\mathrm E}\|_2},
  \qquad
  \widehat\lambda_{\mathrm E}
  :=\frac{\lambda_0}{\|b_{\mathrm E}\|_2}.
  \label{eq:epi-anchor}
\end{equation}
Corollary~\ref{cor:reward-temperature-scaling} shows that this
joint rescaling preserves the policy of the projected reward.
The projection itself may change \(\widehat\pi_{\mathrm E}\).
If \(\|b_{\mathrm E}\|_2\le\delta_{\mathrm E}\), we instead
choose a random unit vector for \(\widehat\beta_{\mathrm E}\)
and take \(\widehat\lambda_{\mathrm E}=\lambda_0\).

\paragraph{Reward initial values}
For \(2\le K\le p\), choose \(J_\beta\) angular offsets
\(\alpha_j\), \(j=1,\ldots,J_\beta\), and \(R\) orientations
of a regular simplex with \(K\) unit vertices in the subspace
orthogonal to \(\widehat\beta_{\mathrm E}\).
For each orientation \(r=1,\ldots,R\), the vertices satisfy
\[
\begin{aligned}
  v_k^{(r)}
  &\in
  \left\{v\in\mathbb R^p:
    \widehat\beta_{\mathrm E}^{\mathsf T}v=0,\;
    \|v\|_2=1
  \right\},
  && k=1,\ldots,K,\\
  (v_k^{(r)})^{\mathsf T}v_{k'}^{(r)}
  &=-\frac{1}{K-1},
  && 1\le k<k'\le K.
\end{aligned}
\]
For each pair \((j,r)\), set
\begin{equation}
  \beta_k^{(j,r)}
  =\cos(\alpha_j)\widehat\beta_{\mathrm E}
   +\sin(\alpha_j)v_k^{(r)},
  \qquad k=1,\ldots,K.
  \label{eq:reward-head-starts}
\end{equation}
These coefficients lie symmetrically around the EPI direction on
the unit sphere and are distinct whenever \(\sin\alpha_j\ne0\).
The angle \(\alpha_j\) sets their deviation from the EPI direction,
while \(r\) changes the orientation of the configuration.
This gives \(J_\beta R\) reward configurations.
For \(K=1\), we use \(\widehat\beta_{\mathrm E}\) alone.

When \(K>p\), the orthogonal subspace is too small for this
simplex.
For \(p\ge2\), we instead place the coefficients along a
circular arc on the unit sphere.
Choose a unit vector \(v^{(r)}\) orthogonal to
\(\widehat\beta_{\mathrm E}\), and define
\(t_k^{(j)}:=\alpha_j(2k-K-1)/(K-1)\), where
\(0<\alpha_j<\pi\).
Set
\begin{equation}
  \beta_k^{(j,r)}
  :=\cos(t_k^{(j)})\widehat\beta_{\mathrm E}
   +\sin(t_k^{(j)})v^{(r)},
  \qquad k=1,\ldots,K.
  \label{eq:overcomplete-reward-starts}
\end{equation}
These coefficients are distinct and have unit norm.%
\footnote{When \(p=1\), the normalized coefficient space is
\(\{-1,1\}\). For \(K=2\), we use
\((\widehat\beta_{\mathrm E},-\widehat\beta_{\mathrm E})\);
no regular model with \(K>2\) exists in this space.}

\paragraph{Temperature initial values}
We place the temperatures at equal intervals on the log scale,
with their center as close as possible to the EPI reference while
respecting the bounds \(u_{\min}<u_{\max}\) in
Appendix~\ref{app:gem-numerics}.
For \(H\ge2\), choose \(J_\lambda\) positive half-spans satisfying
\(0<b_1<\cdots<b_{J_\lambda}
\le(u_{\max}-u_{\min})/2\), and let
\(u_{\mathrm E}:=\log\widehat\lambda_{\mathrm E}\).
For each \(\ell\), define
\(c_\ell:=\min\{\max(u_{\mathrm E},u_{\min}+b_\ell),
u_{\max}-b_\ell\}\).
This shifts the center by the smallest amount needed to keep
the configuration within the bounds.
The initial temperatures are
\begin{equation}
  \lambda_h^{(\ell)}
  :=\exp\!\left\{
    c_\ell+b_\ell\frac{2h-H-1}{H-1}
  \right\},
  \qquad h=1,\ldots,H.
  \label{eq:temperature-head-starts}
\end{equation}
The \(J_\lambda\) configurations \(\lambda_{1:H}^{(\ell)}\)
have equally spaced log temperatures and spans \(2b_\ell\).
For \(H=1\), there is a single temperature configuration
(\(J_\lambda=1\)):
\begin{equation}
  \lambda_1^{(1)}
  :=\exp\!\left\{
    \min\{\max(u_{\mathrm E},u_{\min}),u_{\max}\}
  \right\}.
  \label{eq:single-temperature-start}
\end{equation}

We combine each reward configuration with each temperature
configuration and assign uniform weights
\(\omega_{kh}^{(0)}=1/(KH)\).
Adding the baseline EPI initialization with the same weights gives
\(N_{\mathrm{init}}=1+J_\beta J_\lambda R\) indexed starts,
collected as
\(\mathcal B_{K,H}=(\vartheta_q^{(0)})_{q=1}^{N_{\mathrm{init}}}\).
Algorithm~\ref{alg:epi-bank} summarizes the construction.

\begin{algorithm}[!htbp]
\caption{Construction of the initialization bank by EPI and MIS}
\label{alg:epi-bank}
\small
\begin{flushleft}
\textbf{Input:} \(\mathcal D_n,P,\gamma,K,H\), reward specification,
and the initialization settings defined above.
\end{flushleft}
\begin{enumerate}[leftmargin=1.6em,itemsep=2pt,topsep=4pt]
\item Compute \(\widehat\pi_{\mathrm E}\) using \eqref{eq:epi-policy}.
Solve \eqref{eq:epi-value-inversion}, construct
\eqref{eq:epi-reward-inversion}, and project and normalize using
\eqref{eq:epi-anchor}, with the treatment of zero or negligible
projections described above.
\item For \(2\le K\le p\), construct rewards using
\eqref{eq:reward-head-starts}. For \(K>p\), use
\eqref{eq:overcomplete-reward-starts} when \(p\ge2\), or the
\(p=1\) case described above. For \(K=1\),
use \(\widehat\beta_{\mathrm E}\).
\item Construct temperature configurations using
\eqref{eq:temperature-head-starts}, or
\eqref{eq:single-temperature-start} when \(H=1\).
\item Cross the configurations with \(\omega_{kh}^{(0)}=1/(KH)\)
and append the baseline EPI start.
\end{enumerate}
\begin{flushleft}
\textbf{Return:} \(\mathcal B_{K,H}\), the collection of
\(N_{\mathrm{init}}=1+J_\beta J_\lambda R\) indexed starts.
\end{flushleft}
\end{algorithm}

\subsection{Numerical Implementation}
\label{app:gem-numerics}

Starting from the initial values in
Appendix~\ref{app:epi-multistart}, we apply the GEM updates in
Section~\ref{sec:generalized-m-step}.
We specify below the parameter constraints, acceptance checks,
and stopping criteria, together with the screening of initial fits.

\paragraph{Parameter constraints.}
To keep mixture weights from becoming too small during the early
iterations, we impose a lower bound that decreases to a baseline
value.
For a positive integer \(B_{\mathrm w}\), define
\begin{equation}
  \underline\omega_m
  :=\max\left\{
    \underline\omega_{\mathrm{base}},
    \frac{\rho_0}{KH}
    \left(1-\frac{m}{B_{\mathrm w}}\right)
  \right\},
  \qquad m=0,1,\ldots,
  \label{eq:warmup-weight-floor}
\end{equation}
where \(0\le\underline\omega_{\mathrm{base}}<1/(KH)\) and
\(0\le\rho_0<1\).
The baseline bound applies for \(m\ge B_{\mathrm w}\).

For a given lower bound \(\underline\omega\), maximizing the
weight block subject to
\(\omega_{kh}\ge\underline\omega\) and
\(\sum_{k,h}\omega_{kh}=1\) gives the candidate
\begin{equation}
  \omega_{kh}^{\mathrm{prop}}(\underline\omega)
  :=\max\{\underline\omega,N_{kh}^{(m)}/\xi_m\},
  \qquad
  \sum_{k,h}\omega_{kh}^{\mathrm{prop}}(\underline\omega)=1,
  \label{eq:trimmed-weight-update}
\end{equation}
where \(\xi_m>0\) is chosen to satisfy the sum constraint.
We use \(\underline\omega=\underline\omega_m\) at iteration \(m\).
Setting \(\underline\omega=0\) recovers
\eqref{eq:mixture-weight-update}.

Temperatures are optimized on the log scale.
We write \(u_h:=\log\lambda_h\) and restrict \(u_h\) to a
prespecified interval \([u_{\min},u_{\max}]\).
These numerical constraints do not enforce the distinctness of
reward or temperature parameters required for regularity in
Section~\ref{sec:mixture-identifiability-definitions}.

\paragraph{Acceptance of parameter updates.}
Within each M-step, we hold the E-step responsibilities fixed
and update the weights, rewards, and temperatures in turn.
Let \(\vartheta^{\mathrm{cur}}\) and
\(\vartheta^{\mathrm{trial}}\) denote the parameter values before
and after a proposed block update.
We accept the update only if
\begin{equation}
  v(\vartheta^{\mathrm{trial}})
  \ge
  v(\vartheta^{\mathrm{cur}})
  -10^{-10}\{1+|v(\vartheta^{\mathrm{cur}})|\}
  \label{eq:block-acceptance-tolerance}
\end{equation}
holds for both
\(v\in\{Q_n(\,\cdot\mid\vartheta^{(m)}),\ell_n\}\).
The tolerance in \eqref{eq:block-acceptance-tolerance} permits
small numerical decreases.
Updates satisfying the exact condition \eqref{eq:gem-ascent}
retain the GEM likelihood ascent property.

For reward and temperature updates, backtracking can be used
to reduce the step size when a proposed update fails the
acceptance criterion in \eqref{eq:block-acceptance-tolerance}.

\paragraph{Stopping criteria.}
We assess convergence under the baseline weight bound using
both changes in the objectives and residuals of the constrained
first-order conditions.
For each completed iteration, define
\begin{equation}
\begin{aligned}
  r_Q^{(m)}
  &:=
  \frac{
    \left|
      Q_n(\vartheta^{(m+1)}\mid\vartheta^{(m)})
      -Q_n(\vartheta^{(m)}\mid\vartheta^{(m)})
    \right|
  }{
    1+\left|Q_n(\vartheta^{(m)}\mid\vartheta^{(m)})\right|
  },\\
  r_\ell^{(m)}
  &:=
  \frac{
    \left|\ell_n(\vartheta^{(m+1)})
          -\ell_n(\vartheta^{(m)})\right|
  }{
    1+\left|\ell_n(\vartheta^{(m)})\right|
  }.
\end{aligned}
\label{eq:gem-relative-changes}
\end{equation}
Both Q-function values use the responsibilities from the same
E-step.

To assess stationarity at the updated estimate \(\vartheta\),
we recompute its responsibilities by
\eqref{eq:gem-responsibility} and set
\(N_{kh}(\vartheta):=\sum_i r_{i,kh}(\vartheta)\).
Let
\(\mathcal W_{\underline\omega_{\mathrm{base}}}(N(\vartheta))\)
denote the weight update in \eqref{eq:trimmed-weight-update}
evaluated with these counts and the baseline bound.
The weight residual measures the change implied by this update,
while the reward residual measures the gradient along the
unit sphere:
\[
\begin{aligned}
  R_\omega(\vartheta)
  &:=
  \left\|
    \omega-
    \mathcal W_{\underline\omega_{\mathrm{base}}}(N(\vartheta))
  \right\|_\infty,\\
  R_{\mathrm R,k}(\vartheta)
  &:=
  \left\|
    (I-\beta_k\beta_k^{\mathsf T})
    \nabla_{\beta_k}\ell_n(\vartheta)
  \right\|_2.
\end{aligned}
\]
Likelihood derivatives are evaluated using the recomputed
responsibilities.

For temperature \(h\), define
\(g_h(\vartheta):=\partial\ell_n(\vartheta)/\partial u_h\).
Its residual accounts for the direction in which the parameter
can move at either bound:
\[
  R_{u,h}(\vartheta)
  :=
  \begin{cases}
    |g_h(\vartheta)|,
      & u_{\min}<u_h<u_{\max},\\
    \max\{g_h(\vartheta),0\},
      & u_h=u_{\min},\\
    \max\{-g_h(\vartheta),0\},
      & u_h=u_{\max}.
  \end{cases}
\]
We combine these quantities as
\(R_{\mathrm{KKT}}(\vartheta)
:=\max\{R_\omega(\vartheta),
        \max_k R_{\mathrm R,k}(\vartheta),
        \max_h R_{u,h}(\vartheta)\}\).
For prespecified positive tolerances, we stop when
\begin{equation}
  r_Q^{(m)}\le\varepsilon_Q,\qquad
  r_\ell^{(m)}\le\varepsilon_\ell,\qquad
  R_{\mathrm{KKT}}(\vartheta^{(m+1)})
  \le\varepsilon_{\mathrm{KKT}}.
  \label{eq:gem-stopping-criterion}
\end{equation}

\paragraph{Screening and final selection.}
For multistart fitting, we screen the initial runs after a
prespecified number \(B_{\mathrm s}\) of GEM iterations.
Among runs with finite likelihoods and no numerical failures,
we retain the \(L\) fits with the largest training log likelihoods.
The retained runs continue from their current estimates under
the same weight-bound schedule until the stopping criteria hold.
We select the converged fit with the largest training log
likelihood, as in Section~\ref{sec:epi-initialization}.

For reporting, we order the temperatures increasingly.
Any permutation of reward or temperature labels is applied
consistently to the mixture weights and posterior probabilities.

\begin{algorithm}[!htbp]
\caption{Generalized EM from a fixed initialization}
\label{alg:gem-single}
\small
\begin{flushleft}
\textbf{Input:} an initial estimate \(\vartheta^{(0)}\) and
the numerical settings specified above.

For \(m=0,1,\ldots\), perform the following steps:
\end{flushleft}
\begin{enumerate}[leftmargin=1.6em,itemsep=2pt,topsep=4pt]
\item Compute the responsibilities in
\eqref{eq:gem-responsibility} and hold them fixed during
the parameter updates.
\item Form the weight candidate in
\eqref{eq:trimmed-weight-update} with lower bound
\(\underline\omega_m\), and apply
\eqref{eq:block-acceptance-tolerance}.
\item Update each reward and then each temperature, accepting
a proposed update only if
\eqref{eq:block-acceptance-tolerance} holds.
\item Compute the relative changes in
\eqref{eq:gem-relative-changes}.
Recompute the responsibilities at the updated estimate and
evaluate the first-order residuals.
Once the baseline weight bound applies, stop if
\eqref{eq:gem-stopping-criterion} holds.
\end{enumerate}
\begin{flushleft}
\textbf{Return:} the final parameter estimate and its posterior
class probabilities.
\end{flushleft}
\end{algorithm}
\FloatBarrier

\section{Simulation details and additional results}
\label{app:simulation-details}

Section~\ref{app:higher-order} gives the simulation specifications,
performance metrics, and complete recovery results.
Section~\ref{app:rare} examines zero and small cell weights.
Sections~\ref{app:rtro} and~\ref{app:bic-details} report additional
RT--RO comparisons and model-selection results.

\subsection{Simulation specifications and additional recovery results}
\label{app:higher-order}
\makeatletter
\edef\simSavedFloatTop{\the\@fptop}
\setlength{\@fptop}{0pt}
\makeatother

The recovery experiments include \(\RT(2,2)\), \(\RT(2,3)\),
\(\RT(2,4)\), and \(\RT(3,2)\), with the sample sizes,
trajectory lengths, and evaluation procedure in
Section~\ref{sec:recovery}.
In each setting, the fitted order is fixed at its generating value.
Hats denote fitted quantities, and a superscript \(0\) denotes
a generating value.

\subsubsection{Generating parameters}
\label{app:sim-reward-spec}

The experiments use an \(8\times10\) grid in the unit square,
with state space
\(\mathcal S:=\{0,1/9,\ldots,1\}\times\{0,1/7,\ldots,1\}\)
and \(|\mathcal S|=80\).
The nine actions in \(\mathcal A\) are to stay in place or move
in one of the eight compass directions, as in
Figure~\ref{fig:grid-trajectory}(b).
All moves are deterministic, and an attempted move beyond the
grid boundary leaves the state unchanged.
Let \(s^a\) denote the state reached from \(s\) under action \(a\).
We set \(\gamma=0.95\) and take \(\mu_0\) uniform on
\(\{(1/9,1/7),(8/9,1/7),(4/9,3/7),(1/9,6/7),(8/9,6/7)\}\).

The generating rewards favor moves toward three targets:
\(g_1:=(0,0)\), \(g_2:=(1,1)\), and \(g_3:=(1,0)\).
For target \(g_k\), the quantity
\(-\lVert s^a-g_k\rVert_2^2\) assigns a higher value to an action
whose next state is closer to the target.
We obtain \(f_k^0\) by subtracting the mean over actions at each
state and normalizing to unit \(\nu\)-norm, with
\(\nu(s,a)=1/720\).
A generating model with \(K\) reward classes uses
\(f_1^0,\ldots,f_K^0\).
For example, generating models with \(K=2\) use \(f_1^0\) and \(f_2^0\),
whereas those with \(K=3\) use all three rewards.

The next-state coordinates and their squares give a linear reward
representation with \(p=4\).
We center and whiten these four functions to obtain the features
in \eqref{eq:linear-reward-class}.
For \(s^a=(x^a,y^a)\), define
\(z(s,a):=(x^a,y^a,(x^a)^2,(y^a)^2)^\top\)
and its mean over actions,
\(\bar z(s):=9^{-1}\sum_{b\in\mathcal A}z(s,b)\).
Using the uniform reference weights
\(\eta(s)=1/80\), \(\omega_s(a)=1/9\), and
\(\nu(s,a)=1/720\), define the Gram matrix and transformed
features by
\begin{equation}
\begin{aligned}
  G
  &:=
  \frac{1}{720}
  \sum_{s\in\mathcal S}\sum_{a\in\mathcal A}
  \{z(s,a)-\bar z(s)\}
  \{z(s,a)-\bar z(s)\}^{\top},\\
  \phi(s,a)
  &:=
  G^{-1/2}\{z(s,a)-\bar z(s)\}.
\end{aligned}
\label{eq:sim-features}
\end{equation}
Here \(G^{1/2}\) denotes the symmetric positive definite
square root of \(G\).
The resulting features satisfy
\eqref{eq:linear-feature-normalization}.

For each target \(g_k=(g_{kx},g_{ky})\), define
\(\xi_k:=(2g_{kx},2g_{ky},-1,-1)^\top\).
Expanding the negative squared distance shows that its
statewise centered form is
\(\xi_k^\top\{z(s,a)-\bar z(s)\}\), whose squared
\(\nu\)-norm is \(\xi_k^\top G\xi_k\).
The normalized rewards and their coefficient vectors are
therefore
\[
\begin{aligned}
  f_k^0(s,a)
  &:=
  \frac{\xi_k^\top\{z(s,a)-\bar z(s)\}}
       {\sqrt{\xi_k^\top G\xi_k}}
  =\phi(s,a)^\top\beta_k^0,\\
  \beta_k^0
  &:=
  \frac{G^{1/2}\xi_k}{\sqrt{\xi_k^\top G\xi_k}},
  \qquad k=1,2,3.
\end{aligned}
\]
Thus \(\|f_k^0\|_\nu=\|\beta_k^0\|_2=1\).

In the recovery and RT--RO comparison experiments, all
\(KH\) components have equal generating weights,
\(\omega_{kh}^0:=1/(KH)\).
The generating temperatures are equally spaced from \(0.5\)
to \(1.5\), with
\(\lambda_h^0:=0.5+(h-1)/(H-1)\)
for \(h=1,\ldots,H\) and \(H\in\{2,3,4\}\).

\subsubsection{Performance metrics and Monte Carlo summaries}
\label{app:sim-metrics}

All trajectories within a simulated setting have the same
length, \(T_i=T\).
We fit the models using the procedure in Section~\ref{sec:gem}
and denote the fitted component policies by
\(\widehat\pi_{kh}:=\pi^*_{\widehat f_k,\widehat\lambda_h}\).

We evaluate the fitted models on \(N_{\mathrm{eva}}\) independent
held-out trajectories generated under the same setting.
For evaluation trajectory \(i\), let \(Z_i\) be its generating cell
and \(\widehat Z_i\) its maximum a posteriori (MAP) assignment,
the cell with the largest fitted posterior probability in
\eqref{eq:gem-responsibility}.
For generating cell \(u\) and fitted cell \(v\), define \(n_{uv} = \#\{i:Z_i=u,\widehat Z_i=v\}, \; n_{u\cdot}=\sum_v n_{uv}, \; n_{\cdot v}=\sum_u n_{uv}.\)
We measure cell recovery by the adjusted Rand index (ARI)
\citep{hubert1985},
\[
\ARI
=
\frac{
\displaystyle
\sum_{u,v}\binom{n_{uv}}{2}
-
\displaystyle
\frac{
\left(\sum_u\binom{n_{u\cdot}}{2}\right)
\left(\sum_v\binom{n_{\cdot v}}{2}\right)
}{
\binom{N_{\mathrm{eva}}}{2}
}
}{
\displaystyle
\frac{1}{2}
\left\{
\sum_u\binom{n_{u\cdot}}{2}
+
\sum_v\binom{n_{\cdot v}}{2}
\right\}
-
\displaystyle
\frac{
\left(\sum_u\binom{n_{u\cdot}}{2}\right)
\left(\sum_v\binom{n_{\cdot v}}{2}\right)
}{
\binom{N_{\mathrm{eva}}}{2}
}
}.
\]
An ARI of \(1\) indicates perfect agreement up to label permutation.
Values of \(0\) indicate chance-level agreement, and negative
values indicate less agreement than expected by chance.

Reward error compares the fitted and generating rewards after
optimal label matching:
\begin{equation}
D_f
=
\left\{
\frac{1}{K}
\min_{\sigma\in\mathfrak S_K}
\sum_{k=1}^K
\left\lVert
\widehat f_{\sigma(k)}-f_k^0
\right\rVert_\nu^2
\right\}^{1/2},
\label{eq:reward-error}
\end{equation}
where \(\mathfrak S_K\) is the set of permutations of
\(\{1,\ldots,K\}\).

We evaluate predictive performance using the mixture negative log-likelihood
per action on the held-out trajectories:
\begin{equation*}
\NLL_{\mathrm{eva}}
=
-\frac{1}{N_{\mathrm{eva}}T}
\sum_{i=1}^{N_{\mathrm{eva}}}
\log
\left\{
\sum_{k,h}\widehat\omega_{kh}
\prod_{t=0}^{T-1}
\widehat\pi_{kh}(a_{it}\mid s_{it})
\right\}.
\end{equation*}
Parameter-independent factors from \(\mu_0\) and \(P\) are omitted.

State--action occupancy describes the expected proportion of
steps spent at each state--action pair over a length-\(T\)
trajectory generated by a policy mixture: \(d_T(s,a) = \frac{1}{T} \sum_{t=0}^{T-1} \Pr(S_t=s,A_t=a).\)
We measure occupancy error by \(D_{\Occ} = \left\lVert\widehat d_T-d_T^0\right\rVert_1 = \sum_{s,a} \left| \widehat d_T(s,a)-d_T^0(s,a) \right|,\)
where \(\widehat d_T\) and \(d_T^0\) are induced by the fitted
and generating mixtures, respectively.
Smaller values of \(D_f\), \(\NLL_{\mathrm{eva}}\), and \(D_{\Occ}\)
indicate better reward recovery, predictive fit, and occupancy
recovery, respectively.

For each metric \(M\), let \(M^{(b)}\) denote its value in Monte Carlo
replication \(b\), \(b=1,\ldots,n_{\mathrm{rep}}\).
The sample mean and variance are
\(\overline M=n_{\mathrm{rep}}^{-1}\sum_{b=1}^{n_{\mathrm{rep}}}M^{(b)}\)
and
\(s_M^2=(n_{\mathrm{rep}}-1)^{-1}\sum_{b=1}^{n_{\mathrm{rep}}}
(M^{(b)}-\overline M)^2\).
Figure bands are pointwise \(95\%\) Monte Carlo
confidence intervals for the mean performance,
\(\overline M \pm t_{n_{\mathrm{rep}}-1,0.975}
s_M/\sqrt{n_{\mathrm{rep}}}\),
where \(t_{n_{\mathrm{rep}}-1,0.975}\) is the \(0.975\) quantile of the
\(t\)-distribution with \(n_{\mathrm{rep}}-1\) degrees of freedom.
We use \(n_{\mathrm{rep}}=50\) throughout the simulation study.

For the paired RT--RO comparisons, we also report medians and
interquartile intervals of the performance differences.
Here \(Q_{25}\) and \(Q_{75}\) are the 25th and 75th percentiles.
The interval \([Q_{25},Q_{75}]\) contains the middle 50\% of
the paired differences, and its width \(Q_{75}-Q_{25}\) is
the interquartile range (IQR).

\subsubsection{Additional recovery results}
\label{app:sim-higher-results}

At \((n,T)=(300,1000)\), mean reward error remains below
\(0.02\) in all three settings with \(K=2\), while mean ARI
decreases as \(H\) increases
(Table~\ref{tab:higher-summary}).
In this design, increasing \(H\) reduces both the spacing
between generating temperatures and the weight of each
mixture component.
For \(\RT(3,2)\), both the mean and standard deviation of
reward error are larger than in the \(K=2\) settings.

\begin{table}[!htbp]
\centering
\caption{Recovery under correctly specified models at
$(n,T)=(300,1000)$. Entries are means over 50 replications,
with standard deviations in parentheses.}
\label{tab:higher-summary}
\scriptsize
\setlength{\tabcolsep}{3pt}
\begin{tabular}{lccccc}
\toprule
Model & Cell ARI & Reward $D_f$
& \makecell[c]{Ordered temperature estimates}
& NLL/action & Occupancy $D_{\Occ}$ \\
\midrule
$\RT(2,2)$ & .9692 (.0148) & .0170 (.0088)
& .5006 (.0078), 1.4960 (.0303)
& 2.1693 (.0005) & .0538 (.0346) \\
$\RT(2,3)$ & .6330 (.0400) & .0194 (.0087)
& .4982 (.0160), .9842 (.0964), 1.5010 (.0998)
& 2.1704 (.0004) & .0439 (.0327) \\
$\RT(2,4)$ & .4644 (.0365) & .0192 (.0097)
& \makecell[c]{.4925 (.0313), .7898 (.1473)\\
1.1546 (.1669), 1.5723 (.1895)}
& 2.1706 (.0004) & .0491 (.0299) \\
$\RT(3,2)$ & .8773 (.1642) & .1593 (.2362)
& .5139 (.0275), 1.5386 (.0865)
& 2.1716 (.0033) & .0781 (.0406) \\
\bottomrule
\end{tabular}
\end{table}

\begin{figure}[!htbp]
\centering
\includegraphics[width=\linewidth]{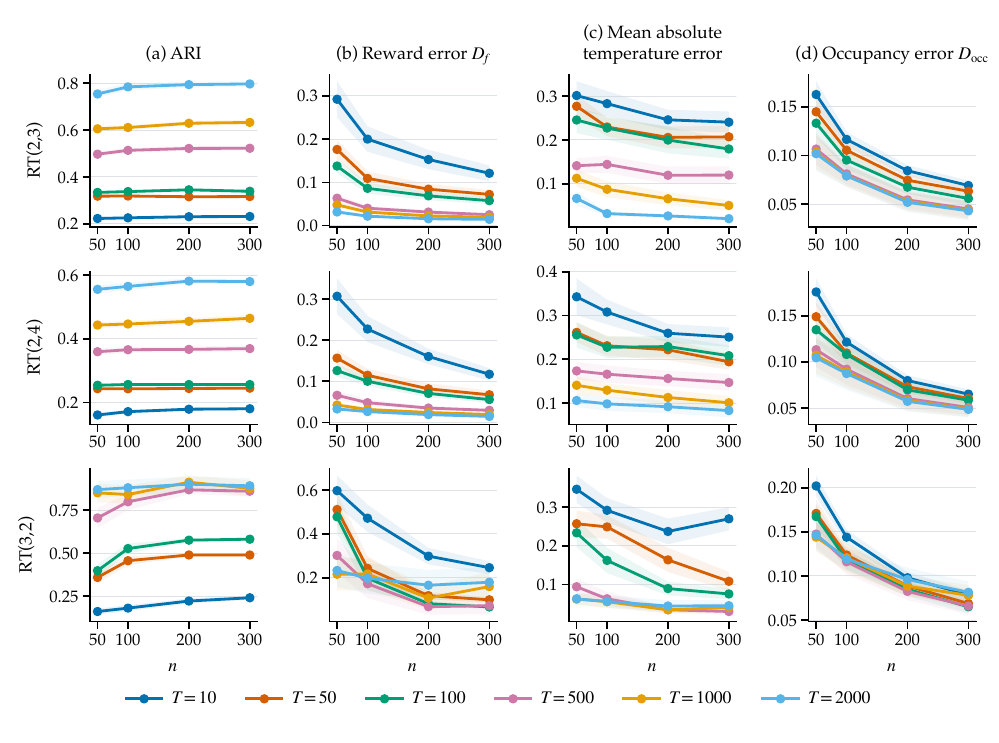}
\caption{Higher-order recovery over \(n\) and \(T\). Lines are means over
50 replications. Bands are pointwise 95\% intervals for the means.
The temperature panel shows mean absolute
error \(D_\lambda\); full ordered temperature estimates appear in
Tables~\ref{tab:exp1-rt2_3-complete}--\ref{tab:exp1-rt3_2-complete}.}
\label{fig:appendix-exp1-higher}
\end{figure}

We summarize temperature recovery in
Figure~\ref{fig:appendix-exp1-higher} using the mean absolute
error across temperature classes.
For each replication, we arrange the fitted and generating
temperatures in increasing order and calculate
\(D_\lambda:=H^{-1}\sum_{h=1}^H
|\widehat\lambda_h-\lambda_h^0|\).
Tables~\ref{tab:exp1-rt2_2-complete}--\ref{tab:exp1-rt3_2-complete}
report means and standard deviations for all recovery
settings, including those with \(T=2000\).

\begin{table}[!htbp]
\centering
\caption{Complete correctly specified recovery for $\RT(2,2)$. Entries are mean (SD).}
\label{tab:exp1-rt2_2-complete}
\label{tab:rt22-recovery}
\scriptsize
\setlength{\tabcolsep}{3pt}
\begin{tabular}{rrccccc}
\toprule
$n$ & $T$ & ARI & $D_f$ & Ordered temperatures & NLL/action & Occupancy $D_{\Occ}$ \\
\midrule
50 & 10 & .3237 (.0486) & .2825 (.1161) & .4662 (.2055), 1.2916 (.4915) & 2.1465 (.0093) & .1693 (.0470) \\
50 & 50 & .4683 (.0227) & .1695 (.0855) & .4985 (.1667), 1.5311 (.3391) & 2.1647 (.0027) & .1598 (.0663) \\
50 & 100 & .5309 (.0410) & .1165 (.0678) & .5092 (.1291), 1.5402 (.2618) & 2.1680 (.0015) & .1481 (.0695) \\
50 & 500 & .8681 (.0382) & .0615 (.0342) & .5042 (.0388), 1.5073 (.1074) & 2.1695 (.0005) & .1323 (.0827) \\
50 & 1000 & .9667 (.0171) & .0397 (.0230) & .4979 (.0239), 1.4983 (.0705) & 2.1694 (.0005) & .1301 (.0842) \\
50 & 2000 & .9986 (.0033) & .0292 (.0129) & .5003 (.0177), 1.4992 (.0472) & 2.1692 (.0005) & .1286 (.0853) \\
100 & 10 & .3319 (.0467) & .2141 (.0995) & .4910 (.1977), 1.4464 (.4826) & 2.1420 (.0079) & .1254 (.0332) \\
100 & 50 & .4660 (.0284) & .1134 (.0511) & .4984 (.1179), 1.5375 (.2918) & 2.1632 (.0024) & .1094 (.0397) \\
100 & 100 & .5336 (.0461) & .0827 (.0432) & .4971 (.0934), 1.4943 (.2446) & 2.1672 (.0012) & .1010 (.0426) \\
100 & 500 & .8705 (.0323) & .0408 (.0195) & .5023 (.0323), 1.5047 (.0933) & 2.1694 (.0005) & .0935 (.0500) \\
100 & 1000 & .9688 (.0172) & .0278 (.0123) & .4990 (.0193), 1.4970 (.0530) & 2.1693 (.0005) & .0912 (.0515) \\
100 & 2000 & .9986 (.0033) & .0198 (.0109) & .5007 (.0128), 1.4971 (.0410) & 2.1692 (.0005) & .0905 (.0523) \\
200 & 10 & .3374 (.0473) & .1446 (.0684) & .4787 (.1736), 1.4855 (.4351) & 2.1385 (.0076) & .0845 (.0227) \\
200 & 50 & .4629 (.0311) & .0790 (.0394) & .4875 (.1073), 1.5218 (.2381) & 2.1626 (.0021) & .0750 (.0286) \\
200 & 100 & .5467 (.0437) & .0510 (.0224) & .4916 (.0671), 1.4959 (.1610) & 2.1669 (.0012) & .0682 (.0310) \\
200 & 500 & .8750 (.0360) & .0271 (.0149) & .5040 (.0230), 1.4991 (.0591) & 2.1693 (.0005) & .0627 (.0366) \\
200 & 1000 & .9679 (.0150) & .0209 (.0108) & .5002 (.0114), 1.4914 (.0365) & 2.1693 (.0005) & .0616 (.0375) \\
200 & 2000 & .9987 (.0032) & .0127 (.0079) & .5009 (.0074), 1.4994 (.0278) & 2.1692 (.0005) & .0606 (.0382) \\
300 & 10 & .3377 (.0414) & .1167 (.0535) & .4578 (.1497), 1.4545 (.4052) & 2.1377 (.0077) & .0719 (.0195) \\
300 & 50 & .4698 (.0298) & .0620 (.0312) & .5023 (.0837), 1.5290 (.1978) & 2.1624 (.0022) & .0647 (.0249) \\
300 & 100 & .5632 (.0401) & .0450 (.0189) & .4990 (.0462), 1.4959 (.1184) & 2.1668 (.0012) & .0607 (.0279) \\
300 & 500 & .8793 (.0305) & .0228 (.0118) & .5043 (.0148), 1.5042 (.0437) & 2.1693 (.0005) & .0548 (.0336) \\
300 & 1000 & .9692 (.0148) & .0170 (.0088) & .5006 (.0078), 1.4960 (.0303) & 2.1693 (.0005) & .0538 (.0346) \\
300 & 2000 & .9986 (.0033) & .0111 (.0065) & .5008 (.0058), 1.4994 (.0219) & 2.1692 (.0005) & .0532 (.0351) \\
\bottomrule
\end{tabular}
\end{table}

\begin{table}[!htbp]
\centering
\caption{Complete correctly specified recovery for $\RT(2,3)$. Entries are mean (SD).}
\label{tab:exp1-rt2_3-complete}
\scriptsize
\setlength{\tabcolsep}{3pt}
\begin{tabular}{rrccccc}
\toprule
$n$ & $T$ & ARI & $D_f$ & Ordered temperatures & NLL/action & Occupancy $D_{\Occ}$ \\
\midrule
50 & 10 & .2225 (.0251) & .2921 (.1468) & .4009 (.1534), .9387 (.4081), 1.4996 (.4790) & 2.1472 (.0085) & .1626 (.0481) \\
50 & 50 & .3180 (.0137) & .1760 (.0771) & .4611 (.2163), .9267 (.3661), 1.5337 (.3873) & 2.1653 (.0023) & .1449 (.0509) \\
50 & 100 & .3340 (.0196) & .1377 (.0569) & .4383 (.1687), .8994 (.3652), 1.5542 (.3377) & 2.1684 (.0015) & .1332 (.0485) \\
50 & 500 & .4973 (.0471) & .0633 (.0296) & .4785 (.0719), .9820 (.2318), 1.5384 (.2125) & 2.1704 (.0004) & .1066 (.0584) \\
50 & 1000 & .6056 (.0468) & .0480 (.0246) & .4975 (.0397), .9768 (.2068), 1.5156 (.1865) & 2.1705 (.0004) & .1034 (.0605) \\
50 & 2000 & .7546 (.0603) & .0316 (.0169) & .5000 (.0227), .9630 (.1374), 1.4971 (.1210) & 2.1703 (.0004) & .1019 (.0622) \\
100 & 10 & .2254 (.0207) & .1999 (.1030) & .4445 (.2196), .8587 (.3020), 1.4054 (.4348) & 2.1420 (.0083) & .1165 (.0274) \\
100 & 50 & .3185 (.0192) & .1092 (.0561) & .4215 (.1640), .8809 (.2653), 1.5181 (.3413) & 2.1640 (.0020) & .1054 (.0315) \\
100 & 100 & .3374 (.0221) & .0862 (.0441) & .4545 (.1597), .9123 (.3313), 1.5275 (.3249) & 2.1678 (.0014) & .0953 (.0336) \\
100 & 500 & .5136 (.0334) & .0402 (.0200) & .4940 (.0487), 1.0341 (.2159), 1.6024 (.2567) & 2.1703 (.0005) & .0811 (.0379) \\
100 & 1000 & .6112 (.0357) & .0314 (.0158) & .4953 (.0354), .9940 (.1540), 1.5500 (.1735) & 2.1705 (.0004) & .0793 (.0391) \\
100 & 2000 & .7849 (.0536) & .0218 (.0120) & .5004 (.0145), .9958 (.0478), 1.4909 (.0598) & 2.1703 (.0004) & .0790 (.0400) \\
200 & 10 & .2302 (.0200) & .1529 (.0787) & .4393 (.1704), .8626 (.2107), 1.5324 (.4209) & 2.1397 (.0075) & .0844 (.0201) \\
200 & 50 & .3155 (.0176) & .0842 (.0422) & .4314 (.1414), .8958 (.2648), 1.5816 (.2932) & 2.1635 (.0019) & .0746 (.0220) \\
200 & 100 & .3450 (.0256) & .0688 (.0352) & .4540 (.1326), .9884 (.3301), 1.5398 (.2818) & 2.1675 (.0014) & .0675 (.0247) \\
200 & 500 & .5219 (.0392) & .0313 (.0157) & .4960 (.0376), 1.0129 (.2005), 1.5449 (.2230) & 2.1702 (.0005) & .0544 (.0327) \\
200 & 1000 & .6294 (.0355) & .0225 (.0105) & .4984 (.0200), .9811 (.1164), 1.5148 (.1324) & 2.1704 (.0004) & .0530 (.0338) \\
200 & 2000 & .7945 (.0439) & .0159 (.0072) & .5001 (.0099), .9942 (.0394), 1.4935 (.0511) & 2.1703 (.0004) & .0519 (.0353) \\
300 & 10 & .2311 (.0192) & .1210 (.0645) & .3957 (.1575), .8345 (.1942), 1.5295 (.3988) & 2.1389 (.0072) & .0692 (.0181) \\
300 & 50 & .3162 (.0181) & .0722 (.0351) & .4177 (.1340), .9346 (.2724), 1.5678 (.3021) & 2.1633 (.0019) & .0633 (.0236) \\
300 & 100 & .3387 (.0257) & .0577 (.0281) & .4614 (.1013), .9780 (.2693), 1.5657 (.2786) & 2.1674 (.0014) & .0558 (.0249) \\
300 & 500 & .5226 (.0359) & .0253 (.0123) & .4909 (.0402), .9521 (.2100), 1.5337 (.2094) & 2.1702 (.0005) & .0451 (.0315) \\
300 & 1000 & .6330 (.0400) & .0194 (.0087) & .4982 (.0160), .9842 (.0964), 1.5010 (.0998) & 2.1704 (.0004) & .0439 (.0327) \\
300 & 2000 & .7974 (.0411) & .0144 (.0063) & .4995 (.0069), .9913 (.0292), 1.4863 (.0377) & 2.1703 (.0004) & .0432 (.0336) \\
\bottomrule
\end{tabular}
\end{table}

\begin{table}[!htbp]
\centering
\caption{Complete correctly specified recovery for $\RT(2,4)$. Entries are mean (SD).}
\label{tab:exp1-rt2_4-complete}
\scriptsize
\setlength{\tabcolsep}{3pt}
\begin{tabular}{rrccccc}
\toprule
$n$ & $T$ & ARI & $D_f$ & Ordered temperatures & NLL/action & Occupancy $D_{\Occ}$ \\
\midrule
50 & 10 & .1603 (.0260) & .3069 (.1529) & \makecell[c]{.4475 (.2225), .9206 (.4647)\\1.2266 (.4988), 1.4306 (.4754)} & 2.1524 (.0089) & .1757 (.0516) \\
50 & 50 & .2430 (.0090) & .1563 (.0682) & \makecell[c]{.4930 (.2096), .7835 (.2721)\\1.0605 (.3723), 1.5735 (.3751)} & 2.1660 (.0025) & .1489 (.0525) \\
50 & 100 & .2537 (.0146) & .1259 (.0523) & \makecell[c]{.4880 (.2189), .8702 (.3243)\\1.1432 (.3534), 1.6694 (.3187)} & 2.1686 (.0012) & .1348 (.0511) \\
50 & 500 & .3593 (.0375) & .0658 (.0374) & \makecell[c]{.4848 (.0938), .8341 (.2472)\\1.1385 (.2454), 1.5474 (.2745)} & 2.1707 (.0006) & .1132 (.0653) \\
50 & 1000 & .4432 (.0314) & .0422 (.0244) & \makecell[c]{.4985 (.0673), .8870 (.2427)\\1.1782 (.1656), 1.5791 (.2456)} & 2.1708 (.0004) & .1071 (.0685) \\
50 & 2000 & .5555 (.0431) & .0326 (.0160) & \makecell[c]{.4809 (.0388), .7794 (.1399)\\1.1092 (.1787), 1.5132 (.1736)} & 2.1707 (.0004) & .1045 (.0710) \\
100 & 10 & .1708 (.0195) & .2271 (.1116) & \makecell[c]{.4326 (.2045), .7997 (.2706)\\1.2219 (.5270), 1.5986 (.4110)} & 2.1447 (.0063) & .1215 (.0302) \\
100 & 50 & .2427 (.0098) & .1145 (.0551) & \makecell[c]{.5145 (.2006), .8471 (.2212)\\1.1823 (.3542), 1.5876 (.3429)} & 2.1647 (.0022) & .1096 (.0413) \\
100 & 100 & .2563 (.0178) & .1000 (.0414) & \makecell[c]{.4880 (.1743), .8957 (.2774)\\1.1467 (.3197), 1.6103 (.3277)} & 2.1679 (.0010) & .1080 (.0460) \\
100 & 500 & .3656 (.0311) & .0478 (.0268) & \makecell[c]{.4879 (.0819), .8763 (.2297)\\1.1444 (.2108), 1.5902 (.2918)} & 2.1706 (.0005) & .0921 (.0588) \\
100 & 1000 & .4465 (.0350) & .0313 (.0165) & \makecell[c]{.4799 (.0558), .7864 (.1696)\\1.1516 (.1896), 1.5748 (.2281)} & 2.1707 (.0004) & .0891 (.0603) \\
100 & 2000 & .5648 (.0453) & .0262 (.0132) & \makecell[c]{.4929 (.0245), .7902 (.1240)\\1.1472 (.1892), 1.5080 (.1714)} & 2.1707 (.0004) & .0874 (.0624) \\
200 & 10 & .1785 (.0173) & .1602 (.0665) & \makecell[c]{.4506 (.2041), .8013 (.2157)\\1.0381 (.3566), 1.5123 (.4124)} & 2.1412 (.0063) & .0799 (.0200) \\
200 & 50 & .2436 (.0096) & .0815 (.0482) & \makecell[c]{.4262 (.1561), .7976 (.2200)\\1.1801 (.3510), 1.6001 (.3392)} & 2.1642 (.0021) & .0731 (.0260) \\
200 & 100 & .2560 (.0154) & .0704 (.0331) & \makecell[c]{.4499 (.1574), .8586 (.2888)\\1.1761 (.3473), 1.6151 (.2968)} & 2.1676 (.0010) & .0697 (.0301) \\
200 & 500 & .3667 (.0325) & .0348 (.0185) & \makecell[c]{.4796 (.0713), .8530 (.2148)\\1.1666 (.1846), 1.6639 (.2483)} & 2.1705 (.0005) & .0603 (.0364) \\
200 & 1000 & .4548 (.0375) & .0240 (.0119) & \makecell[c]{.4877 (.0438), .7889 (.1607)\\1.1778 (.1672), 1.5805 (.1918)} & 2.1707 (.0004) & .0583 (.0376) \\
200 & 2000 & .5813 (.0306) & .0188 (.0107) & \makecell[c]{.4939 (.0194), .7809 (.1100)\\1.1273 (.1671), 1.5422 (.1754)} & 2.1707 (.0004) & .0573 (.0386) \\
300 & 10 & .1799 (.0161) & .1171 (.0533) & \makecell[c]{.4795 (.2068), .8128 (.2072)\\1.1111 (.3705), 1.5478 (.4080)} & 2.1402 (.0064) & .0651 (.0178) \\
300 & 50 & .2447 (.0105) & .0671 (.0347) & \makecell[c]{.4407 (.1586), .7328 (.1947)\\1.1352 (.2348), 1.5907 (.3146)} & 2.1640 (.0021) & .0604 (.0217) \\
300 & 100 & .2561 (.0169) & .0554 (.0282) & \makecell[c]{.4511 (.1466), .8330 (.2653)\\1.1175 (.3025), 1.6056 (.2897)} & 2.1675 (.0010) & .0586 (.0236) \\
300 & 500 & .3691 (.0318) & .0292 (.0146) & \makecell[c]{.4888 (.0589), .8472 (.2056)\\1.1539 (.1903), 1.6395 (.2313)} & 2.1705 (.0005) & .0507 (.0292) \\
300 & 1000 & .4644 (.0365) & .0192 (.0097) & \makecell[c]{.4925 (.0313), .7898 (.1473)\\1.1546 (.1669), 1.5723 (.1895)} & 2.1706 (.0004) & .0491 (.0299) \\
300 & 2000 & .5800 (.0278) & .0144 (.0079) & \makecell[c]{.4964 (.0114), .7917 (.0946)\\1.1197 (.1557), 1.5422 (.1680)} & 2.1707 (.0004) & .0486 (.0304) \\
\bottomrule
\end{tabular}
\end{table}

\begin{table}[!htbp]
\centering
\caption{Complete correctly specified recovery for $\RT(3,2)$. Entries are mean (SD).}
\label{tab:exp1-rt3_2-complete}
\scriptsize
\setlength{\tabcolsep}{3pt}
\begin{tabular}{rrccccc}
\toprule
$n$ & $T$ & ARI & $D_f$ & Ordered temperatures & NLL/action & Occupancy $D_{\Occ}$ \\
\midrule
50 & 10 & .1600 (.0619) & .5970 (.2421) & .3563 (.1632), 1.2732 (.5020) & 2.1654 (.0105) & .2020 (.0414) \\
50 & 50 & .3586 (.1220) & .5109 (.2653) & .4925 (.2071), 1.4893 (.3927) & 2.1766 (.0038) & .1709 (.0537) \\
50 & 100 & .3980 (.1345) & .4777 (.2683) & .5162 (.1774), 1.5765 (.3694) & 2.1765 (.0037) & .1673 (.0484) \\
50 & 500 & .7052 (.2046) & .3015 (.2716) & .5188 (.0635), 1.5879 (.1621) & 2.1737 (.0038) & .1475 (.0515) \\
50 & 1000 & .8516 (.1897) & .2157 (.2595) & .5087 (.0314), 1.5516 (.1278) & 2.1722 (.0038) & .1440 (.0554) \\
50 & 2000 & .8702 (.1748) & .2328 (.2636) & .5200 (.0352), 1.5611 (.1139) & 2.1722 (.0039) & .1462 (.0609) \\
100 & 10 & .1798 (.0600) & .4714 (.2196) & .4043 (.1700), 1.3265 (.4300) & 2.1560 (.0084) & .1440 (.0301) \\
100 & 50 & .4560 (.0899) & .2425 (.1934) & .4614 (.1682), 1.4993 (.3919) & 2.1723 (.0035) & .1237 (.0367) \\
100 & 100 & .5269 (.1032) & .1980 (.1915) & .4887 (.1255), 1.5267 (.2855) & 2.1725 (.0027) & .1180 (.0379) \\
100 & 500 & .7996 (.1701) & .1723 (.2321) & .5137 (.0354), 1.5629 (.1322) & 2.1720 (.0033) & .1160 (.0493) \\
100 & 1000 & .8416 (.1846) & .2169 (.2571) & .5151 (.0296), 1.5623 (.1178) & 2.1724 (.0038) & .1205 (.0567) \\
100 & 2000 & .8812 (.1656) & .1976 (.2452) & .5199 (.0343), 1.5679 (.1029) & 2.1719 (.0037) & .1192 (.0443) \\
200 & 10 & .2210 (.0537) & .2982 (.1336) & .4398 (.1444), 1.4729 (.4092) & 2.1488 (.0074) & .0979 (.0229) \\
200 & 50 & .4895 (.0316) & .1179 (.0579) & .5006 (.1151), 1.5517 (.2928) & 2.1701 (.0022) & .0887 (.0311) \\
200 & 100 & .5759 (.0374) & .0813 (.0369) & .4996 (.0662), 1.4852 (.1611) & 2.1710 (.0011) & .0849 (.0283) \\
200 & 500 & .8692 (.0868) & .0675 (.1255) & .5049 (.0179), 1.5168 (.0754) & 2.1706 (.0018) & .0825 (.0347) \\
200 & 1000 & .9136 (.1360) & .1069 (.1875) & .5078 (.0223), 1.5213 (.0744) & 2.1709 (.0027) & .0884 (.0395) \\
200 & 2000 & .9019 (.1587) & .1656 (.2439) & .5138 (.0269), 1.5536 (.0974) & 2.1715 (.0034) & .0960 (.0468) \\
300 & 10 & .2400 (.0366) & .2456 (.1039) & .4215 (.1356), 1.3962 (.4366) & 2.1468 (.0075) & .0786 (.0176) \\
300 & 50 & .4892 (.0322) & .0990 (.0431) & .5040 (.0778), 1.5203 (.2100) & 2.1697 (.0021) & .0687 (.0229) \\
300 & 100 & .5814 (.0308) & .0665 (.0321) & .4972 (.0496), 1.4678 (.1388) & 2.1708 (.0011) & .0650 (.0250) \\
300 & 500 & .8613 (.1057) & .0721 (.1384) & .5054 (.0168), 1.5136 (.0674) & 2.1708 (.0019) & .0659 (.0338) \\
300 & 1000 & .8773 (.1642) & .1593 (.2362) & .5139 (.0275), 1.5386 (.0865) & 2.1716 (.0033) & .0781 (.0406) \\
300 & 2000 & .8926 (.1577) & .1791 (.2447) & .5179 (.0290), 1.5567 (.0890) & 2.1717 (.0034) & .0815 (.0430) \\
\bottomrule
\end{tabular}
\end{table}
\FloatBarrier

\subsection{Robustness to zero and rare mixture cells}
\label{app:rare}

We examine recovery when one reward--temperature combination
has zero or small mixing weight.
Both settings use the generating rewards \(f_1^0\) and \(f_2^0\)
from Appendix~\ref{app:sim-reward-spec}. Both settings use generating temperatures \((\lambda_1^0,\lambda_2^0)=(0.5,1.5)\).
In each replication, we fit \(\RT(2,2)\) to \(n=300\)
training trajectories of length \(T=100\) 
with \(N_{eva}=300\).

The generating weight matrices are
\[
W_{\mathrm{zero}}^0
:=
\begin{pmatrix}
1/3 & 1/3\\[2pt]
1/3 & 0
\end{pmatrix},
\qquad
W_{\mathrm{rare}}^0
:=
\begin{pmatrix}
0.98/3 & 0.98/3\\[2pt]
0.98/3 & 0.02
\end{pmatrix},
\]
where rows index reward classes and columns index temperature
classes.
Both reward classes and both temperature classes have positive
marginal probabilities in each setting.
In the rare setting, the \((2,2)\) component contributes six
training trajectories in expectation.

The zero setting has three components with positive generating
weight and falls outside the regular family covered by
Theorem~\ref{thm:mixture-identification}.
We retain four components in the fitted model and calculate
ARI from all evaluation trajectories, following
Section~\ref{app:sim-metrics}.

Both settings have mean reward errors below \(0.05\)
and similar mean temperature estimates
(Table~\ref{tab:rare-endpoints}).
The mean estimate of \(\omega_{22}\) is \(0.0018\) when its
generating value is zero and \(0.0197\) when its generating
value is \(0.02\) (Table~\ref{tab:rare-weights}).

\begin{table}[!htbp]
\centering
\caption{Recovery with a zero or rare mixture cell at
\((n,T)=(300,100)\). Entries are means over 50 replications,
with standard deviations in parentheses.}
\label{tab:rare-endpoints}
\scriptsize
\setlength{\tabcolsep}{3pt}
\renewcommand{\arraystretch}{1.12}
\begin{tabular}{@{}lcccccc@{}}
\toprule
Setting & Cell ARI & Reward \(D_f\)
& \(\widehat\lambda_1\) & \(\widehat\lambda_2\)
& NLL/action & Occupancy \(D_{\Occ}\) \\
\midrule
Zero & .6623 (.0412) & .0473 (.0276) & .4999 (.0286) & 1.4948 (.1405) & 2.1619 (.0013) & .0536 (.0225) \\
Rare & .6492 (.0410) & .0435 (.0215) & .4928 (.0321) & 1.4970 (.1195) & 2.1623 (.0011) & .0564 (.0273) \\
\bottomrule
\end{tabular}
\end{table}

\begin{table}[!htbp]
\centering
\caption{Recovery of mixture weights after matching reward
and temperature labels separately. Estimated entries are
means over 50 replications, with standard deviations
in parentheses.}
\label{tab:rare-weights}
\scriptsize
\setlength{\tabcolsep}{3pt}
\renewcommand{\arraystretch}{1.12}
\begin{tabular}{@{}llcccc@{}}
\toprule
Setting & Quantity & \(\omega_{11}\) & \(\omega_{12}\)
& \(\omega_{21}\) & \(\omega_{22}\) \\
\midrule
Zero & Generating & 0.333333 & 0.333333 & 0.333333 & 0.0 \\
Zero & Estimated & .3424 (.0576) & .3250 (.0531) & .3307 (.0275) & .0018 (.0040) \\
\addlinespace[3pt]
Rare & Generating & 0.326667 & 0.326667 & 0.326667 & 0.02 \\
Rare & Estimated & .3144 (.0448) & .3464 (.0474) & .3196 (.0354) & .0197 (.0143) \\
\bottomrule
\end{tabular}
\end{table}
\FloatBarrier

\subsection{Additional RT--RO comparisons}
\label{app:rtro}

We compare the recovery of latent classes and rewards by
\(\RT(K,H)\) and \(\RO(KH,1)\) at \((n,T)=(300,1000)\),
using the paired experiments described in
Section~\ref{sec:separation}.

Each reward in the generating \(\RT(K,H)\) model is shared
by \(H\) temperature classes, whereas \(\RO(KH,1)\) estimates
a separate reward for each of its \(KH\) components.
We therefore match the fitted RO rewards one-to-one with
\(H\) copies of each generating reward.
Writing \(\widehat f^{\RO}_1,\ldots,\widehat f^{\RO}_{KH}\)
for the fitted RO rewards, we define their error using the
same root-mean-square form as \eqref{eq:reward-error}:
\begin{equation}
D_f^{\RO}
:=
\left\{
\frac{1}{KH}
\min_{\sigma\in\mathfrak S_{KH}}
\sum_{k=1}^K\sum_{h=1}^H
\left\lVert
\widehat f^{\RO}_{\sigma((k-1)H+h)}-f_k^0
\right\rVert_\nu^2
\right\}^{1/2}.
\label{eq:ro-reward-error}
\end{equation}

Across the four generating models, RT has higher mean cell ARI
and lower mean reward error than RO
(Table~\ref{tab:rtro-structural}).
Mean occupancy error is also lower for RT in the three settings
with \(K=2\), but slightly lower for RO under \(\RT(3,2)\).
The occupancy comparison here uses mean errors;
Table~\ref{tab:rt-ro-gaps} reports medians and interquartile
intervals of paired differences.

\begin{table}[!htbp]
\centering
\caption{Recovery for \(\RT(K,H)\) and \(\RO(KH,1)\)
at \((n,T)=(300,1000)\). Entries are means over 50 paired
replications, with standard deviations in parentheses.
Reward errors use \eqref{eq:reward-error} for RT and
\eqref{eq:ro-reward-error} for RO.}
\label{tab:rtro-structural}
\scriptsize
\setlength{\tabcolsep}{3pt}
\renewcommand{\arraystretch}{1.12}
\begin{tabular}{@{}lcccccc@{}}
\toprule
& \multicolumn{2}{c}{Cell ARI}
& \multicolumn{2}{c}{Reward \(D_f\)}
& \multicolumn{2}{c}{Occupancy \(D_{\Occ}\)} \\
\cmidrule(lr){2-3}
\cmidrule(lr){4-5}
\cmidrule(lr){6-7}
Model & RT & RO & RT & RO & RT & RO \\
\midrule
$\RT(2,2)$ & .9737 (.0144) & .4998 (.0040) & .0170 (.0084) & .7067 (.0305) & .0423 (.0317) & .0590 (.0246) \\
$\RT(2,3)$ & .6394 (.0327) & .3341 (.0032) & .0175 (.0108) & .7828 (.0398) & .0366 (.0213) & .0455 (.0173) \\
$\RT(2,4)$ & .4599 (.0394) & .2508 (.0034) & .0175 (.0100) & .8360 (.0293) & .0455 (.0289) & .0499 (.0276) \\
$\RT(3,2)$ & .9221 (.1341) & .5711 (.0041) & .1050 (.1975) & .7867 (.0299) & .0762 (.0397) & .0747 (.0241) \\
\bottomrule
\end{tabular}
\end{table}
\FloatBarrier

\subsection{Additional model-selection results}
\label{app:bic-details}

\subsubsection{Generating parameters}
\label{app:sim-order-design}

Table~\ref{tab:sim-order-design} gives the mixture weights and
temperatures for the model-selection experiment in
Section~\ref{sec:selection}.
In \(W_{K,H}^0:=(\omega_{kh}^0)\), rows index reward classes
and columns index temperature classes.
The rewards follow the construction in
Appendix~\ref{app:sim-reward-spec}.

\begin{table}[!htbp]
\centering
\caption{Generating mixture weights and temperatures for model selection.
All mixture weights are positive. The \(\RT(2,3)\) temperatures differ
from those in the recovery and RT--RO experiments.}
\label{tab:sim-order-design}
\scriptsize
\setlength{\tabcolsep}{3pt}
\renewcommand{\arraystretch}{1.12}
\begin{tabular}{@{}lcc@{}}
\toprule
Generating model & \(W_{K,H}^0\) & \(\lambda_{1:H}^0\) \\
\midrule
\(\RT(1,1)\)
& \((1)\)
& \((0.8)\) \\
\addlinespace[4pt]
\(\RO(3,1)\)
& \(\begin{pmatrix}.45\\.35\\.20\end{pmatrix}\)
& \((1.0)\) \\
\addlinespace[4pt]
\(\RT(2,2)\)
& \(\begin{pmatrix}.34&.21\\.27&.18\end{pmatrix}\)
& \((0.5,1.5)\) \\
\addlinespace[4pt]
\(\RT(2,3)\)
& \(\begin{pmatrix}.24&.17&.10\\.19&.16&.14\end{pmatrix}\)
& \((0.4,1.0,1.8)\) \\
\addlinespace[4pt]
\(\RT(3,2)\)
& \(\begin{pmatrix}.23&.14\\.18&.12\\.20&.13\end{pmatrix}\)
& \((0.5,1.5)\) \\
\bottomrule
\end{tabular}
\end{table}

\subsubsection{BIC comparisons}

We compare each candidate order with the generating order
\((K^0,H^0)\) through the difference
\(\Delta_{\mathrm{BIC}}(K,H)
:=\mathrm{BIC}(K,H)-\mathrm{BIC}(K^0,H^0)\).
Both BIC values are calculated using \eqref{eq:bic} from models
fitted to the same training sample within each replication.
Table~\ref{tab:bic-gap} summarizes these differences.

\begin{table}[!htbp]
\centering
\caption{BIC differences from the generating order over 50 independent
Monte Carlo replications. Panels correspond to the generating models;
rows and columns give candidate reward and temperature orders.
Entries are medians, with standard deviations in parentheses.
Positive differences favor the generating order, marked in bold.}
\label{tab:bic-gap}
\scriptsize
\setlength{\tabcolsep}{3pt}
\renewcommand{\arraystretch}{1.12}

\textbf{(a) \(\RT(1,1)\)}\\[-2pt]
\begin{tabular}{@{}c*{4}{c}@{}}
\toprule
\(K\backslash H\) & \(1\) & \(2\) & \(3\) & \(4\) \\
\midrule
\(1\)
& \(\mathbf{0.000\;(0.000)}\)
& \(11.407\;(1.818)\)
& \(22.706\;(1.896)\)
& \(34.039\;(1.890)\) \\
\(2\)
& \(22.815\;(2.710)\)
& \(38.459\;(3.386)\)
& \(55.209\;(3.950)\)
& \(72.051\;(3.143)\) \\
\(3\)
& \(45.612\;(3.720)\)
& \(66.416\;(3.062)\)
& \(87.925\;(4.185)\)
& \(111.209\;(4.268)\) \\
\(4\)
& \(67.685\;(3.615)\)
& \(93.491\;(3.800)\)
& \(121.739\;(3.965)\)
& \(150.469\;(4.058)\) \\
\bottomrule
\end{tabular}

\vspace{0.7em}

\textbf{(b) \(\RO(3,1)\)}\\[-2pt]
\begin{tabular}{@{}c*{4}{c}@{}}
\toprule
\(K\backslash H\) & \(1\) & \(2\) & \(3\) & \(4\) \\
\midrule
\(1\)
& \(40771.388\;(2283.185)\)
& \(40782.796\;(2283.185)\)
& \(40794.204\;(2283.185)\)
& \(40805.611\;(2283.185)\) \\
\(2\)
& \(3472.587\;(645.095)\)
& \(3393.695\;(611.954)\)
& \(3416.531\;(596.240)\)
& \(3427.066\;(573.995)\) \\
\(3\)
& \(\mathbf{0.000\;(0.000)}\)
& \(19.922\;(487.464)\)
& \(41.965\;(691.743)\)
& \(64.463\;(487.366)\) \\
\(4\)
& \(22.815\;(487.700)\)
& \(48.438\;(487.460)\)
& \(76.367\;(487.394)\)
& \(104.325\;(487.355)\) \\
\bottomrule
\end{tabular}

\vspace{0.7em}

\textbf{(c) \(\RT(2,2)\)}\\[-2pt]
\begin{tabular}{@{}c*{4}{c}@{}}
\toprule
\(K\backslash H\) & \(1\) & \(2\) & \(3\) & \(4\) \\
\midrule
\(1\)
& \(52566.391\;(3432.421)\)
& \(52577.799\;(3432.421)\)
& \(52589.207\;(3432.421)\)
& \(52600.614\;(3432.421)\) \\
\(2\)
& \(1362.960\;(100.186)\)
& \(\mathbf{0.000\;(0.000)}\)
& \(15.896\;(1.996)\)
& \(32.541\;(2.532)\) \\
\(3\)
& \(1385.774\;(100.186)\)
& \(28.519\;(0.001)\)
& \(49.772\;(2.359)\)
& \(72.219\;(2.732)\) \\
\(4\)
& \(1408.589\;(100.186)\)
& \(57.038\;(0.001)\)
& \(84.077\;(2.084)\)
& \(112.321\;(2.749)\) \\
\bottomrule
\end{tabular}

\vspace{0.7em}

\textbf{(d) \(\RT(2,3)\)}\\[-2pt]
\begin{tabular}{@{}c*{4}{c}@{}}
\toprule
\(K\backslash H\) & \(1\) & \(2\) & \(3\) & \(4\) \\
\midrule
\(1\)
& \(55991.872\;(2440.134)\)
& \(56003.280\;(2440.134)\)
& \(56014.687\;(2440.134)\)
& \(56026.095\;(2440.134)\) \\
\(2\)
& \(1774.044\;(114.797)\)
& \(73.824\;(30.996)\)
& \(\mathbf{0.000\;(0.000)}\)
& \(15.440\;(1.854)\) \\
\(3\)
& \(1796.860\;(114.796)\)
& \(102.343\;(30.996)\)
& \(34.223\;(7.438)\)
& \(55.358\;(7.396)\) \\
\(4\)
& \(1819.674\;(114.796)\)
& \(130.861\;(30.996)\)
& \(68.445\;(7.438)\)
& \(95.018\;(7.377)\) \\
\bottomrule
\end{tabular}

\vspace{0.7em}

\textbf{(e) \(\RT(3,2)\)}\\[-2pt]
\begin{tabular}{@{}c*{4}{c}@{}}
\toprule
\(K\backslash H\) & \(1\) & \(2\) & \(3\) & \(4\) \\
\midrule
\(1\)
& \(37191.852\;(2953.855)\)
& \(37203.260\;(2953.855)\)
& \(37214.667\;(2953.855)\)
& \(37226.075\;(2953.855)\) \\
\(2\)
& \(5629.133\;(1503.432)\)
& \(4740.499\;(1518.047)\)
& \(4738.030\;(1512.595)\)
& \(4736.320\;(1494.417)\) \\
\(3\)
& \(1347.397\;(2122.059)\)
& \(\mathbf{0.000\;(0.000)}\)
& \(20.997\;(2235.592)\)
& \(42.584\;(2313.356)\) \\
\(4\)
& \(1362.697\;(1929.128)\)
& \(28.519\;(2079.600)\)
& \(54.898\;(1708.725)\)
& \(82.165\;(1566.842)\) \\
\bottomrule
\end{tabular}
\end{table}

All alternative candidate orders have positive median BIC
differences. These differences are especially large for candidates
with fewer reward classes than the generating model.

All selection errors involve one additional reward or temperature
class. Under \(\RO(3,1)\), BIC selects \(\RT(3,2)\) in one
replication. Under \(\RT(2,3)\), it selects \(\RT(3,3)\) in one
replication. Under \(\RT(3,2)\), it selects \(\RT(3,3)\) in three
replications and \(\RT(4,2)\) in two.

\FloatBarrier

\section{White-Stork Data Processing and Supplementary Results}
\label{app:stork-details}

This appendix gives the data preparation and model-fitting details for
Section~\ref{sec:white-stork-application}. It also defines the trajectory
summaries and specifies the regressions used to examine associations
with migration age and relative migration position.

\subsection{Data preparation and state space}
\label{app:stork-anchors}

We analyze observations from August~1 to October~31 in 2013--2023.
We retain records marked as visible in the source data, with valid
timestamps and coordinates, complete bird and tracking identifiers,
and no outlier flag. Duplicate records for the same bird and timestamp
are removed. For each date, we select the GPS location nearest
12:00 UTC within a window of $\pm30$ minutes. This daily location,
called an anchor, is missing if the window contains no record.

A bird--year enters the sample if it has at least 14 daily locations,
at least seven observed transitions between dates one day apart, and
locations on at least half the dates between its first and last
observations, inclusive. These criteria give 140 bird--years from
79 birds, with 10,532 daily locations and 10,343 observed transitions.

We project the locations onto a Lambert conformal conic coordinate
system and assign each to the $20$ km grid cell containing it.
Table~\ref{tab:stork-spatial-settings} gives the projection and grid
settings. The observed cells and intermediate grid locations along
observed daily transitions form a corridor containing 1,527 cells.
We add all cells within two grid widths in Chebyshev distance and
fill enclosed holes under eight-neighbor connectivity. This gives
the 7,893 states in Figure~\ref{fig:stork-state-space}.
The added cells define the state space; only observed transitions
enter the likelihood.

\begin{table}[!htbp]
\centering
\caption{Projection and grid used for the white-stork state space.
Projected distances are expressed in kilometres.}
\label{tab:stork-spatial-settings}
\small
\begin{tabular}{@{}ll@{}}
\toprule
Quantity & Setting \\
\midrule
Reference datum & WGS84 \\
Projection & Lambert conformal conic \\
Standard parallels & $17.25^\circ\mathrm N$, $46.5^\circ\mathrm N$ \\
Latitude of origin; central meridian & $32^\circ\mathrm N$; $2^\circ\mathrm W$ \\
False easting; false northing & $0$; $0$ \\
Grid origin $(x_0,y_0)$ & $(-1550,-2040)$ km \\
Cell width $w$ & $20$ km \\
\bottomrule
\end{tabular}
\end{table}

For bird--year $i$ on date $t$, write $X_{it}$ for its GPS location,
$\boldsymbol p_{it}$ for its projected location, and $S_{it}$ for
its grid state. Let $\mathcal O_i$ contain the dates $t$ for which
locations on both $t$ and $t+1$ are observed within the same tracking
deployment, and set $T_i:=|\mathcal O_i|$.

\subsection{Transition model and reward specification}
\label{app:stork-kernel}
\label{app:stork-features}

The transition model allows a directional action to cover several
grid cells in one day. Write $\boldsymbol x(s)$ for the center of
state $s$, in kilometres. An observed action is Stay if the two
daily locations occupy the same state. Otherwise, it is the compass
direction nearest to
$\boldsymbol x(S_{i,t+1})-\boldsymbol x(S_{it})$.
For each directional action $a$, let $\mathcal C_a$ be its
$45^\circ$ sector and define the displacement radius in grid units
as $q(s,s'):=\|\boldsymbol x(s')-\boldsymbol x(s)\|_\infty/w$.

We estimate radial weights by pooling observed transitions across
bird--years. Let $N_a(q)$ count transitions with direction $a$ and
radius $q$, and let $N_a:=\sum_q N_a(q)$. We smooth the directional
frequencies toward the pooled frequency $\widehat p_0(q)$ among all
non-Stay transitions, using
$\widehat p_a(q):=\{N_a(q)+10\widehat p_0(q)\}/(N_a+10)$.
Each possible destination receives the weight
$W_a(s,s'):=\widehat p_a\{q(s,s')\}
\mathbf 1\{\boldsymbol x(s')-\boldsymbol x(s)\in\mathcal C_a\}$.
Writing $Z_a(s):=\sum_{u\in\mathcal S}W_a(s,u)$, we obtain
\[
\widehat P(s'\mid s,a):=
\begin{cases}
W_a(s,s')/Z_a(s), & a\ne\mathrm{Stay},\ Z_a(s)>0,\\
\mathbf 1\{s'=s\}, & a=\mathrm{Stay}
\text{ or }[a\ne\mathrm{Stay},\ Z_a(s)=0].
\end{cases}
\]
Destinations in the same sector and at the same radius receive equal
weights, so their total probability also depends on the number of
available states. A directional action with no destination of positive
weight leaves the state unchanged but retains its directional label.
The kernel is held fixed throughout mixture fitting and model selection.

The reward features describe spatial preferences and an additional
preference for Stay. Write $\boldsymbol x(s)=(x_s,y_s)^\top$, and
standardize each coordinate by its mean and standard deviation under
the uniform distribution on $\mathcal S$. Denote the standardized
coordinates by $\widetilde x_s$ and $\widetilde y_s$. The raw features
are the conditional first and second moments of the next-day
coordinates, together with a Stay indicator:
\[
\psi(s,a):=
\begin{pmatrix}
\displaystyle\sum_{s'}\widehat P(s'\mid s,a)
\bigl(\widetilde x_{s'},\widetilde x_{s'}^2,
      \widetilde y_{s'},\widetilde y_{s'}^2\bigr)^\top\\[3pt]
\mathbf 1\{a=\mathrm{Stay}\}
\end{pmatrix}.
\]

We center these features over the nine actions at each state,
$\widetilde\psi(s,a):=\psi(s,a)-9^{-1}\sum_{b\in\mathcal A}\psi(s,b)$,
and whiten them under uniform state--action weights:
\[
\Sigma:=\frac{1}{9|\mathcal S|}\sum_{s,a}
\widetilde\psi(s,a)\widetilde\psi(s,a)^\top,
\qquad
\phi(s,a):=\Sigma^{-1/2}\widetilde\psi(s,a),
\]
where $\Sigma^{-1/2}$ is the symmetric inverse square root.
These features satisfy \eqref{eq:linear-feature-normalization} with
$\omega_s(a)=1/9$ and $\nu(s,a)=1/(9|\mathcal S|)$.
We fit $f_k(s,a)=\phi(s,a)^\top\beta_k$ with $\|\beta_k\|_2=1$,
as in \eqref{eq:linear-reward-class}. The feature map remains fixed
across candidate models.

\subsection{Likelihood and model fitting}
\label{app:stork-segments}
\label{app:stork-fitting}

Each bird--year may contain several observed segments, separated by
missing dates or deployment boundaries. Let $\mathcal E_i$ denote
these segments, each a maximal sequence of daily locations recorded
one day apart within a deployment. We condition on each segment's
starting state and assume that all segments of a bird--year share
one latent reward--temperature pair. Its component likelihood is
\begin{equation}
L_{i,kh}(\vartheta):=
\prod_{t\in\mathcal O_i}
\pi^*_{f_k,\lambda_h}(A_{it}\mid S_{it})
\widehat P(S_{i,t+1}\mid S_{it},A_{it}).
\label{eq:stork-segment-likelihood}
\end{equation}
No transition is inserted across a gap or deployment boundary.
The working likelihood treats bird--years as independent units
and uses these component likelihoods in
\eqref{eq:observed-loglikelihood}, with $n=140$.
GEM uses the undiscounted action counts over $\mathcal O_i$;
the discount factor $\gamma=0.95$ enters the policy calculation.

The shared transition factors cancel from the posterior probabilities
in \eqref{eq:gem-responsibility}. For the selected fit, write
$r_{i,kh}$ for these probabilities and define
$\rho_{ik}:=\sum_h r_{i,kh}$.
The marginal reward assignment
$\widehat C_i:=\operatorname*{arg\,max}_k\rho_{ik}$ determines
the trajectory groups in Figure~\ref{fig:stork-reward-maps}.
The joint MAP assignments in Table~\ref{tab:stork-temperature}
maximize $r_{i,kh}$ over both $k$ and $h$.

We fit the candidate orders in Section~\ref{sec:stork-model-selection}
using GEM and the initialization procedure in
Appendix~\ref{app:gem-computation}.
The initialization bank contains one start for $(K,H)=(1,1)$,
four for $K=1,H>1$, 13 for $K>1,H=1$, and 37 for $K,H>1$.
After 20 GEM cycles, we retain the two starts with the largest
likelihoods, or the single available start for $(1,1)$, and continue
from their current estimates. For each candidate order, we select
the converged fit with the largest likelihood.

\begin{table}[!htbp]
\centering
\caption{Constraints and convergence settings for the white-stork fits.
The weight bound applies at update $m=0,1,\ldots$, with
$(z)_+:=\max(z,0)$. All three convergence criteria must hold for
the specified number of consecutive cycles.}
\label{tab:stork-fitting-settings}
\small
\begin{tabular}{@{}ll@{}}
\toprule
Quantity & Setting \\
\midrule
Temperature range & $0.3\le\lambda_h\le3.0$ \\
Weight lower bound at update $m$
& $\max\{10^{-4},\,0.5(1-m/50)_+/(KH)\}$ \\
Relative change in the Q-function & $10^{-5}$ \\
Relative change in the policy log likelihood & $10^{-5}$ \\
Maximum scaled KKT residual & $10^{-2}$ \\
Consecutive cycles required & $5$ \\
Maximum total GEM cycles & $1,000$ \\
\bottomrule
\end{tabular}
\end{table}

Table~\ref{tab:stork-fitting-settings} gives the numerical settings.
For convergence assessment, both objectives omit the shared transition
terms. The Q-function at each iteration uses posterior probabilities
evaluated at that iteration's parameter estimate. Each relative change
divides the absolute increment by the larger of one and the absolute
previous value. We use the KKT residuals in
Appendix~\ref{app:gem-numerics}, dividing each reward and
log-temperature residual by the posterior mass of its corresponding
class. The weight residual is evaluated under the baseline bound
$10^{-4}$.

For BIC in \eqref{eq:bic}, the sample size is the number of bird--years,
$n=140$, and the five reward features give
$d_{K,H}=4K+H+KH-1$.
The shared transition contribution is the same for all candidate
orders and therefore does not affect the BIC differences reported
in Table~\ref{tab:stork-bic}.

\subsection{Trajectory summaries and reference policies}
\label{app:stork-metrics}

We calculate movement summaries for each bird--year before averaging
them within reward or reward--temperature groups.
Let $d$ denote great-circle distance in kilometres between GPS
locations. Mean daily displacement and the Stay proportion are
\[
M_i:=\frac{1}{T_i}\sum_{t\in\mathcal O_i}d(X_{it},X_{i,t+1}),
\qquad
S_i:=\frac{1}{T_i}\sum_{t\in\mathcal O_i}
\mathbf 1\{A_{it}=\mathrm{Stay}\}.
\]
Stay records whether two daily locations occupy the same grid cell,
so a Stay action can still have positive GPS displacement.

Path directness compares the sum of the net distances within observed
segments with the total distance traveled:
\[
D_i:=\frac{\displaystyle\sum_{e\in\mathcal E_i}
d(X_{i,e,\mathrm{first}},X_{i,e,\mathrm{last}})}
{\displaystyle\sum_{t\in\mathcal O_i}d(X_{it},X_{i,t+1})},
\]
where $X_{i,e,\mathrm{first}}$ and $X_{i,e,\mathrm{last}}$ are
the first and last GPS locations of segment $e$.
Larger values indicate more direct paths between segment endpoints.

Directional concentration measures agreement among the non-Stay
action directions. Let $\chi(a)$ denote the central angle of compass
direction $a$, and let
$N_i^{\mathrm{move}}:=\sum_{t\in\mathcal O_i}
\mathbf 1\{A_{it}\ne\mathrm{Stay}\}$.
The concentration is
\[
C_i:=\left\|\frac{1}{N_i^{\mathrm{move}}}
\sum_{\substack{t\in\mathcal O_i\\A_{it}\ne\mathrm{Stay}}}
\begin{pmatrix}\cos\chi(A_{it})\\\sin\chi(A_{it})\end{pmatrix}
\right\|_2,
\]
with larger values indicating more closely aligned directions.
The minimum-latitude summary uses all retained daily locations,
and the final-location indicator records whether the last location
lies north of $35^\circ\mathrm N$.

For any bird--year summary $g_i$, we calculate the posterior-weighted
means
\begin{equation}
\overline g_{kh}:=\frac{\sum_i r_{i,kh}g_i}{\sum_i r_{i,kh}},
\qquad
\overline g_k:=\frac{\sum_i\rho_{ik}g_i}{\sum_i\rho_{ik}}.
\label{eq:stork-weighted-summaries}
\end{equation}
Both sums in each ratio include only bird--years for which $g_i$ is
defined. Path directness requires positive total distance, and
directional concentration requires at least one non-Stay transition.
These means use posterior probabilities, whereas the MAP counts
report discrete assignments. The three groups marked by
$\dagger$ in Table~\ref{tab:stork-temperature} have no joint MAP
assignments and each has total posterior mass below $0.01$ bird--years.

\paragraph{Policies at a common temperature.}
\label{app:stork-displays}
Figure~\ref{fig:stork-reward-maps} compares the fitted rewards at
$\lambda_\star=\widehat\lambda_2=0.801$.
Writing $\pi_{k,\star}:=\pi^*_{\widehat f_k,\lambda_\star}$,
the Stay probability and expected daily displacement vector are
\begin{equation}
\begin{aligned}
p_k^{\mathrm{stay}}(s)
&:=\pi_{k,\star}(\mathrm{Stay}\mid s),\\
\boldsymbol v_k(s)
&:=\sum_a\pi_{k,\star}(a\mid s)
\sum_{s'}\widehat P(s'\mid s,a)
\{\boldsymbol x(s')-\boldsymbol x(s)\}.
\end{aligned}
\label{eq:stork-reference-policy-maps}
\end{equation}
The vectors average displacements between grid-cell centers over
actions and destinations and are expressed in km/day.

\subsection{Association with migration age}
\label{app:stork-biology}
\label{app:stork-age}

All 79 birds were tagged as nestlings. Migration age, denoted by
$A_i$, is migration year minus birth year, so age zero identifies
the first autumn migration. The sample contains 79 first and
61 later autumn migrations.
Table~\ref{tab:stork-age-summary} summarizes age and first-migration
status within each reward class using
\eqref{eq:stork-weighted-summaries}.

\begin{table}[!htbp]
\centering
\caption{Migration age by reward class. Means and proportions are
weighted by posterior reward probabilities $\rho_{ik}$.}
\label{tab:stork-age-summary}
\small
\setlength{\tabcolsep}{9pt}
\renewcommand{\arraystretch}{1.10}
\begin{tabular}{@{}lrrrr@{}}
\toprule
Quantity & Reward 1 & Reward 2 & Reward 3 & Reward 4 \\
\midrule
Mean migration age (years) & 0.17 & 0.96 & 2.40 & 1.02 \\
First-migration proportion (\%) & 93.5 & 70.2 & 33.6 & 70.5 \\
\bottomrule
\end{tabular}
\end{table}

For each reward class, We examine the association between migration age and reward-class
membership by fitting a separate linear regression for each class:
\begin{equation}
\rho_{ik}
=
\alpha_k+\beta_k^{\mathrm{age}}A_i
+\boldsymbol\eta_k^\top\boldsymbol z_i+\varepsilon_{ik},
\qquad k=1,\ldots,4,
\label{eq:stork-age-regression}
\end{equation}
where $A_i$ is migration age and $\boldsymbol z_i$ contains
indicators for calendar year and sex.
The coefficient $\beta_k^{\mathrm{age}}$ measures the adjusted
difference in mean posterior reward probability per year of age.
We estimate the coefficients by ordinary least squares, giving
each bird--year equal weight.

For both association analyses, we report $95\%$ confidence intervals
using standard errors clustered by bird.
The fitted RT model and front scores are treated as fixed
in these calculations.
P-values for the eight age and first-migration tests are adjusted
using the Benjamini--Hochberg procedure.

\subsection{Temperature and relative migration position}
\label{app:stork-position}

The front score in Section~\ref{sec:stork-action-regularity} requires
a common migration direction for each reward class.
We estimate this direction from the net displacements within observed
segments. Define
\[
\boldsymbol d_i:=\sum_{e\in\mathcal E_i}
\bigl(\boldsymbol p_{i,e,\mathrm{last}}
      -\boldsymbol p_{i,e,\mathrm{first}}\bigr),
\qquad
\boldsymbol u_k:=
\frac{\sum_i\rho_{ik}\boldsymbol d_i}
     {\|\sum_i\rho_{ik}\boldsymbol d_i\|_2},
\]
where the two endpoint vectors are projected GPS locations.
The unit vector $\boldsymbol u_k$ gives the direction of net movement
averaged across bird--years with weights $\rho_{ik}$.

We rank projected positions
$\boldsymbol p_{it}^{\top}\boldsymbol u_{\widehat C_i}$
among birds in the same marginal MAP reward class on the same
calendar date. Only dates with at least three such birds, including
the focal bird, contribute to the front score. Using the notation
of Section~\ref{sec:stork-action-regularity}, we therefore require
$n_{it}\ge3$ on each comparison date and retain bird--years with
$m_i\ge10$. This leaves 121 bird--years from 72 birds.
Ranks, ties, and the averaging of daily percentiles follow the
definition of $F_i$ in that section.

We examine the association between temperature and relative
migration position using the posterior mean temperature
$\overline\lambda_i:=\sum_{k=1}^{4}\sum_{h=1}^{3}
r_{i,kh}\widehat\lambda_h$.
The regression model is
\begin{equation}
F_i
=
\alpha+\tau^{front}\overline\lambda_i
+\boldsymbol\zeta^\top\boldsymbol v_i+\epsilon_i,
\label{eq:stork-front-regression}
\end{equation}
where $\boldsymbol v_i$ contains indicators for reward class,
calendar year, and sex, together with migration age $A_i$ and
first-migration status $\mathbf 1\{A_i=0\}$.
The coefficient $\tau^{front}$ measures the adjusted difference in mean
front score per unit of posterior mean temperature.
We estimate the coefficients by ordinary least squares, giving
each bird--year equal weight.

The reported $p$-value is adjusted using the Benjamini--Hochberg
procedure across 18 tests of temperature associations with
relative migration position.

\FloatBarrier

\section{Continuous-State Reward--Temperature IRL with Finite Actions}
\label{app:continuous-state}

Movement data often record continuous spatial coordinates, and replacing
these coordinates by grid cells can obscure variation within cells.
Continuous-state IRL allows such observations to retain their original
state representation \citep{dexter2021,finn2016}.
We extend the reward--temperature model to this setting while retaining
finitely many actions and discrete decision times.
The extension preserves the shared reward and temperature structure,
with numerical Bellman solutions providing a basis for GEM estimation.

\subsection{Continuous-state model}
\label{app:continuous-mdp}

Let \(\mathcal S\subseteq\mathbb R^d\) be compact and let
\(\mathcal A:=\{1,\ldots,A\}\), where \(A\geq2\).
We retain the notation of the main text, with \(P\) now denoting
a transition kernel and \(\mu_0\) an initial-state probability measure.
Both are fixed and common to all mixture components.
For a continuous function \(v\), write
\(P_av(s):=\int_{\mathcal S}v(s')P(ds'\mid s,a)\).
Assume that \(P\) is Feller, meaning that \(P_av\) is continuous
whenever \(v\) is continuous, and fix \(0\leq\gamma<1\).

For a continuous reward \(f\) and temperature \(\lambda>0\),
the entropy-regularized Bellman equations are
\begin{align}
  Q^*_{f,\lambda}(s,a)
  &=f(s,a)+\gamma P_aV^*_{f,\lambda}(s),
  \label{eq:continuous-soft-Q}\\
  V^*_{f,\lambda}(s)
  &=\lambda\log\sum_{b\in\mathcal A}
    \exp\!\left\{\frac{Q^*_{f,\lambda}(s,b)}{\lambda}\right\}.
  \label{eq:continuous-soft-V}
\end{align}
The Bellman operator is a \(\gamma\)-contraction on
\(C(\mathcal S)\), so these equations have a unique continuous
value function \citep{geist2019}.
The corresponding optimal policy is
\begin{equation}
  \pi^*_{f,\lambda}(a\mid s)
  =
  \exp\!\left\{
    \frac{Q^*_{f,\lambda}(s,a)-V^*_{f,\lambda}(s)}{\lambda}
  \right\},
  \label{eq:continuous-soft-policy}
\end{equation}
which is continuous in \(s\) and strictly positive.

Given the current state, an action is drawn from
\(\pi^*_{f,\lambda}\), followed by a transition under \(P\).
Thus a trajectory
\(\tau=(s_0,a_0,\ldots,a_{T-1},s_T)\) has law
\begin{equation}
  d\PP_{f,\lambda}^{(T)}(\tau)
  =
  \mu_0(ds_0)
  \prod_{t=0}^{T-1}
  \pi^*_{f,\lambda}(a_t\mid s_t)
  P(ds_{t+1}\mid s_t,a_t).
  \label{eq:continuous-trajectory-kernel}
\end{equation}
This formulation does not require transition densities.

The crossed model combines \(K\) rewards and \(H\) temperatures.
Each trajectory has a latent class \(Z_i=(k,h)\), with probability
\(\omega_{kh}\), and follows
\(\PP_{f_k,\lambda_h}^{(T_i)}\) conditional on that class.
The weights satisfy \(\omega_{kh}\geq0\) and
\(\sum_{k,h}\omega_{kh}=1\).
As in the main text, \(f_k\) is shared across the \(H\) temperature
classes, and \(\lambda_h\) is shared across the \(K\) reward classes.
Independent trajectories therefore follow the mixture law
\(\PP_{\vartheta}^{(T_i)}
=\sum_{k,h}\omega_{kh}\PP_{f_k,\lambda_h}^{(T_i)}\).

\subsection{Normalized rewards and identification}
\label{app:continuous-canonical}

As in Section~\ref{sec:normalized-reward-class}, we impose statewise
centering to remove the potential ambiguity and unit norm to fix
the common reward--temperature scale.
Choose fixed, continuous weights \(\omega_s(a)>0\) satisfying
\(\sum_a\omega_s(a)=1\), and a probability measure \(\eta\)
with full topological support on \(\mathcal S\).
Here \(\omega_s(a)\) are the centering weights, distinct from
the mixture weights \(\omega_{kh}\).
Define \(\nu(ds,a):=\omega_s(a)\eta(ds)\) and
\(\|f\|_\nu^2
:=\int_{\mathcal S}\sum_a\omega_s(a)f(s,a)^2\eta(ds)\).
The continuous-state reward class is
\begin{equation}
  \mathcal F_{\mathrm{sn}}^{\mathrm c}
  :=
  \left\{
    f\in C(\mathcal S\times\mathcal A):
    \sum_a\omega_s(a)f(s,a)=0\ \text{for every }s,
    \quad \|f\|_\nu=1
  \right\}.
  \label{eq:continuous-Fsn}
\end{equation}
The superscript \(\mathrm c\) distinguishes this class from
the finite-state reward class used in the main text.

\begin{proposition}[Single-component identification]
\label{prop:continuous-inversion}
Under the assumptions above, fix the known transition kernel \(P\)
and discount factor \(\gamma\).
For \(f,\widetilde f\in\mathcal F_{\mathrm{sn}}^{\mathrm c}\)
and \(\lambda,\widetilde\lambda>0\), equality of the optimal policies
at every state and action implies
\((f,\lambda)=(\widetilde f,\widetilde\lambda)\).
If \(\mu_0\) also has full topological support on \(\mathcal S\),
then, for any \(T\geq1\), equality of the population trajectory laws
\(\PP_{f,\lambda}^{(T)}
=\PP_{\widetilde f,\widetilde\lambda}^{(T)}\)
implies the same conclusion.
\end{proposition}

This proposition identifies the reward and temperature within a single
component. Identification of the full continuous-state mixture
additionally requires conditions ensuring uniqueness of its
mixture decomposition.

\subsection{GEM estimation}
\label{app:continuous-mixture-algorithm}

We estimate normalized linear rewards and approximate transition
expectations by a fixed quadrature rule.
Write \(f_\beta(s,a):=\phi(s,a)^{\mathsf T}\beta\), using continuous
features that satisfy \eqref{eq:linear-feature-normalization}
with state sums replaced by integrals.
Then \(\|\beta\|_2=1\) ensures that
\(f_\beta\in\mathcal F_{\mathrm{sn}}^{\mathrm c}\).

Choose quadrature nodes
\(x_1,\ldots,x_{N_{\mathrm q}}\in\mathcal S\) and
nonnegative weights \(w_j(s,a)\) satisfying
\(\sum_{j=1}^{N_{\mathrm q}} w_j(s,a)=1\).
The weights approximate the transition law and remain fixed
throughout estimation.
Define
\begin{equation}
  P_a^{\mathrm q}v(s)
  :=
  \sum_{j=1}^{N_{\mathrm q}} w_j(s,a)v(x_j).
  \label{eq:continuous-quadrature}
\end{equation}
The superscript \(\mathrm q\) distinguishes this numerical operator
from the transition operator \(P_a\).

Let \(V_{\beta,\lambda}^{\mathrm q}\) solve
\eqref{eq:continuous-soft-Q}--\eqref{eq:continuous-soft-V}
with reward \(f_\beta\) and \(P_a^{\mathrm q}\) in place of \(P_a\).
Its values at the \({N_{\mathrm q}}\) nodes can be computed by value iteration.
The fitted action values and policy are
\begin{equation}
\begin{aligned}
  \widehat q_{f_\beta,\lambda}(s,a)
  &:=
  f_\beta(s,a)+\gamma P_a^{\mathrm q}
  V_{\beta,\lambda}^{\mathrm q}(s),\\
  \pi^{\mathrm{fit}}_{f_\beta,\lambda}(a\mid s)
  &:=
  \frac{\exp\{\widehat q_{f_\beta,\lambda}(s,a)/\lambda\}}
       {\sum_{b\in\mathcal A}
        \exp\{\widehat q_{f_\beta,\lambda}(s,b)/\lambda\}}.
\end{aligned}
\label{eq:continuous-fitted-policy}
\end{equation}

For trajectory \(i\), define the component log-policy contribution
\(\ell_{i,kh}^{\pi}(\vartheta)
:=\sum_{t=0}^{T_i-1}
\log\pi^{\mathrm{fit}}_{f_{\beta_k},\lambda_h}
(a_{it}\mid s_{it})\).
At iteration \(m\), the E-step gives
\begin{equation}
  r_{i,kh}^{\mathrm{fit},(m)}
  :=
  \frac{
    \omega_{kh}^{(m)}
    \exp\{\ell_{i,kh}^{\pi}(\vartheta^{(m)})\}
  }{
    \sum_{k',h'}\omega_{k'h'}^{(m)}
    \exp\{\ell_{i,k'h'}^{\pi}(\vartheta^{(m)})\}
  }.
  \label{eq:continuous-responsibility}
\end{equation}
These probabilities remain fixed throughout the M-step.
The numerical observed and conditional log likelihoods,
omitting the common initial-state and transition factors, are
\begin{equation}
\begin{aligned}
  \ell_n^{\mathrm{red}}(\vartheta)
  &:=
  \sum_i\log\left\{
    \sum_{k,h}\omega_{kh}
    \exp\{\ell_{i,kh}^{\pi}(\vartheta)\}
  \right\},\\
  Q_n^{\mathrm{red}}(\vartheta\mid\vartheta^{(m)})
  &:=
  \sum_{i,k,h}r_{i,kh}^{\mathrm{fit},(m)}
  \left\{\log\omega_{kh}
  +\ell_{i,kh}^{\pi}(\vartheta)\right\}.
\end{aligned}
\label{eq:continuous-reduced-objectives}
\end{equation}
Maximizing over the mixture weights gives
\begin{equation}
  \omega_{kh}^{(m+1)}
  =
  \frac{1}{n}\sum_{i=1}^n r_{i,kh}^{\mathrm{fit},(m)}.
  \label{eq:continuous-weight-update}
\end{equation}

\paragraph{Reward updates.}
We first differentiate with respect to the unrestricted coefficient
vector and then impose the unit-norm constraint in the update.
At the current parameter values, write
\(\pi_{kh}^{\mathrm{fit}}
:=\pi^{\mathrm{fit}}_{f_{\beta_k},\lambda_h}\) and define
\(U_{kh}(s)
:=\left.\nabla_\beta V_{\beta,\lambda_h}^{\mathrm q}(s)
\right|_{\beta=\beta_k}\).
The value derivative satisfies
\begin{equation}
  U_{kh}(s)
  =
  \sum_{a\in\mathcal A}\pi_{kh}^{\mathrm{fit}}(a\mid s)
  \left\{\phi(s,a)+\gamma P_a^{\mathrm q}U_{kh}(s)\right\}.
  \label{eq:continuous-value-gradient}
\end{equation}
At the quadrature nodes, this is a linear system with an
\(N_{\mathrm q}\times N_{\mathrm q}\) coefficient matrix.
The derivative at each observed state then follows from the
same equation.

The contribution of component \((k,h)\) to the coefficient gradient is
\begin{equation}
  g_{kh}^{\beta}
  :=
  \frac{1}{\lambda_h}
  \sum_{i=1}^n r_{i,kh}^{\mathrm{fit},(m)}
  \sum_{t=0}^{T_i-1}
  \left\{
    \phi(s_{it},a_{it})
    +\gamma P_{a_{it}}^{\mathrm q}U_{kh}(s_{it})
    -U_{kh}(s_{it})
  \right\}.
  \label{eq:continuous-cell-gradient}
\end{equation}
Because reward \(k\) is shared across temperatures, its gradient
sums over all \(H\) temperature classes.
We project this gradient onto the tangent space of the unit sphere
and normalize the candidate:
\begin{equation}
\begin{aligned}
  g_k^\beta
  &:=\sum_{h=1}^H g_{kh}^{\beta},
  \qquad
  d_k^\beta
  :=(I_p-\beta_k\beta_k^{\mathsf T})g_k^\beta,\\
  \beta_k^{\mathrm{trial}}(\alpha)
  &:=
  \frac{\beta_k+\alpha d_k^\beta}
       {\|\beta_k+\alpha d_k^\beta\|_2},
  \qquad \alpha>0.
\end{aligned}
\label{eq:continuous-reward-update}
\end{equation}

\paragraph{Temperature updates.}
Write \(u_h:=\log\lambda_h\) and restrict \(u_h\) to a fixed
interval \([u_{\min},u_{\max}]\).
For linear rewards, simultaneous scaling of \(\beta\) and
\(\lambda\) leaves the fitted policy unchanged.
The log-temperature gradient therefore follows from the
coefficient gradients in \eqref{eq:continuous-cell-gradient}.
Summing over the \(K\) rewards that share temperature \(h\), we obtain
\begin{equation}
\begin{aligned}
  g_h^u
  &:=-\sum_{k=1}^K\beta_k^{\mathsf T}g_{kh}^{\beta},\\
  u_h^{\mathrm{trial}}(\alpha)
  &:=
  \Pi_{[u_{\min},u_{\max}]}\{u_h+\alpha g_h^u\},
  \qquad
  \lambda_h^{\mathrm{trial}}(\alpha)
  :=\exp\{u_h^{\mathrm{trial}}(\alpha)\}.
\end{aligned}
\label{eq:continuous-temperature-update}
\end{equation}
Here \(\Pi_{[u_{\min},u_{\max}]}\) denotes projection onto the
interval.
Temperature gradients are recomputed using the updated rewards,
with the E-step probabilities still fixed.

For each reward or temperature update, choose the step size by
backtracking.
Recompute the affected policies for each candidate and accept it
when
\begin{equation}
  Q_n^{\mathrm{red}}
  (\vartheta^{\mathrm{trial}}\mid\vartheta^{(m)})
  \geq
  Q_n^{\mathrm{red}}
  (\vartheta^{\mathrm{cur}}\mid\vartheta^{(m)}),
  \label{eq:continuous-block-acceptance}
\end{equation}
where \(\vartheta^{\mathrm{cur}}\) includes earlier accepted updates
in the same iteration.
Retain the current block if no candidate is accepted.
This criterion makes the numerical log likelihood
\(\ell_n^{\mathrm{red}}\) nondecreasing for the fixed quadrature
approximation.

\begin{algorithm}[htbp]
\caption{GEM with fixed quadrature for a continuous-state
\(\RT(K,H)\) model}
\label{alg:continuous-mixture}
\small

\textbf{Input:} trajectories, order \((K,H)\), normalized features,
a fixed quadrature rule, and log-temperature bounds.

\textbf{Initialize:} unit-norm reward coefficients, admissible
temperatures, and positive mixture weights summing to one.

\begin{enumerate}[leftmargin=1.6em,itemsep=3pt,topsep=4pt]

\item Evaluate the policies using
\eqref{eq:continuous-fitted-policy}.

\item Apply \eqref{eq:continuous-responsibility} and
\eqref{eq:continuous-weight-update}.

\item For each reward \(k\), apply
\eqref{eq:continuous-value-gradient}--\eqref{eq:continuous-reward-update}
and \eqref{eq:continuous-block-acceptance}.

\item For each temperature \(h\), apply
\eqref{eq:continuous-value-gradient}--\eqref{eq:continuous-cell-gradient},
\eqref{eq:continuous-temperature-update}, and
\eqref{eq:continuous-block-acceptance}.

\item Repeat Steps 1--4 until changes in
\(\ell_n^{\mathrm{red}}\) and the parameters satisfy the
prescribed tolerances.

\end{enumerate}

\textbf{Output:} fitted rewards, temperatures, mixture weights,
and posterior probabilities evaluated at the final estimates.
\end{algorithm}

\end{document}